# A Disease-Centric Vision-Language Foundation Model for Precision Oncology in Kidney Cancer

Yuhui Tao[1,2†], Zhongwei Zhao[3†], Zilong Wang[4†], Xufang Luo[4†], Feng Chen[5†], Kang Wang[1,2], Chuanfu Wu[6,7], Xue Zhang[8], Shaoting Zhang[9], Jiaxi Yao[10], Xingwei Jin[11], Xinyang Jiang[4], Yifan Yang[4], Dongsheng Li[4], Lili Qiu[4], Zhiqiang Shao[12*], Jianming Guo[13*], Nengwang Yu[3*], Ying Xiong[13,14*], Shuo Wang[1,2*]

[1*]Digital Medical Research Center, School of Basic Medical Sciences, Fudan University, Shanghai, 200032, China.

[2*]Shanghai Key Laboratory of Medical Imaging Computing and Computer Assisted Intervention, Shanghai, 200032, China.

[3*]Department of Urology, Qilu Hospital of Shandong University, Jinan, Shandong, 250012, China.

[4]Microsoft Research Asia, Shanghai, 200232, China.

[5]Department of Radiology, The First Affiliated Hospital, Zhejiang University School of Medicine, Hangzhou, 310006, China.

[6]Center of Health data science, Linyi People's Hospital, Shandong, 276003, China.

[7]Shandong Open Laboratory of Data Innovation Application, Shandong, 276003, China.

[8]Department of Radiology, the First People's Hospital of Lianyungang, Lianyungang, 222002, China.

[9]Department of Radiology, Shandong Cancer Hospital and Institute, Shandong First Medical University and Shandong Academy of Medical Sciences, Jinan, 250117, China.

[10]Department of Urology, Zhangye People's Hospital affiliated to Hexi University, Zhangye, 734000, China.

[11]Department of Urology, Ruijin Hospital, Shanghai Jiao Tong University School of Medicine, Shanghai, 200025, China.

[12*]Department of Urology, Linyi People's Hospital, Shandong, 276003, China.

[13*]Department of Urology, Zhongshan Hospital, Fudan University, Shanghai, 200032, China.

[14*]Department of Urology, Xiamen Branch, Zhongshan Hospital, Fudan University, Xiamen, 361000, China.

*Corresponding author(s). E-mail(s): shaozq2005@163.com; guo.jianming@zs-hospital.sh.cn; qiluyunengwang@hotmail.com; xiong.ying@zs-hospital.sh.cn; shuowang10@fudan.edu.cn;

†These authors contributed equally to this work.

**Abstract**

The non-invasive assessment of renal masses remains a critical challenge in urologic oncology, where diagnostic uncertainty frequently causes overtreatment. Here, we develop RenalCLIP, a vision-language foundation model for precision oncology in kidney cancer. Utilizing 27,866 computed tomography scans from 8,809 patients across diverse multi-center cohorts, we employ a two-stage pre-training strategy to align domain-specific visual and textual representations. RenalCLIP achieves enhanced performance and generalizability across ten core clinical tasks, spanning anatomical assessment, diagnostic classification, and survival prediction, significantly outperforming state-of-the-art general-purpose foundation models. Furthermore, RenalCLIP demonstrates strong data efficiency in diagnostic classification, achieving peak baseline performance using only 20% of the training data. The model also exhibits robust zero-shot diagnostic capabilities, effective image-text retrieval, and high-quality medical report generation. Our findings establish RenalCLIP as a powerful, generalizable tool to enhance diagnostic precision, refine prognostic stratification, and personalize the management of renal masses.

## Introduction

The widespread adoption of medical imaging has dramatically increased the incidental detection of renal masses, posing new clinical dilemmas in kidney cancer management. Despite a surge in radiologically identified renal tumors, kidney cancer-specific mortality has not declined in parallel, underscoring a substantial burden of overtreatment for benign or indolent lesions[1,2]. Notably, up to 20% of surgically resected renal masses are ultimately confirmed as benign, subjecting patients to unnecessary surgical risks, healthcare costs, and potential loss of renal function[3,4]. This clinical dilemma is most pronounced for small renal masses (≤ 4 cm), driven by the size-dependent prevalence of malignancy. As demonstrated by Bhindi et al.[4], the probability of malignancy is 68% for tumors ≤ 1 cm and steadily increases to over 90% for those > 4 cm. Therefore, active surveillance is primarily considered for smaller masses and is rarely contemplated for larger tumors (> 4 cm) unless dictated by limited life expectancy or high surgical risk. Current diagnostic pathways are limited: radiological interpretation remains highly subjective with considerable inter-observer variability[5], while percutaneous biopsy is invasive, carries a significant non-diagnostic rate, and frequently fails to capture the true biological aggressiveness of tumors due to intratumoral heterogeneity[6–9]. A critical unmet need therefore exists for a non-invasive tool that can precisely and comprehensively assess a tumor's malignant potential, aggressiveness, and prognosis from preoperative CT scans[10-16].

While artificial intelligence (AI) has shown promise in medical image analysis, conventional models for kidney cancer have been constrained by critical limitations that have prevented their clinical translation. Most are trained on image data alone, overlooking the rich semantic context within radiology reports that is essential for expert-level reasoning[17]. Furthermore, the prevailing one-model, one-task paradigm is inefficient and results in models that lack generalizability, often failing when tested on external data[18,19]. Recent breakthroughs in general-purpose vision-language foundation models[20–24] offer a path forward but possess a critical flaw for oncology: their broad training across all pathologies lacks the necessary depth for the nuanced, high-stakes decisions required in cancer care[19,25–28]. A general CT model may identify an organ, but it cannot be expected to master the subtle radiological features that distinguish an indolent from an aggressive renal tumor. We reasoned that to solve a specific clinical problem, a disease-centric strategy is essential. Such an approach, which revolves entirely around the clinical and pathological complexities of a single disease, is necessary to bridge the gap between generalist AI and the demands of precision medicine.

Here, we present RenalCLIP, a multi-task, disease-centric vision-language foundation model designed for comprehensive assessment of kidney cancer. RenalCLIP is pre-trained on a large-scale, multi-institutional dataset comprising 21,819 CT series and their corresponding radiology reports from 6,867 patients, collected across four leading medical centers in China. By aligning visual representations from CT scans with clinical-pathological semantics extracted from textual reports, RenalCLIP learns robust, context-aware embeddings of renal tumors. We systematically evaluate its performance across a broad spectrum of clinically relevant tasks, including automated anatomical scoring, benign-versus-malignant discrimination, aggressiveness stratification, multi-endpoint survival prognosis, and cross-modal retrieval. In all benchmarks, RenalCLIP consistently outperforms existing state-of-the-art models. Moreover, the model demonstrates advanced capabilities in zero-shot learning and radiology report generation, underscoring its potential as a foundation model for precision kidney cancer imaging and decision support.

## Results

### Overview of RenalCLIP

A comprehensive assessment of renal masses from CT imaging, spanning their anatomical complexity, diagnosis, and long-term prognosis, is essential for guiding personalized patient management. However, achieving this in a consistently accurate and objective manner remains a significant clinical challenge. To address this, we developed RenalCLIP, a disease-centric vision-language foundation model for renal cancer, by pre-training with a large-scale dataset of 21,819 CT scans from 6,867 patients. The model's performance on a suite of downstream clinical tasks was then rigorously evaluated on a multi-center dataset of 1,942 patients (6,047 CT scans). This dataset was partitioned into: (1) an internal cohort ($n$ = 500; split into $n$ = 400 for fine-tuning and $n$ = 100 for test); (2) a combined external testing cohort ($n$ = 1,017), which aggregated data from five independent centers: Xiamen ($n$ = 251), Shandong ($n$ = 250), Ruijin ($n$ = 229), Lianyun ($n$ = 220), and Zhangye ($n$ = 67); and (3) the public Cancer Imaging Archive (TCIA)[29,30] cohort for independent evaluation (Fig. 1a-b, see Supplementary Tables 1–2 for detailed cohort characteristics and task-specific class distributions).

The pre-training of RenalCLIP is designed to explicitly embed disease-centric clinical knowledge into its encoders before their cross-modal alignment (Fig. 1c, Supplementary Fig. 2c–e). This is achieved through a crucial initial knowledge-enhancement stage. The image encoder is pre-trained within a multi-task learning framework, supervised by a rich set of structured attributes that were systematically extracted from radiology reports. Concurrently, the text encoder, built upon the Llama3 large language model[31], is transformed into a specialized medical language expert using LLM2Vec techniques[32,33]. In the second stage, these knowledge-infused encoders are jointly optimized through a vision-language contrastive objective[34], forcing a deep alignment between fine-grained visual features and their clinical semantics (see Methods for a detailed description of the entire pre-training pipeline).

Following the completion of vision-language pre-training, we then systematically benchmarked RenalCLIP's capabilities across a comprehensive suite of 13 distinct downstream tasks. These ranged from a core set of 10 applications across three major clinical domains—renal mass characterization (five R.E.N.A.L. score components), diagnosis (malignancy and aggressiveness), and prognosis (prediction of recurrence-free survival (RFS), disease-specific survival (DSS), and overall survival (OS))—to 3 tests of advanced capabilities, namely zero-shot classification, cross-modal retrieval, and report generation. As summarized in the performance radar plots (Fig. 1d–f, see Supplementary Fig. 1 for adjusted-scale versions), RenalCLIP consistently demonstrates improved or highly competitive performance against a range of baseline models, including a standard 3D CNN[35] trained from scratch (termed CNN (Rand Init) in figures) and three general-purpose CT foundation models (CT-CLIP[36], Merlin[37], and CT-FM[38]), across the three primary task domains. The detailed evidence for its leading performance across all 13 tasks is presented in the subsequent sections (Figs. 2–5).

### Automated anatomical characterization and report generation

The R.E.N.A.L. nephrometry score serves as a standardized framework for quantifying the anatomical complexity of renal tumors, facilitating preoperative risk stratification and surgical planning[39]. We systematically evaluated RenalCLIP's ability to predict the five constituent anatomical features—R (Radius), E (Exophytic/Endophytic), N (Nearness to collecting system), A (Anterior/Posterior), and L (Location) using ROC AUC as the primary metric (see Supplementary Table 3 for R.E.N.A.L. grading details). RenalCLIP's leading discriminative performance was first evident in the internal test cohort, where it outperformed all comparative models on features R (ROC AUC = 0.908), E (0.646), A (0.857), and L (0.747), while achieving a competitive ROC AUC of 0.713 on feature N (vs. 0.723 for the baseline CNN). This strong performance was largely replicated in the combined external test cohort, where it again attained the highest ROC AUC for R (0.902), E (0.610), N (0.715), and L (0.727), while remaining competitive for feature A (0.754 vs. 0.757 for CT-FM) (Fig. 2a). Further analysis of individual external cohorts confirmed its robustness, with RenalCLIP achieving the highest ROC AUC in the majority of head-to-head comparisons across most features. Taken together, these results establish RenalCLIP's superior and generalizable capability for automated, multi-component anatomical scoring across diverse clinical settings (Fig. 2b; see Supplementary Tables 4–8 for comprehensive results).

To assess the clinical utility of these anatomical descriptors for surgical planning, we stratified performance by renal mass size (Supplementary Tables 9–10) and benchmarked the model against six independent board-certified experts (Supplementary Table 11). RenalCLIP demonstrated a competitive advantage in the clinically actionable cT1a (≤ 4 cm) subgroup, the primary candidates for nephron-sparing surgery. In this critical population, the model achieved the highest discriminative performance across all four anatomical descriptors (E, N, A, and L), consistently outperforming baseline models (Supplementary Table 10). This robust performance extended to the cT1b (4–7 cm) subgroup, confirming reliable support for complex surgical decision-making. Although performance in cT2 (> 7 cm) masses exhibited variability—likely due to the substantial anatomical distortion inherent to large tumors, the blinded multi-reader study confirmed that RenalCLIP's scoring fell well within the range of inter-observer variability (Supplementary Table 11). Notably, the model achieved top-tier consistency in objective features (R, A) and matched expert consensus in complex ones (E, N, L). This indicates that the observed performance constraints likely reflect the intrinsic subjectivity of characterizing complex renal masses rather than model-specific failure.

Beyond discrete anatomical descriptors, the radiology report is the primary communication tool for translating complex imaging findings into actionable clinical insights[40]. We therefore evaluated RenalCLIP on the advanced task of automated report generation, which tests its ability to synthesize learned anatomical features into a coherent narrative, moving beyond the prediction of discrete scores. To this end, we benchmarked its report generation quality against state-of-the-art models, including GPT-4o[41], MedGemma[42], RadFM[43], and CT-CHAT[36]. Across all standard language generation metrics (BLEU-1, BLEU-2, BLEU-4, METEOR, and ROUGE-L), RenalCLIP achieved the highest scores on the internal test cohort, indicating a stronger alignment with ground-truth reports annotated by expert radiologists (Fig. 2c, top). Crucially, this performance translated into superior clinical validity: in a blinded expert evaluation by board-certified radiologists (see Methods for details), RenalCLIP achieved the highest expert ratings for report accuracy (median score 2.67/5), establishing it as the only model to surpass the threshold for clinical acceptability (score > 2; adapted from HistoGPT[44]; Fig. 2c, bottom and Supplementary Table 12.) A representative qualitative example comparing a RenalCLIP-generated report to the human-annotated ground truth is provided in Fig. 2d, while comprehensive comparisons against all baseline models and the quantitative results are detailed in Supplementary Fig. 3 and Supplementary Tables 13–17.

**Robust diagnosis of malignancy and aggressiveness stratifies patient outcomes**

We first assessed RenalCLIP's capability in diagnosing the malignancy of renal masses, a critical prerequisite for determining treatment strategies[45]. Across both the internal and combined external test cohorts, RenalCLIP consistently achieved the highest ROC AUC among all models (Fig. 3a). Its diagnostic precision was further evidenced in the combined external cohort, where RenalCLIP attained leading performance across five key metrics: ROC AUC (0.841), PR AUC (0.941), sensitivity (0.827), specificity (0.735), and F1-score (0.876) (Fig. 3b). This superiority was maintained across the majority of the six individual external cohorts (5 out of 6, Fig. 3c and Supplementary Table 18). Crucially, the performance gap between RenalCLIP and the baseline models widened on the external cohorts, providing compelling evidence for its extended generalizability. Specifically, RenalCLIP outperformed the top baseline (CNN) by 8.2% in the internal test cohort (0.856 vs. 0.791), this margin increased to 17.3% against CNN (0.841 vs. 0.717) and 13.8% against the leading baseline (CT-FM, 0.841 vs. 0.739) in the combined external test cohort (visualized in ROC curves in Fig. 3d–e, and Supplementary Fig. 4). However, we observed a performance degradation in the public TCIA cohort (ROC AUC = 0.680). This decline is attributable to the extreme class imbalance within the TCIA dataset, where benign lesions constitute only 3.3% of the population (Supplementary Table 2), posing a statistical challenge for discrimination tasks[46]. Notably, this trend was consistent across all benchmarked models (e.g., CT-FM dropped to ROC AUC 0.61), yet RenalCLIP continued to outperform these baselines, highlighting its relative stability even under adverse data distributions. To dissect the drivers of this robust generalizability, we conducted controlled ablations on the malignancy diagnosis task, focusing on pre-training data scale and diversity. External performance improved consistently as pre-training volume increased from 0% to 100% (Fig. 3i). Moreover, under matched data volume (~45%), multi-center pre-training yielded better external generalization than single-center pre-training (Fig. 3j), despite the latter's dominance on the internal set. These results confirm that both the scale and multi-center diversity of our pre-training strategy are pivotal for ensuring robust generalizability.

Moving beyond binary malignancy, we evaluated the model's performance on the more nuanced task of aggressiveness stratification, which is paramount for personalized treatment selection[47]. Here again, RenalCLIP demonstrated superior accuracy compared to all baseline models in the internal cohort (Fig. 3a). In the combined external test cohort, RenalCLIP achieved leading performance metrics, including an ROC AUC of 0.703, PR AUC of 0.460, sensitivity of 0.713, specificity of 0.613, and an F1-score of 0.506 (Fig. 3b). A detailed cohort-level analysis further underscored its robust generalizability, showing that RenalCLIP consistently outperformed other models across most external cohorts (Fig. 3c and Supplementary Table 19).

To explicitly verify that these predictions rely on intrinsic imaging phenotypes rather than tumor size, we performed a focused subgroup analysis on small renal masses (SRMs, ≤ 4 cm) within the combined external test cohort (see Supplementary Table 20 for sample sizes and class distributions stratified by tumor size). In this clinically critical subgroup, RenalCLIP maintained superior performance for malignancy diagnosis (ROC AUC = 0.762), substantially outperforming the baseline CNN (0.630), other foundation models, and the classic R.E.N.A.L. score-based nomogram[48] (0.476) (Fig. 3h). Furthermore, for the challenging task of aggressiveness stratification in SRMs, RenalCLIP demonstrated strong robustness (ROC AUC = 0.655), performing comparably to CT-FM (0.653) while outperforming the baseline CNN (0.571), which exhibited performance degradation in smaller lesions. These results confirm that RenalCLIP captures size-independent features essential for early-stage diagnosis (more details can be found in Supplementary Table 21).

To validate the clinical relevance of this aggressiveness classification, we investigated whether the prediction of the model could stratify patients into distinct prognostic groups, as oncologic outcomes are the ultimate manifestation of tumor aggressiveness. In the internal test cohort, tumors classified as aggressive by RenalCLIP exhibited significantly worse RFS outcomes compared to those predicted to be indolent ($P = 0.005$, HR = 10.5), a capability shared by most baseline models in this less challenging setting (Supplementary Fig. 5a–e). In contrast, within the independent TCIA cohort, only the predictions made by RenalCLIP retained statistically significant prognostic power ($P = 0.010$, HR = 2.23) (Fig. 3f). Notably, tumors stratified by any of the baseline models, including CNN ($P = 0.225$, HR = 1.40), CT-CLIP ($P = 0.590$, HR = 1.26), Merlin ($P = 0.178$, HR = 1.44), and CT-FM ($P = 0.094$, HR = 1.64), showed no significant difference in RFS (Fig. 3g, Supplementary Fig. 5f–h). This unique ability to retain prognostic significance in a challenging external cohort provides powerful, clinically meaningful evidence for the robust generalizability of RenalCLIP's aggressiveness characterization.

**Non-invasive prognostic biomarker for kidney cancer survival**

Accurate prognosis prediction is crucial for guiding personalized follow-up strategies and adjuvant therapy decisions in renal cancer. With a lack of reliable non-invasive biomarkers[12], CT-based indicators offer a promising alternative[49,50]. We therefore evaluated whether a risk score derived from RenalCLIP could serve as a reliable prognostic biomarker. Initial analysis showed that RenalCLIP achieved the highest Concordance Index (C-index) for predicting RFS across the internal test cohort (0.864), the combined external test cohort (0.671), and the TCIA cohort (0.726). Acknowledging the limited follow-up periods in some external sites, we focused our subsequent in-depth evaluation primarily on the TCIA cohort. Within this cohort, RenalCLIP demonstrated a substantial advantage over baseline models: it outperformed the top-performing baseline (CT-FM) by 22.6% for RFS (0.726 vs. 0.592), by 6.3% for disease-specific survival (DSS; 0.690 vs. 0.649), and by 4.3% for overall survival (OS; 0.650 vs. 0.623) (Fig. 4a). We refer the reader to Supplementary Tables 22–24 for more detailed results of model performance.

To assess its clinical utility in risk stratification, we categorized patients into high- and low-risk groups based on the cohort-specific median score (see Supplementary Table 25 for specific thresholds). Kaplan-Meier analysis in the TCIA cohort revealed that RenalCLIP provided the most significant and clearest risk stratification for RFS. Patients in the RenalCLIP-defined high-risk group had significantly worse survival than the low-risk group, exhibiting the highest hazard ratio among all models ($P < 0.001$, HR = 3.7) (Fig. 4b–e). Furthermore, RenalCLIP demonstrated strong and statistically significant prognostic power for both DSS and OS (Supplementary Fig. 6a–j). We further validated the model's prognostic value for RFS specifically within the SRMs subgroup of the TCIA cohort, an indolent population where risk stratification is difficult. Despite the limited number of recurrence events, RenalCLIP successfully identified high-risk patients with statistical significance ($P = 0.032$, HR = 2.28), whereas baseline models failed to generalize to this subgroup (Fig. 4i, Supplementary Fig. 6k–o, and Supplementary Table 26).

To confirm that this prognostic value was independent of established clinical standards, we next performed multivariate Cox regression analysis. After adjusting for standard pathological indicators such as TNM stage and WHO/ISUP grade, the RenalCLIP risk score remained an independent adverse prognostic factor for RFS in the TCIA cohort ($P = 0.016$, HR = 2.27), while the risk scores derived from all baseline models failed to retain statistical significance (see Supplementary Table 27 for details).

Furthermore, we performed time-dependent analyses to evaluate the stability of the model's predictive accuracy over time. For RFS, RenalCLIP consistently outperformed all baseline models at all evaluated time points, as measured by ROC AUC (Fig. 4f), C-index (Fig. 4g), and Brier score for prediction error (Fig. 4h). Similar improvements over baselines were observed for DSS and OS across the same time range (Supplementary Fig. 7a–f). Collectively, these comprehensive analyses establish RenalCLIP as a robust and independent prognostic biomarker. Its ability to non-invasively predict patient outcomes from pre-operative imaging offers significant potential to refine clinical trial enrichment, guide personalized management, and improve patient counseling.

**Zero-shot generalization and data efficiency**

A critical barrier to the widespread clinical deployment of deep learning models is their heavy reliance on large, expert-annotated datasets, which are scarce and labor-intensive to create[51,52]. To overcome this, we designed RenalCLIP to excel in low-data regimes. We rigorously evaluated two of its hallmark capabilities: first, its ability to generalize to novel tasks via zero-shot transfer without any fine-tuning[34], and second, its data efficiency when adapted to new tasks with limited labeled samples.

To quantitatively validate the quality of the shared embedding space learned by RenalCLIP, we assessed its cross-modal retrieval performance. This task serves as a direct probe of the model's vision-language alignment and has practical applications in cohort identification and medical education. We benchmarked RenalCLIP against other vision-language models (CT-CLIP and Merlin) using Recall@K (K = 1, 3, 5) on five cohorts containing radiology reports. RenalCLIP substantially outperformed both baselines across all cohorts in both text-to-image (t2i) and image-to-text (i2t) retrieval (Fig. 5a and Supplementary Tables 28–32). For instance, in the t2i task, RenalCLIP achieved a mean Recall@5 of 0.120 across the five cohorts. This performance exceeds the random guess baseline (0.039) and represents an improvement over existing state-of-the-art medical VLMs, such as Merlin (0.051) and CT-CLIP (0.004).

Building on this strong cross-modal alignment, we next evaluated the model's zero-shot classification performance on malignancy and aggressiveness diagnosis. To ensure a robust evaluation against linguistic variations, we employed a prompt ensemble methodology and assessed performance under two distinct strategies: a deterministic approach using maximum similarity and a stochastic prompt sampling strategy (see Supplementary Tables 33–35 for prompts used for each class in zero-shot classification and Methods for details). Our results show that RenalCLIP is robust to the choice of evaluation strategy and consistently outperforms baseline models. Using the primary maximum similarity ensemble strategy, RenalCLIP substantially outperformed all baselines on both tasks across all three test cohorts (Fig. 5b–c and Supplementary Tables 36–37). On the challenging TCIA cohort, for example, RenalCLIP achieved a ROC AUC of 0.730 for malignancy prediction, significantly outperforming Merlin (0.664) and CT-CLIP (0.605). A similar margin of performance gap was observed for aggressiveness prediction, where RenalCLIP (ROC AUC = 0.657) again surpassed both Merlin (0.591) and CT-CLIP (0.412). Notably, in a stratified analysis of challenging SRMs ($\leq$ 4 cm), RenalCLIP retained its lead in zero-shot malignancy detection (ROC AUC 0.681 vs. 0.490–0.622 for baselines, Supplementary Table 38), though zero-shot capability for aggressiveness was limited across all models due to the lack of macroscopic correlates. This underscores that while disease-centric pre-training captures fundamental diagnostic features, explicit supervision remains indispensable for resolving the subtle pathological grading of small lesions, as evidenced by the substantial performance recovery achieved with fine-tuning (Malignancy ROC AUC 0.762; Aggressiveness ROC AUC 0.655, Supplementary Table 21).

Finally, we investigated the value of RenalCLIP's pre-trained features for downstream data efficiency. When fine-tuned, RenalCLIP demonstrated strong data efficiency, achieving performance comparable to fully-trained baseline models while using only 20–40% of the labeled images on both malignancy and aggressiveness tasks (Fig. 5d–g). Exceptions were observed in the TCIA malignancy task (Supplementary Fig. 8a) due to extreme class imbalance (Supplementary Table 2). Additionally, for aggressiveness prediction, external evaluation (combined external test and TCIA) performance peaked at intermediate data fractions (40–80%). This suggests that utilizing the full source dataset may induce overfitting to source-specific characteristics, whereas a curated subset yields better generalization[53]. However, even with the observed performance drop at 100% data usage, RenalCLIP consistently outperformed all comparative baselines. The power of its pre-trained representations was most strikingly evident when comparing its zero-shot performance directly against fine-tuned models. For malignancy prediction, RenalCLIP's zero-shot model surpassed the peak performance of all baseline models—even after they were fully fine-tuned on 100% of the data—across both internal and combined external cohorts. When compared against itself on the external cohort, its zero-shot generalization was so effective that it outperformed the RenalCLIP model fine-tuned on up to 80% of the data and was comparable to the fully fine-tuned version (Fig. 5d–e). For the more challenging aggressiveness task, although fine-tuning on RenalCLIP improved performance over its zero-shot baseline (even with only 10% of the data), this zero-shot performance itself remained highly competitive against the other fine-tuned models; it exceeded the peak fine-tuned results of the CNN, CT-CLIP, and Merlin models in the internal cohort, and its performance in the external cohort was comparable to that of the CNN, Merlin, and CT-FM models when fine-tuned on 40% of the data (Fig. 5f–g). The detailed analysis for the TCIA cohort, which showed similar trends of strong zero-shot performance, is provided in Supplementary Fig. 8. We refer readers to Supplementary Tables 39–40 for more detailed reporting of model performance.

## Discussion

In this study, we developed and validated RenalCLIP, a disease-centric vision-language foundation model that presents a framework for the comprehensive, non-invasive assessment of renal masses from CT imaging. By uniquely integrating deep visual features with the rich semantic context of radiological reports through a specialized knowledge-enhancement strategy, RenalCLIP demonstrates robust performance, generalizability, and data efficiency across a wide spectrum of clinical tasks compared to both conventional AI models[35] and existing general-purpose CT foundation models[36–38].

The management of renal masses presents a cascade of clinical challenges where precision is paramount[5–16]. For younger, otherwise healthy patients with a long-life expectancy, the non-invasive differentiation of benign from malignant tumors remains the critical diagnostic objective to prevent the overtreatment of benign lesions. However, the concept of overtreatment is highly dependent on patient demographics. For older patients with significant comorbidities and limited life expectancy, overtreatment frequently involves the unnecessary management of frankly malignant but indolent neoplasms. In this demographic, distinguishing aggressive neoplasms from benign and indolent malignant masses is arguably the more consequential clinical priority[54]. Furthermore, these clinical challenges extend to the accurate prediction of postoperative prognosis to personalize follow-up and adjuvant treatment decisions. Current imaging interpretation struggles to meet these needs, and the utility of percutaneous biopsy is limited by significant non-diagnostic rates and potential complications.

Aiming to address these challenges, prior AI research has largely focused on radiomics and conventional deep learning. For instance, in malignancy diagnosis, one of the largest CT-based radiomics models achieved an ROC AUC of 0.75 in a cohort of 735 patients[55]. Similarly, for aggressiveness assessment, a notable radiomics signature developed to distinguish low-grade from high-grade clear cell renal cell carcinoma (RCC) reached an ROC AUC of 0.70 in 587 patients[56]. However, the clinical translation of such models faces multifaceted challenges, including data governance and privacy constraints, regulatory compliance, and integration into clinical workflows. In addition, from a methodological standpoint, the field has often been limited by single-task development paradigms, relatively small cohorts, and insufficient independent external validation, which together can hinder robustness and real-world generalization; indeed, in the few studies where external evaluation was performed, these models often exhibited poor generalizability.

More recent general CT foundation models, while powerful, lack the disease-centric focus required for high-stakes, nuanced clinical questions in oncology. Although models like CT-FM[38] and Merlin[37] demonstrate versatility in broad tasks such as whole-body segmentation, and CT-CLIP[36] excels in chest imaging, their pre-training corpora are not enriched with sufficient, high-quality examples of renal cancer. Our head-to-head comparisons confirmed this limitation: their performance on fine-grained tasks like aggressiveness prediction and prognostication was significantly inferior to RenalCLIP. In contrast, RenalCLIP is explicitly designed to fill this gap by integrating deep visual features with renal-specific clinical-pathological semantics. This disease-centric design, which infuses the model with specialized knowledge through a unique pre-training strategy, enables RenalCLIP to excel in nuanced oncology tasks. It offers a level of precision in aggressiveness stratification and prognostication that general-purpose foundation models, which lack this targeted training, cannot match.

It is also important to position RenalCLIP relative to highly optimized, task-specific deep learning models, such as the recent work by Xiong et al. on CNN-based classification of renal mass malignancy and aggressiveness[57]. While fully supervised specialist models trained on extensive downstream datasets (e.g., ~2,400 cases in prior work) can achieve high peak performance on singular endpoints, direct numerical comparisons are confounded by differing evaluation protocols. The present study intentionally adopts a more rigorous, clinically realistic setting—incorporating highly heterogeneous multi-center cohorts and strictly retaining cases with missing contrast phases that were previously excluded. Under these challenging real-world conditions, RenalCLIP achieves comparable diagnostic accuracy (ROC AUC 0.84 vs. 0.87 for malignancy diagnosis) while utilizing less than 17% of the fine-tuning data (~400 cases). This highlights the exceptional data efficiency and feature transferability of our disease-centric pre-training paradigm, overcoming the common bottleneck of data scarcity in clinical AI. Ultimately, aligning with the emerging consensus on generalist medical AI[58,59], RenalCLIP represents a transition from a specialist to a generalist: it consciously trades marginal peak optimization on a single task for broad generalizability, zero-shot inference capabilities, and multi-modal proficiency (e.g., complex report generation and cross-modal retrieval) that traditional CNNs fundamentally cannot achieve.

A key advantage of RenalCLIP is its robust generalizability, which we attribute to our disease-centric strategy coupled with the large scale of our multi-center pre-training dataset. It was demonstrated through three key findings. First, the performance gap between RenalCLIP and baseline models widened substantially in external cohorts. For malignancy diagnosis, RenalCLIP's margin over the top CNN baseline expanded from 8.2% in the internal cohort to 17.3% in the combined external evaluation, underscoring its robustness against domain shifts. Second, while predictions from both RenalCLIP and most baseline models could significantly stratify patient RFS in the internal test cohort, only those from RenalCLIP retained this statistically significant prognostic capability in the independent TCIA cohort, where all baselines failed. Finally, despite being pre-trained predominantly on data from Chinese medical centers, RenalCLIP consistently outperformed all baseline models across all tasks in the independent TCIA cohort, supporting its generalizability across datasets with different population compositions.

Beyond its robust generalizability, the power of RenalCLIP's pre-trained representations was most evident in its strong data-efficient learning capability, which addresses a fundamental barrier to the translation of AI in medicine: the scarcity of large, expert-annotated datasets. As observed, conventional 3D CNNs trained from random initialization stagnated even when the sample size increased tenfold (40 to 400 subjects, Fig. 5f). Interestingly, while 400 subjects begin to trigger a discernible learning trend for the relatively macroscopic task of malignancy diagnosis (Fig. 5e), this volume falls short of the data threshold required to extract the subtle features necessary for aggressiveness stratification, resulting in a flat learning curve. In contrast, RenalCLIP effectively circumvents this bottleneck, achieving peak baseline performance with minimal annotations and demonstrating strong zero-shot capabilities. This is not merely a technical achievement; it signifies a potential paradigm shift in how specialized medical AI tools are developed. This capability opens a feasible pathway for rapidly extending the model to other critical but data-scarce problems in renal oncology. For instance, developing classifiers for rare but aggressive histologic subtypes (e.g., collecting duct carcinoma, medullary carcinoma), predicting response to targeted therapies, or identifying genomic markers from imaging currently suffers from a lack of sufficient training examples. The strong few-shot and zero-shot performance of RenalCLIP demonstrates its capability to learn robust representations from limited data. While specific evaluation on rare subtypes remains a subject for future work, these results suggest that RenalCLIP holds promise as a foundational platform to address data-scarce, long-tail clinical challenges where collecting large-scale training datasets is impractical.

Finally, the potential clinical value of these advanced capabilities was explored in the evaluation of prognostic endpoints. For aggressiveness stratification, many models, including our baselines, could separate prognostic groups in the internal cohort. However, the independent TCIA cohort provided a more rigorous test of generalizability, where RenalCLIP's predictions retained statistically significant prognostic power. These findings highlight the potential future utility of such approaches. If further validated and refined, a non-invasive imaging marker for aggressiveness, available at the time of initial diagnosis, could eventually support patient management by assisting in the consideration of active surveillance for low-risk tumors, and informing decisions on adjuvant therapy trials for high-risk patients post-nephrectomy.

Despite these promising results, our study has several limitations that must be acknowledged. First, while RenalCLIP demonstrated improved performance over baseline models in characterizing SRMs, arguably the most challenging yet clinically impactful subgroup, its current diagnostic accuracy in this specific scenario remains insufficient to independently guide clinical decisions. Further algorithmic refinements and larger, targeted training datasets are necessary to achieve the reliability required for routine clinical deployment in SRMs. Second, the analysis was retrospective, which may introduce selection and spectrum biases; prospective validation in a real-world clinical workflow is essential to confirm its utility. Third, although the independent TCIA cohort provided external evaluation beyond the proprietary Chinese cohorts, our pre-training data were predominantly derived from Chinese medical centers. Further training and evaluation using more geographically and demographically diverse datasets are warranted. Fourth, while RenalCLIP's report generation outperformed baselines, it is not yet at human expert level and can still exhibit minor factual inconsistencies, highlighting that high-fidelity medical report generation remains a significant challenge. Finally, our evaluation, while comprehensive, did not extend to predicting genomic profiles (e.g., BAP1, PBRM1 mutations) from imaging, which is a critical frontier in radiogenomics and an important avenue for future work.

In conclusion, RenalCLIP provides a robust, multi-task foundation for renal cancer assessment that bridges the gap between radiological imaging and clinical semantics. This work establishes the necessity and effectiveness of a disease-centric pre-training strategy for building high-performing, generalizable, and trustworthy medical foundation models. This disease-centric paradigm serves as a blueprint for developing similar specialized AI systems in other areas of oncology, potentially accelerating the integration of artificial intelligence into the core of personalized patient care.

## Methods

The study followed the Declaration of Helsinki and complied with all relevant ethical regulations. The study was approved by the ethics committees of Zhongshan Hospital, Fudan University; the First Affiliated Hospital of Zhejiang University; Qilu Hospital of Shandong University; Linyi People's Hospital; Xiamen Branch of Zhongshan Hospital; Shandong Cancer Hospital & Institute; Ruijin North Hospital of Shanghai Jiao Tong University; the First People's Hospital of Lianyungang; and Zhangye People's Hospital of Hexi University. The requirement for written informed consent was waived by the ethics committees due to the retrospective nature of the analysis, which involved pre-existing anonymized CT images and posed no additional risk to the patients. Information on race, ethnicity and ancestry was not collected for the proprietary cohorts and was not used for model development, validation or testing.

### Study cohorts and data preprocessing

#### Study population

We screened patients who underwent nephrectomy or biopsy for suspicious renal masses between January 1, 2009, and October 30, 2023, from nine medical centers in China and the public Cancer Imaging Archive (TCIA). The key inclusion criterion was the availability of preoperative multi-phase CT scans. Patients were excluded if they had received neoadjuvant therapy, had a diagnosis of hereditary renal cell carcinoma (RCC), or if their CT scans exhibited severe artifacts. Sex information, reported as male sex (%) in Supplementary Table 1, was determined based on self-report. Sex was summarized as a cohort characteristic but was not used as a stratification variable in the study design. Sex-stratified performance analyses were not performed because this study was designed to evaluate model performance across predefined renal imaging tasks and cohorts rather than to assess sex-specific effects. This screening process yielded a final study population of 8,809 patients providing a total of 27,866 preoperative CT scans, which were subsequently divided into pre-training and downstream task datasets. Crucially, this division was conducted strictly at the patient level. This ensured that all scans associated with a single patient were assigned exclusively to either the training, validation, or test cohort, thereby preventing any data leakage.

The pre-training dataset was sourced from four major medical centers (Zhongshan Hospital of Fudan University, The First Affiliated Hospital of Zhejiang University, Qilu Hospital of Shandong University, and Linyi People's Hospital) and comprised a total of 6,867 patients (21,819 CT scans). This dataset was further partitioned into a training set of 6,367 patients (20,278 scans) and a validation set of 500 patients (1,541 scans). The validation set was used exclusively for selecting the best-performing model checkpoint during the pre-training phase.

The downstream task dataset, used for fine-tuning and evaluation, consisted of a separate cohort of 1,942 patients (6,047 CT scans). This included an internal cohort sourced from the pre-training institutions, except for Linyi People's Hospital, which was partitioned into a training set of 400 patients (1,240 scans) and a test set of 100 patients (316 scans). For comprehensive external evaluation, we assembled five proprietary cohorts from the following institutions: Xiamen Branch of Zhongshan Hospital (termed the Xiamen cohort; 251 patients, 796 scans), Shandong Cancer

Hospital & Institute (termed the Shandong cohort; 250 patients, 991 scans), Ruijin North Hospital of Shanghai Jiao Tong University (termed the Ruijin cohort; 229 patients, 727 scans), the First People's Hospital of Lianyungang (termed the Lianyun cohort; 220 patients, 671 scans), and Zhangye People's Hospital of Hexi University (termed the Zhangye cohort; 67 patients, 259 scans). These five cohorts were aggregated to form the combined external test cohort used for pooled analyses. The public TCIA cohort (425 patients, 1,047 scans) was used as an additional, independent evaluation set.

Radiology reports were unavailable for the Lianyun and TCIA cohorts, while such reports were accessible for all other cohorts. The detailed clinicopathological characteristics of all cohorts are provided in Supplementary Table 1.

**Histologic classification**

Renal tumors were classified into three categories—benign indolent, malignant indolent, and malignant aggressive—based on previous research[4,57]. Benign indolent renal masses included angiomyolipoma, complex renal cyst, oncocytoma, mixed epithelial and stromal tumor, metanephric adenoma, among others. Malignant indolent renal tumors encompassed indolent clear cell renal cell carcinoma (ccRCC, defined as ccRCC without invasion into major veins or perinephric tissues, grade 3–4 components, necrosis, or sarcomatoid differentiation), indolent papillary RCC (papillary RCC without invasion into major veins or perinephric tissues, grade 3–4 components, or sarcomatoid differentiation), chromophobe RCC, clear cell papillary renal cell tumors, multilocular cystic renal neoplasm of low malignant potential, epithelioid angiomyolipoma, and others. Malignant aggressive subtypes included aggressive ccRCC (non-indolent ccRCC), aggressive papillary RCC (non-indolent papillary RCC), TFE3-rearranged renal cell carcinoma, renal cell carcinoma NOS, and other high-risk variants. Tumor stage at the time of surgery was determined according to the 8th edition of the AJCC cancer staging system. For cases with available slides, three pathologists re-evaluated tumor subtypes in accordance with the 2022 WHO classification for renal tumors and tumor nuclear grades based on the WHO/ISUP grading system for cases with available slides. The detailed distribution of these histologic subtypes and tumor grades for the entire cohort is provided in Supplementary Table 1.

**CT image acquisition and preprocessing**

To ensure spatial correspondence across the multi-phase CT scans (non-contrast (N), arterial (A), venous (V), and delayed (D)), we performed inter-phase rigid alignment within each patient based on DICOM geometry. Specifically, all available phases were mapped to the coordinate system of the arterial phase and resampled onto the arterial reference grid using linear interpolation with the Resample Image (BRAINS) module in 3D Slicer (v4.11.20210226). If the arterial phase was unavailable, the venous phase was used as the reference (all cases included at least one contrast-enhanced phase).

Following this alignment, we employed a semi-automated pipeline to localize the region of interest (ROI) (Supplementary Fig. 2a). A pre-trained nnU-Net[60] model was first utilized to automatically segment the kidneys, renal cysts, and tumors on the contrast-enhanced phase (arterial, or venous if arterial was unavailable), where lesion boundaries are most distinct. The resulting high-quality segmentation masks were then propagated to the spatially corresponding non-contrast phase of the same patient.

The subsequent localization strategy differed by task. This distinction was intentional, reflecting the different objectives of each stage: the pre-training phase was designed to maximize data diversity and learn robust features from the entire kidney, where precise lesion-centric localization was less critical, whereas the downstream tasks required highly accurate ROI definition centered on the lesion to learn task-specific diagnostic features. For the pre-training dataset, the ROI was automatically defined as the geometric center of the segmented foreground; in the rare event that the segmentation model produced no foreground mask, the kidney center was then manually annotated with a single point by an expert radiologist. For the downstream task datasets, an initial ROI center was proposed by identifying the axial slice with the largest cross-sectional area of the primary lesion (tumors prioritized over cysts) and calculating the lesion's geometric center on that slice. To ensure localization accuracy, an expert radiologist reviewed the quality of the underlying nnU-Net segmentations for all 1,942 cases in the downstream datasets. The automatically calculated center point was used for all accurately segmented lesions (approximately 90% of cases). In the remaining 10% of cases with suboptimal segmentation, the radiologist instead manually placed a single point to define the lesion center. All manual point annotations were performed using 3D Slicer.

Following the established protocol[57,61], a 3D volume centered on the localized ROI was cropped to a fixed physical size of 140 × 140 × 160 mm. These volumes were then resampled to an anisotropic spacing of 1.0 × 1.0 × 5.0 mm, yielding a uniform input size of 140 × 140 × 32 voxels. Finally, image intensities were normalized by windowing with a window level of 50 Hounsfield Units (HU) and a window width of 500 HU.

The strategy for selecting input volumes also varied by task. For pre-training, cropped volumes from both kidneys of each patient were utilized. For downstream tasks, only the kidney containing the primary lesion was included; in cases with bilateral lesions, the lesion determined to be most aggressive via pathology reports was selected. During model training, a 128 × 128 × 32 voxel patch was randomly cropped from the larger 140 × 140 × 32 volume for data augmentation. During inference, a deterministic center crop of the same size was used.

For all comparison models, we utilized their officially provided preprocessing functions and default configurations. The input for these models was based on the same lesion-centered ROIs generated by our pipeline to ensure a fair comparison.

**Radiology report processing**

The original free-text radiology reports, sourced in Mandarin Chinese, underwent a two-step processing pipeline to align with the single-kidney image inputs. First, given that our image encoder processes individual kidneys, we developed a large language model (LLM)-based method[41] to parse each report and partition the descriptive text into left kidney and right kidney components. Second, following this anatomical separation, the partitioned Chinese texts were programmatically translated into English to facilitate model development (Supplementary Fig. 2b). The specific prompts designed for both the report partitioning and translation tasks are detailed in the Supplementary Tables 41–42.

**Multi-phase CT data handling and fusion strategy**

Our core strategy for handling multi-phase CT scans treats each available phase from a patient's scan as an independent 3D sample. This approach serves two primary objectives: first, to substantially augment the effective size of our training dataset, and second, to implicitly regularize the model by exposing it to a wide variety of imaging contexts.

During the pre-training stage, our data loader randomly samples a single existing phase (selected from all available phases: non-contrast, arterial, venous, or delayed) for each patient within a batch. This forces the model to learn a robust feature representation that is agnostic to the specific enhancement characteristics of any single phase. For the primary downstream validation experiments, however, we adopted a standardized single-phase input to ensure fair comparison against baseline models, which typically utilize a single 3D volume (Supplementary Fig. 10a). Specifically, we prioritized the arterial phase, followed by the venous phase if the former was unavailable. This hierarchy reflects established clinical practice, as the arterial phase optimally demonstrates tumor vascularity and the venous phase is crucial for assessing both tumor margin and contrast washout patterns.

Finally, to quantify the performance benefit of integrating multi-phase information, we conducted an ablation study. For this study, a model was first fine-tuned on the downstream tasks using a random single-phase sampling strategy, analogous to the method used in pre-training stage. At inference time, this trained model was then evaluated using a late-fusion approach: each of a patient's available 3D phase volumes was independently passed through the model to generate phase-specific logits. These logits were then combined via element-wise averaging to produce a single, fused prediction.

**RenalCLIP pre-training**

RenalCLIP is a self-supervised learning model specifically designed for the RCC CT imaging domain. It builds upon the success of established language-image pre-training (e.g., CLIP[34]) in the natural image domain. RenalCLIP integrates two core components: an image encoder $f(\cdot;\theta_f)$ and a text encoder $g(\cdot;\theta_g)$. The image encoder $f(\cdot;\theta_f)$ is composed of a backbone $f_{\text{backbone}}(\cdot)$, specifically a 3D ResNet18[35] adapted for volumetric CT data that yields a 512-dimensional feature vector, and a projection head $f_{\text{proj}}(\cdot)$, where $\theta_f$ collectively denotes the parameters of both modules. The projection head maps the high-dimensional features extracted by the image backbone directly to a single global image representation of 4,096 dimensions, specifically designed with the same dimensionality as the text features to enable cross-modal alignment through contrastive learning. The text encoder $g(\cdot;\theta_g)$ is built upon the Llama3 8B architecture[31] and leverages LLM2Vec[32] technology. It comprises 32 transformer layers, featuring an embedding dimension of 4,096 and a hidden dimension of 14,336. A global representation of each caption is extracted by taking the mean of the last_hidden_state outputs across all its tokens. The text encoder does not employ an additional projection head and remains frozen during this phase, motivated by recent evidence[62,63] that text embeddings from a fixed pre-trained text encoder are highly effective for guiding visual representation learning.

**Uni-modal pre-training via knowledge enhancement**

Previous research[22,64] suggests that sequentially pre-training uni-modal modules prior to joint visual-language pre-training, significantly improves zero-shot transfer performance. Following this strategy, both our image and text encoders undergo independent pre-training.

**Image encoder pre-training**

We pre-trained the image encoder using a multi-task learning approach, leveraging structured labels derived from original radiology reports. Clinical experts designed 14 critical questions, each with several options, focusing on key aspects relevant to kidney cancer imaging interpretation. An LLM (GPT-4o[41]) then parsed the free-text reports based on these question-option pairs to extract the required structured attributes, capturing the semantic hierarchy. To validate the fidelity of this automated extraction, a board-certified radiologist manually audited a random subset of 100 kidneys, yielding an accuracy of 93.6% (786/840 verifiable items). This high concordance confirms that the extracted labels are reliable for pre-training. These extracted results were converted into a one-hot encoding format for model training. The image backbone was pre-trained by minimizing a classification loss (categorical cross-entropy), defined as:

$$\mathcal{L}_{\text{CLS}} = -\sum_{i=1}^{N}\sum_{j=1}^{C}\mathbf{1}[y^{(i)} = j]log(\hat{y}_j^{(i)}) \quad (1)$$

where N denotes the number of samples, and C represents the number of classes for each question. $\mathbf{1}[\cdot]$ is the indicator function, which equals 1 if the specified condition is satisfied and 0 otherwise. $y^{(i)}$ is the ground truth label of the $i$-th sample, and $\hat{y}_j^{(i)}$ denotes the predicted probability that sample $i$ belongs to class $j$. This process was crucial for injecting domain-specific clinical knowledge into the image encoder, enhancing its ability to capture features vital for RCC diagnosis and characterization from CT images. Specific questions, options, and prompts for structured attribute extraction are in Supplementary Tables 43-44.

**Text encoder pre-training**

Our text encoder is built upon the Llama3 8B architecture and leverages LLM2Vec technology, which transforms state-of-the-art LLMs into potent text feature extractors. Its core methodology involves three key steps: first, replacing the LLM's causal attention mechanism with a bidirectional structure to enhance contextual semantic understanding. Second, masked language modeling[65,66] (MLM) is applied. In this unsupervised framework, parts of words in reports are masked, and the model is trained to predict their original identity based on their bidirectional context, thereby capturing rich contextual representations. This step fine-tunes the LLM using LoRA[67] (Low-Rank Adaptation) on the MIMIC-CXR dataset[68] (227,973 chest X-ray reports) to significantly improve general medical domain comprehension. The MLM loss is formulated as:

$$\mathcal{L}_{\text{MLM}} = -\frac{1}{N}\frac{1}{M}\sum_{i=1}^{N}\sum_{t\in\mathcal{M}}\sum_{v=1}^{V} 1\,[y_t = v]\, log\left(\hat{p}_v^{(t)}\right) \tag{2}$$

where $N$ denotes the number of samples, $\mathcal{M}$ represents the set of masked tokens, $N$ is the number of elements in $\mathcal{M}$, $V$ is the vocabulary size, $y_t$ is the ground truth label for token $t$, and $\hat{p}_v^{(t)}$ denotes the predicted probability that token t corresponds to the $v$-th vocabulary item. Third, contrastive learning with SimCSE[69] further refines the text encoder's understanding of kidney cancer-related terminology. Here, an input sentence is passed twice through the model with independently sampled dropout masks to generate representations $\mathbf{h}_i$ and its distinct positive representations $\mathbf{h}_i^+$ , which are then optimized to maximize their similarity, while simultaneously minimizing their similarity with negative samples (representations of other sentences in the batch). This SimCSE objective function is formulated as follows:

$$\mathcal{L}_{\text{SimCSE}} = -\frac{1}{N}\sum_{i=1}^{N} \log\left(\frac{\exp(\text{sim}(\mathbf{h}_i, \mathbf{h}_i^+)/\tau)}{\sum_{j=1}^{N} \mathbf{1}[j \neq i]\exp\left(\text{sim}\left(\mathbf{h}_i, \mathbf{h}_j\right)/\tau\right)}\right) \tag{3}$$

where $N$ is the number of samples in a mini-batch, $\tau$ is a temperature hyperparameter, and $\text{sim}\left(\mathbf{h}_i, \mathbf{h}_j\right)$ denotes the cosine similarity between $\mathbf{h}_i$ and $\mathbf{h}_j$, which is defined as:

$$\text{sim}(\mathbf{h}_i, \mathbf{h}_j) = \frac{\mathbf{h}_i^T \mathbf{h}_j}{\|\mathbf{h}_i\|\|\mathbf{h}_j\|} \tag{4}$$

where $\|\mathbf{h}_i\|$ and $\|\mathbf{h}_j\|$ denote the L2 norms of $\mathbf{h}_i$ and $\mathbf{h}_j$, respectively. This fine-tuning is also conducted using LoRA and is applied to the training set of our internal pre-training dataset.

**Cross-modal visual-language pre-training**

Following uni-modal pre-training, RenalCLIP undergoes cross-modal visual-language pre-training[34]. A mini-batch is constructed from $N$ image-report pairs, denoted as $(\mathbf{x}_i, \mathbf{r}_i)_{i=1}^N$, where $\mathbf{x}_i$ is the $i$-th CT volume and $\mathbf{r}_i$ is a sequence of $T$ word tokens representing its corresponding radiological report. For each pair $(\mathbf{x}_i, \mathbf{r}_i)$, the L2-normalized image representation $\mathbf{u}_i$ is obtained from the cascaded backbone and projection head of the image encoder $f\left(\cdot; \theta_f\right)$, and the L2-normalized report representation $\mathbf{v}_i$ is derived from the text encoder $g\left(\cdot; \theta_g\right)$.

The model is trained to align positive (matching) image-report pairs and push apart negative (non-matching) pairs within the shared embedding space using a symmetric InfoNCE loss. The complete objective is given by:

$$\mathcal{L}_{\text{CLIP}} = -\frac{1}{N}\sum_{i=1}^{N}\left(\log\frac{\exp(\text{sim}(\mathbf{u}_i, \mathbf{v}_i)/\tau)}{\sum_{k=1}^{N}\exp(\text{sim}(\mathbf{u}_i, \mathbf{v}_k)/\tau)} + \log\frac{\exp(\text{sim}(\mathbf{u}_i, \mathbf{v}_i)/\tau)}{\sum_{k=1}^{N}\exp(\text{sim}(\mathbf{u}_k, \mathbf{v}_i)/\tau)}\right) \tag{5}$$

where $\text{sim}(\mathbf{u}, \mathbf{v})$ denotes the cosine similarity between representations $\mathbf{u}$ and $\mathbf{v}$, and $\tau$ is a learnable temperature parameter that scales the similarity scores. This dual-sided objective ensures that both image-to-text and text-to-image alignments are optimized simultaneously, driving the model to learn semantically rich and aligned representations for renal CT volumes and their corresponding reports.

**Pre-training configuration**

Our uni-modal pre-training involved separate stages for the image backbone and the text encoder. The image backbone was pre-trained using an AdamW optimizer[70] on a single NVIDIA A100 80GB GPU. Training spanned up to 200 epochs with a batch size of 300 and a learning rate of $5 \times 10^{-4}$. After each epoch, the model's performance was validated on the pre-training validation set, and the best-performing model checkpoint was saved based on the highest macro-averaged ROC AUC across 14 key attributes, reflecting its strong multi-task learning capabilities. The text encoder pre-training, utilizing LLM2Vec, consisted of two sequential steps, both optimized with AdamW on a single NVIDIA A100 80GB GPU. To reduce computational overhead, both steps incorporated gradient checkpointing[71] and mixed-precision training (bfloat16). Each step was trained for 1,000 iterations with a batch size of 32. The initial phase, leveraging MLM, used a learning rate of $5 \times 10^{-5}$ and an MLM probability of 0.2. This was followed by the SimCSE phase, which employed a learning rate of $3 \times 10^{-5}$ and a SimCSE dropout rate of 0.2. Detailed hyperparameters for uni-modal pre-training are provided in Supplementary Tables 45-47.

Our visual-language pre-training experiments were conducted on four NVIDIA A100 80GB GPUs, utilizing a local batch size of 1,024 per GPU and mixed-precision training (bfloat16) for enhanced efficiency. We employed AdamW as the optimizer. The image encoder's backbone was trained with an initial learning rate of $1 \times 10^{-5}$, while its projection head used $5 \times 10^{-4}$. The model was trained for 100 epochs, incorporating a 10-epoch warmup phase. Following each training epoch, we evaluated cross-modality retrieval performance on the pre-training validation set. The checkpoint demonstrating the best retrieval performance was selected as the final model weights. For more detailed cross-modal pre-training hyperparameter configurations, please refer to Supplementary Table 48.

For image data, augmentation strategies included random crop, random affine transformations, and variations in signal intensity. For more details about data augmentation techniques for 3D CT volumes, please refer to Supplementary Table 49. Regarding text data, features derived from the frozen text encoder were pre-computed and saved offline to disk to minimize computational overhead during training. For text data augmentation, we used sentence shuffling: for each original radiological report, we generated and saved four shuffled versions alongside the original.

**Evaluation of downstream tasks**

**Baseline models and comparisons**

We benchmarked RenalCLIP against a set of state-of-the-art baseline models. The specific models used for comparison varied by the downstream task, reflecting a comprehensive evaluation against both vision-only architectures and vision-language models (VLMs). Please refer to Supplementary Table 50 for a comprehensive summary of model sources, architectures, and training protocols across all tasks.

For end-to-end fine-tuning on characterization, diagnosis, and prognosis tasks, we included four 3D models. A standard 3D ResNet-18[35], trained from scratch with random weight initialization (CNN (Rand Init)), served as a baseline to quantify the performance gains attributable to pre-training.

We also included CT-FM[38], a 3D vision-only foundation model with a Seg-ResNet backbone[72], which was pre-trained on 148,000 whole-body CT scans from the Imaging Data Commons. The set of baselines was completed with two 3D VLMs: CT-CLIP36, a 3D VLM that pairs a Vision Transformer[73] encoder with a text encoder and was pre-trained on over 50,000 chest CTs, and Merlin[37], a 3D VLM designed for abdominal imaging that was pre-trained on approximately 15,000 abdominal CTs paired with EHR data and reports.

For zero-shot cross-modal retrieval and classification, comparisons were limited to the VLMs CT-CLIP and Merlin evaluated in inference-only mode, as these tasks require pre-trained, paired image and text encoders. The vision-only 3D CNN and CT-FM models were not applicable for these evaluations.

For the image-to-text report generation task, we selected four established models evaluated in inference-only mode to serve as robust baselines. RadFM[43] is a VLM specifically designed for medical report generation, pre-trained on a large-scale multi-modal medical dataset of 16 million medical scans with paired text. CT-CHAT[36] is another specialized VLM that combines the CT-CLIP vision encoder with a Llama 3.1 8B model, fine-tuned on over 2.7 million question-answer pairs from chest CTs. Additionally, we included two leading 2D VLMs: OpenAI's GPT-4o[41], a general-purpose multi-modal model, and Google's MedGemma[42], a healthcare-specialized model. For these 2D models, the center axial slice of each 3D CT volume was used as the image input. For these baseline models, official pre-trained weights were used directly for inference. In contrast, since RenalCLIP is natively a contrastive model lacking text generation capability, we integrated it with a multi-modal projector and a BioMistral-7B LLM, which were then trained (via projector alignment and LoRA fine-tuning) on the training set (see Methods, Radiology report generation, for details).

**Human expert benchmarking**

**R.E.N.A.L. nephrometry scoring**

To benchmark the model against human inter-observer variability, we conducted a blinded multi-reader study. An expert panel comprising six board-certified clinicians (three urologists and three radiologists) independently evaluated a randomly selected subset of 50 cases from the internal test cohort. Raters were strictly blinded to both the original consensus ground truth and the model's predictions.

**Clinical evaluation of generated reports**

To assess the clinical validity of the generated text, we conducted an independent evaluation on the full internal test cohort. Three board-certified radiologists evaluated the reports. To prevent assessment bias, the reports (generated by models alongside the human ground truth) were presented in a randomized order for each case, with raters strictly blinded to the source.

**Evaluation metrics**

For the renal mass characterization task (i.e., the five R.E.N.A.L. score components), which are multi-class (three-class) classification tasks, we primarily recorded the Area Under the Receiver

Operating Characteristic curve (ROC AUC). For this, we calculated the macro-averaged ROC AUC by treating each class as a positive against the others and then averaging the results. For the human benchmarking study, inter-rater agreement against the consensus ground truth was quantified using accuracy and Cohen's kappa. The relative rank of each rater was determined by their respective kappa value to contextualize the model's performance within the spectrum of human variability. For the binary-class diagnosis tasks (malignancy and aggressiveness), a more comprehensive set of metrics was used: ROC AUC, Precision-Recall Area Under the Curve (PRAUC), sensitivity, specificity, and the F1-score. For these threshold-dependent metrics, the optimal classification threshold for each cohort was determined by maximizing Youden's J statistic. For prognosis tasks, we used Harrell's concordance index (C-index)[74] to evaluate how well a model's predicted risk aligned with the actual order of events. We also employed time-dependent ROC AUC, C-index, and the Brier score[75] to assess predictive accuracy at various follow-up times. The Brier score specifically measures the accuracy of probabilistic predictions, with lower scores indicating better calibration. For zero-shot and retrieval tasks, we used ROC AUC for zero-shot classification and Recall@K (with K = 1, 3, 5) for retrieval, which measures whether a correct item is found among the top K retrieved results. For the report generation task, we reported BLEU[76], METEOR[77], and ROUGE-L[78] scores to evaluate different aspects of generation quality. Specifically, we calculated BLEU-1/-2/-4 to measure the precision of n-gram overlap against human-written references; METEOR, which incorporates semantic matching (e.g., synonyms) to better capture meaning; and ROUGE-L, which measures the longest common subsequence to reflect sentence-level structure and fluency. For the expert qualitative evaluation, reports were scored using a 6-point clinical accuracy scale adapted from HistoGPT[44]: (5) beyond expectation; (4) highly accurate; (3) generally accurate with minor variations lacking clinical impact; (2) partially accurate with variations of potential clinical impact; (1) minimally accurate; and (0) completely inaccurate. As defined by this protocol, a score > 2 indicates a diagnosis that is considered correct or within an acceptable range of subjectivity. The final score for each case was calculated as the average of the ratings from the three experts and inter-rater reliability was assessed using Fleiss' kappa.

**Fine-tuning for clinical tasks**

To evaluate RenalCLIP's utility in key renal cancer clinical applications, we fine-tuned the model on various downstream tasks related to RCC characterization, diagnosis, and prognosis. These experiments were conducted using our internal downstream dataset, comprising 400 training samples and 100 validation samples. Model performance was subsequently validated on independent external and TCIA cohorts. The AdamW optimizer was consistently employed, and each model was trained for 100 epochs on a single A100 80GB GPU.

Optimal hyperparameters (learning rate and batch size) for each model and task were determined through a comprehensive grid search. The best-performing model for each comparison was identified based on its performance on the internal test set: macro-averaged AUC for characterization and diagnosis tasks (specifically macro-AUC for multi-class classification), and C-index for prognosis tasks. These hyperparameter sets were then used to evaluate the models on the external and TCIA cohorts. Due to variations in GPU memory footprint and model architecture, the hyperparameter search ranges differed across models. For 3D ResNet18 (randomly initialized) and RenalCLIP's backbone (35,087,424 parameters), batch sizes were searched from [50, 100, 150], and learning rates from [$5 \times 10^{-4}$, $1 \times 10^{-4}$, $5 \times 10^{-5}$, $1 \times 10^{-5}$]. CT-CLIP (176,885,144 parameters) used a fixed batch size of 8, with its learning rate mirroring that of Merlin due to their comparable parameter counts. Merlin (121,882,204 parameters) used a fixed batch size of 16, and its learning rates were searched from [$1 \times 10^{-4}$, $5 \times 10^{-5}$, $1 \times 10^{-5}$]. For CT-FM (77,760,992 parameters), batch sizes were searched from [50, 100], and learning rates from [$5 \times 10^{-4}$, $1 \times 10^{-4}$, $5 \times 10^{-5}$, $1 \times 10^{-5}$].

For characterization and diagnosis tasks, features extracted by the model's backbone were passed through a Multi-Layer Perceptron (MLP) classification head to map them to the class dimension. These models were trained using a cross-entropy loss function. For prognosis tasks, the same backbone architecture was used, but the extracted features were mapped to a single continuous risk score (representing the log-hazard ratio, or linear predictor) via an MLP. Prognostic models were trained using the Cox Proportional Hazards (CoxPH) model[79], leveraging a partial log-likelihood loss function to account for censored data. While our dataset included three prognostic labels (recurrence-free survival (RFS), defined as the interval from surgery to local recurrence, distant metastasis, or renal cancer-related death; disease-specific survival (DSS), the interval from surgery to death from renal cancer; and overall survival (OS), the interval from surgery to death from any cause), we exclusively utilized the RFS label for model training. This decision was driven by the fact that RFS events directly reflect intrinsic tumor aggressiveness, whereas OS is confounded by non-cancer competing risks and DSS events are limited by effective systemic therapies. The trained RFS model was then used to calculate risk scores across all three survival endpoints (RFS, DSS, and OS) for quantitative evaluation. Consequently, the predicted risk scores are identical for a given patient across these tasks. For risk stratification, patients were categorized into high- and low-risk groups based on the cohort-specific median risk score (see Supplementary Table 25 for specific thresholds).

**Radiology report generation**

We developed a caption generation model, adapted from the LLaVA framework[80,81], designed to interpret 3D renal CT volumes. This model integrates the frozen RenalCLIP image backbone $f_{\text{backbone}}(\cdot)$, a multi-modal projector, and an LLM. The image backbone processes 3D renal CT volumes, compressing them into a feature space. These features are then passed to our multi-modal projector, which is implemented as a simple linear layer. This layer linearly transforms the image features, converting them from their original dimension (e.g., 512) into a dimension compatible with the LLM's hidden size (e.g., 4,096). This effectively functions as a visual tokenizer, aligning the image features with the LLM's word embedding space. For the LLM component, we employed BioMistral-7B[82], a version of Mistral-7B further pre-trained on medical textual data from PubMed Central Open Access, making it well-suited for the medical domain; its architecture includes 32 transformer layers, 32 attention heads, an embedding dimension of 4,096, a hidden dimension of 14,336, and a vocabulary size of 32,000.

The captioning model undergoes a two-step training strategy. In both steps, the model is guided by a consistent instruction prompt structured as follows:

*<s>[INST]*

*Generate a radiology report for the given CT scan of suspected renal cancer.*

*Use the following format:*

*Findings: [Detailed observations based on the CT scan.] Impression: [Summary of the diagnostic conclusions.]*

*[/INST]*

First, in the initial alignment step, only the multi-modal projector is trained while both the image backbone and the LLM remain frozen, utilizing the train set of our internal pre-training dataset. This step is crucial for aligning the multi-modal projector in biomedical concept space and enhancing subsequent fine-tuning performance. Second, in the fine-tuning step, training continues for the multi-modal projector (image backbone remains frozen), but the LLM is fine-tuned using LoRA[67], utilizing the internal training cohort. All training for the captioning model, across both steps, was conducted on a single A100-80GB GPU with a batch size of 30. In the initial alignment step, the projector was trained for 20 epochs using a learning rate of $1 \times 10^{-3}$. After each epoch, performance was validated on the validation set of the pre-training dataset, and the checkpoint with the lowest validation loss was saved. For the subsequent fine-tuning step, the projection layer's learning rate was set to $5 \times 10^{-4}$, while the LoRA weights for the LLM were fine-tuned with a learning rate of $1 \times 10^{-6}$ over 2 epochs. More hyperparameters for both pre-training and fine-tuning in radiology report generation are detailed in Supplementary Tables 51-52.

**Zero-shot classification**

For zero-shot classification, we employed a method that adapts CLIP's original framework[34]. For each class, we generated a text prompt by combining a class prompt (e.g., benign for malignancy assessment) with a pre-defined sentence template. For example, a template might be "A soft tissue mass that is {...} is visible in the kidney." In total, 20 distinct templates were used and each class was associated with five different class prompts, resulting in 100 unique text prompts generated for every class. A comprehensive list of all class names and templates used is provided in Supplementary Tables 33-35.

To rigorously evaluate model performance while accounting for the variability arising from this diverse prompt set, we implemented and compared two distinct zero-shot evaluation strategies.

First, to determine a robust, high-performance metric, we employed a maximum similarity ensemble strategy. For each image embedding $\mathbf{u}_i$, we computed its cosine similarity against the entire set of 100 pre-computed text embeddings for each class $c \in \{0, 1\}$, denoted as $\{\mathbf{v}_{c,k}\}_{k=1}^{100}$, where $k$ indexes the unique prompts for a given class. The maximum similarity score within each class set was then taken as the definitive logit for that class, i.e., $s_c = \max_k \left(\text{sim}(\mathbf{u}_i, \mathbf{v}_{c,k})\right)$. An image was assigned to the class with the highest logit score.

Second, to specifically quantify the model's sensitivity to prompt selection, we employed a stochastic prompt sampling strategy. In this analysis, we performed $B$ bootstrap iterations (e.g., $B = 1000$). For each iteration, we randomly sampled a single prompt embedding for the positive class and one for the negative class to form a classifier pair. This pair was then used to evaluate the model's ROC AUC across the entire test cohort. This procedure yielded a distribution of performance scores, from which we report the mean and 95% CI to visualize the model's sensitivity.

**Cross-modal retrieval**

For zero-shot retrieval, we employed a method analogous to our zero-shot classification approach, operating within the same aligned latent space. For text-to-image retrieval, a specific text query (e.g., a radiology report) was used to find the top-K closest CT volumes. Conversely, image-to-text retrieval was performed by querying with a CT volume to find the nearest text entries.

**Ablation studies**

**Impact of uni-modal pre-training components**

We next sought to dissect the contributions of the uni-modal pre-training stages that precede the final cross-modality alignment in RenalCLIP. We conducted an ablation study comparing the RenalCLIP (Full) model against four variants, all trained using the same CLIP contrastive learning framework (except the random baseline): (1) a model using the CLIP framework initialized with our domain-specific image encoder only (termed RenalCLIP (Image Pre-train)); (2) a model using the CLIP framework initialized with our domain-specific text encoder only (termed RenalCLIP (Text Pre-train)); and (3) a baseline model using the standard CLIP framework initialized with generic weights (BioClinicalBERT[83] text encoder and random image encoder, termed CLIP Baseline), and (4) a 3D ResNet-18 initialized with random weights (termed CNN (Rand Init)). The comprehensive comparison across cross-modality retrieval, zero-shot classification, and fine-tuning tasks is presented in Supplementary Fig. 9.

We found that for cross-modality retrieval, RenalCLIP (Full) achieved the best performance, followed by RenalCLIP (Text Pre-train) and RenalCLIP (Image Pre-train), which demonstrated comparable results with cohort-specific advantages, and finally the CLIP Baseline. For fine-tuning tasks, which are primarily driven by visual features, a different trend was observed. The RenalCLIP (Full) and RenalCLIP (Image Pre-train) performed comparably and were superior to the other models. Crucially, the CLIP Baseline achieved performance comparable to or only marginally better than the CNN (Rand Init). This demonstrates that generic pre-training pipelines are insufficient without domain-specific adaptation, whereas our disease-centric image pre-training provides the decisive performance gain for visual tasks.

The zero-shot classification results presented a more nuanced picture that was dependent on the evaluation strategy. Using the deterministic maximum similarity ensemble strategy, RenalCLIP (Full) again achieved the highest performance, followed by RenalCLIP (Image Pre-train). Conversely, when evaluated using the stochastic prompt sampling strategy to assess prompt sensitivity, the RenalCLIP (Image Pre-train) slightly outperformed the full RenalCLIP (Full). This finding suggests that while our domain-specific text pre-training is highly effective for the deterministic ensemble approach, it may also introduce an increased sensitivity to prompt phrasing variations, a phenomenon that warrants further investigation.

Overall, these findings establish that domain-specific image pre-training provides the most substantial and consistent performance foundation across all tasks. Our renal-specific text pre-training offers a crucial synergistic benefit, particularly for fine-grained retrieval and robust zero-shot classification.

### Impact of multi-phase CT inputs

To investigate the marginal benefit of integrating multi-phase data, we conducted an ablation study evaluating four cumulative input settings: single-phase (A), dual-phase (A and V, termed AV), tri-phase (N, A, and V, termed NAV), and the full quad-phase (N, A, V, and D, termed NAVD) (Supplementary Fig. 10c-e).

Our results reveal a nuanced trade-off. While multi-phase fusion strategies generally yielded a modest performance gain over the single-phase input across R.E.N.A.L scoring, diagnosis, and survival prognosis, this single-phase model itself established a remarkably strong baseline. Consequently, we observed no universally optimal phase combination; the best-performing configuration varied depending on the specific downstream task and patient cohort. For instance, while the quad-phase input occasionally yielded the highest performance, the dual-phase (AV) or tri-phase (NAV) inputs proved more effective in other settings.

We hypothesize that this complex relationship stems from a trade-off between accumulating diagnostic signal and introducing potential noise, a dynamic amplified by our simple logit-averaging fusion method. Although additional phases offer supplementary information (e.g., washout patterns), they can also introduce confounding features that dilute the primary signal, particularly when imaging characteristics vary across cohorts. These findings suggest that while multi-phase data holds clear value, the optimal strategy is not a simple aggregation of all available information. Instead, future work on more sophisticated, learnable fusion mechanisms is warranted, and for current applications, the selection of CT phases should be tailored to the specific clinical question.

### Statistical analysis

To assess statistical uncertainty, 95% confidence intervals (CIs) for all reported performance metrics were estimated using a non-parametric bootstrap method with 1,000 resamples, unless otherwise specified. Specifically, for experiments evaluating model performance with varying proportions of training data, results are presented as mean ± standard deviation (SD) derived from $n = 5$ independent runs with different random seeds. A two-sided $P$-value of less than 0.05 was considered statistically significant for all hypothesis tests. To evaluate the statistical significance of observed differences in model performance across diagnostic and scoring tasks (including malignancy, aggressiveness, and R.E.N.A.L. score predictions), a two-sided paired permutation test with 1,000 permutations was employed. In each iteration, patient-level predictions from the two compared models were randomly swapped to generate a null distribution of performance differences. The $P$-value was calculated as the proportion of permutations where the absolute simulated difference exceeded the observed absolute difference. The analyses focused on three prognostic endpoints: RFS, DSS, and OS. For all survival analyses, the cohort was restricted to patients with malignant tumors. For risk stratification, each model's continuous risk score was first dichotomized into high- and low-risk groups based on the cohort-specific median. Kaplan-Meier curves were plotted to visualize survival differences, which were assessed for significance using the log-rank test. These univariate survival analyses were performed using the lifelines library (version 0.30.0) in Python. To determine the independent prognostic value of each derived risk score, a multivariate Cox proportional hazards regression model was constructed using SPSS Statistics 21.0. The significance of each variable in the model was assessed using the Wald test. In this multivariate analysis, each model's risk score was included as a continuous variable, and the model was adjusted for established clinical and pathological risk factors such as TNM stage and WHO/ISUP grade.

**Computing hardware and software**

Scripts for 3D CT pre-processing were written in Python (version 3.10.14) using the libraries Nibabel (version 5.3.1), Numpy (version 1.26.4), and Scipy (version 1.13.1). The multi-stage pre-training of RenalCLIP was conducted on NVIDIA A100 80GB GPUs using PyTorch (version 2.3.0) with CUDA 12.1. The initial image encoder pre-training utilized a single GPU, with data augmentation implemented via MONAI (version 1.3.1). The subsequent text encoder pre-training was also performed on a single GPU, leveraging the LLM2Vec (version 0.2.2), Transformers (version 4.44.2), flash-attn (version 2.7.2), and PEFT (version 0.10.0) libraries. The final cross-modal pre-training was performed on four GPUs configured for multi-GPU, single-node training using PyTorch's Distributed Data Parallel (DDP) framework. All downstream task fine-tuning experiments were conducted on a single 80GB NVIDIA A100 GPU. The software environment was consistent with pre-training (Python 3.10.14, PyTorch 2.3.0, CUDA 12.1, MONAI 1.3.1). Specific tasks required additional libraries: PyCox (version 0.3.0) was used for prognosis tasks, while PEFT (version 0.10.0) and Transformers (version 4.40.2) were used for report generation. Performance metrics were calculated using several Python libraries. Standard classification metrics, including ROC AUC, PRAUC, sensitivity, specificity, and the F1-score, were computed using Scikit-learn (version 1.5.0). C-index was calculated using PyCox (version 0.3.0), while the time-dependent versions of the ROC AUC, C-index, and Brier score were calculated using scikit-survival (version 0.24.1). Natural language generation metrics were evaluated using NLTK (version 3.9.1) for BLEU and METEOR, and rouge-score (version 0.1.2) for ROUGE. Implementation of the comparative baseline models was based on the official configurations provided in their respective repositories: CT-CLIP (https://github.com/ibrahimethemhamamci/CT-CLIP), Merlin (https://github.com/StanfordMIMI/Merlin), CT-FM (https://github.com/project-lighter/CT-FM), RadFM (https://github.com/chaoyi-wu/RadFM), CT-CHAT (https://github.com/ibrahimethemhamamci/CT-CHAT), GPT-4o (https://ai.azure.com/catalog/models/gpt-4o), and MedGemma (https://huggingface.co/google/medgemma-4b-it).

**Data availability**

The public TCIA data used in this study are available in the Cancer Imaging Archive [https://www.cancerimagingarchive.net/collections/] and the GDC Data Portal [https://portal.gdc.cancer.gov/]. The multi-center imaging datasets analyzed in this study are available under restricted access due to patient privacy laws and institutional review board policies. Access can be obtained by submitting a formal request to the corresponding author, S.W.

(shuowang10@fudan.edu.cn). Access to the data is restricted exclusively to researchers for non-commercial academic purposes. Requests will be reviewed and forwarded to each respective center for consent, with an expected response timeframe of within 4 weeks. Once approved, access to the requested data will be granted for a period of 1 year. Source data are provided with this paper.

**Code availability**

The source code for all experiments, pre-trained model weights for RenalCLIP, and tutorial Jupyter notebooks for key applications are publicly available on GitHub to support reproducibility and further research (https://github.com/dt-yuhui/RenalCLIP)[84]. The complete codebase and models are released under the MIT License. For reproducibility, we archived the exact version used in this study on Zenodo with the DOI [10.5281/zenodo.19565967] and tagged release [v1.0.0].

## References

1 Wong, M. C. S. et al. Incidence and mortality of kidney cancer: temporal patterns and global trends in 39 countries. *Sci. Rep.* **7**, 15698 (2017).

2 Turner, R. M., Morgan, T. M. & Jacobs, B. L. Epidemiology of the small renal mass and the treatment disconnect phenomenon. *Urol. Clin. North. Am.* **44**, 147–154 (2017).

3 Gill, I. S. et al. Comparison of 1,800 laparoscopic and open partial nephrectomies for single renal tumors. *J. Urol.* **178**, 41–46 (2007).

4 Bhindi, B. et al. The Probability of aggressive versus indolent histology based on renal tumor size: implications for surveillance and treatment. *Eur. Urol.* **74**, 489–497 (2018).

5 Wentland, A. L. et al. Differentiation of benign from malignant solid renal lesions using CT-based radiomics and machine learning: comparison with radiologist interpretation. *Abdom. Radiol.* **48**, 642–648 (2023).

6 Silverman, S. G., Gan, Y. U., Mortele, K. J., Tuncali, K. & Cibas, E. S. Renal masses in the adult patient: the role of percutaneous biopsy. *Radiology* **240**, 6–22 (2006).

7 Tomaszewski, J. J., Uzzo, R. G. & Smaldone, M. C. Heterogeneity and renal mass biopsy: a review of its role and reliability. *Cancer Biol. Med.* **11**, 162–172 (2014).

8 Patel, H. D. et al. Diagnostic accuracy and risks of biopsy in the diagnosis of a renal mass suspicious for localized renal cell carcinoma: systematic review of the literature. *J. Urol.* **195**, 1340–1347 (2016).

9 Bjurlin, M. A., Elkin, E. B., Atoria, C. L., Russo, P., Taneja, S. S. & Huang, W. C. Influence of renal biopsy results on the management of small kidney cancers in older patients: results from a population-based cohort. *Urol. Oncol.* **35**, 604.e1–604.e7 (2017).

10 Leibovich, B. C. et al. Predicting oncologic outcomes in renal cell carcinoma after surgery. *Eur. Urol.* **73**, 772–780 (2018).

11 Klatte, T., Rossi, S. H. & Stewart, G. D. Prognostic factors and prognostic models for renal cell carcinoma: a literature review. *World J. Urol.* **36**, 1943–1952 (2018).

12 Rosellini, M. et al. Prognostic and predictive biomarkers for immunotherapy in advanced renal cell carcinoma. *Nat. Rev. Urol.* **20**, 133–157 (2023).

13 Choueiri, T. K. et al. Adjuvant pembrolizumab after nephrectomy in renal-cell carcinoma. *N. Engl. J. Med.* **385**, 683–694 (2021).

14 Braun, D. A. et al. Clinical validation of PBRM1 alterations as a marker of immune checkpoint inhibitor response in renal cell carcinoma. *JAMA Oncol.* **5**, 1631–1633 (2019).

15 Ged, Y. et al. DNA damage repair pathway alterations in metastatic clear cell renal cell carcinoma and implications on systemic therapy. *J. Immunother. Cancer* **8**, e000230 (2020).

16 Xiong, Y. et al. Identification and validation of dichotomous immune subtypes based on intratumoral immune cells infiltration in clear cell renal cell carcinoma patients. *J. Immunother. Cancer* **8**, e000447 (2020).

17 Paschali, M. et al. Foundation models in radiology: what, how, why, and why not. *Radiology* **314**, e240597 (2025).

18 Aggarwal, R. et al. Diagnostic accuracy of deep learning in medical imaging: a systematic review and meta-analysis. *npj Digit. Med.* **4**, 65 (2021).
19 Zhang, K. et al. A generalist vision–language foundation model for diverse biomedical tasks. *Nat. Med.* **30**, 3129–3141 (2024).
20 Wang, Z., Wu, Z., Agarwal, D. & Sun, J. MedCLIP: contrastive learning from unpaired medical images and text. In *Proc. 2022 Conf. Empirical Methods Nat. Lang. Process.* 3876–3887 (2022).
21 You, K. et al. CXR-CLIP: toward large scale chest x-ray language-image pre-training. In *Proc. Int. Conf. Med. Image Comput. Comput. Assist. Interv.* 101–111 (Springer, 2023).
22 Lu, M. Y. et al. A visual-language foundation model for computational pathology. *Nat. Med.* **30**, 863–874 (2024).
23 Lu, Z., Li, H., Parikh, N. A., Dillman, J. R. & He, L. RadCLIP: enhancing radiologic image analysis through contrastive language–image pretraining. *IEEE Trans. Neural Netw. Learn. Syst.* **36**, 17613–17622 (2025).
24 Shi, D. et al. A multimodal visual–language foundation model for computational ophthalmology. *npj Digit. Med.* **8**, 381 (2025).
25 Zhou, Y. et al. A foundation model for generalizable disease detection from retinal images. *Nature* **622**, 156–163 (2023).
26 Qiu, J. et al. VisionFM: a multi-modal multi-task vision foundation model for generalist ophthalmic artificial intelligence. Preprint at *arXiv* https://doi.org/10.48550/arXiv.2310.04992 (2023).
27 Xu, H. et al. A whole-slide foundation model for digital pathology from real-world data. *Nature* **630**, 181–188 (2024).
28 Chen, R. J. et al. Towards a general-purpose foundation model for computational pathology. *Nat. Med.* **30**, 850–862 (2024).
29 Clark, K. et al. The Cancer Imaging Archive (TCIA): maintaining and operating a public information repository. *J. Digit. Imaging* **26**, 1045–1057 (2013).
30 Liu, J. et al. An integrated TCGA pan-cancer clinical data resource to drive high-quality survival outcome analytics. *Cell* **173**, 400–416 (2018).
31 Grattafiori, A. et al. The llama 3 herd of models. Preprint at *arXiv* https://doi.org/10.48550/arXiv.2407.21783 (2024).
32 BehnamGhader, P. et al. Llm2vec: large language models are secretly powerful text encoders. Preprint at *arXiv* https://doi.org/10.48550/arXiv.2404.05961 (2024).
33 Huang, W. et al. LLM2CLIP: Powerful language model unlocks richer visual representation. Preprint at *arXiv* https://doi.org/10.48550/arXiv.2411.04997 (2024).
34 Radford, A. et al. Learning transferable visual models from natural language supervision. In *Proc. 38th Int. Conf. Mach. Learn.* 8748–8763 (2021).
35 He, K., Zhang, X., Ren, S. & Sun, J. Deep residual learning for image recognition. In *Proc. IEEE Conf. Comput. Vis. Pattern Recognit.* 770–778 (2016).
36 Hamamci, I. E. et al. Generalist foundation models from a multimodal dataset for 3D computed tomography. *Nat. Biomed. Eng.* https://doi.org/10.1038/s41551-025-01599-y (2026).

37 Blankemeier, L. et al. Merlin: a computed tomography vision–language foundation model and dataset. *Nature* https://doi.org/10.1038/s41586-026-10181-8 (2026).
38 Pai, S. et al. Vision foundation models for computed tomography. Preprint at *arXiv* https://doi.org/10.48550/arXiv.2501.09001 (2025).
39 Kutikov, A. & Uzzo, R. G. The RENAL nephrometry score: a comprehensive standardized system for quantitating renal tumor size, location and depth. *J. Urol.* **182**, 844–853 (2009).
40 Tanno, R. et al. Collaboration between clinicians and vision–language models in radiology report generation. *Nat. Med.* **31**, 599–608 (2025).
41 Hurst, A. et al. Gpt-4o system card. Preprint at *arXiv* https://doi.org/10.48550/arXiv.2410.21276 (2024).
42 Sellergren, A. et al. Medgemma technical report. Preprint at arXiv https://doi.org/10.48550/arXiv.2507.05201 (2025).
43 Wu, C. et al. Towards generalist foundation model for radiology by leveraging web-scale 2d&3d medical data. *Nat. Commun.* **16**, 7866 (2025).
44 Tran, M. et al. Generating dermatopathology reports from gigapixel whole slide images with HistoGPT. *Nat. Commun.* **16**, 4886 (2025).
45 Schieda, N. et al. Active surveillance of renal masses: the role of radiology. *Radiology* **302**, 11–24 (2022).
46 Buda, M., Maki, A. & Mazurowski, M. A. A systematic study of the class imbalance problem in convolutional neural networks. *Neural Netw.* **106**, 249–259 (2018).
47 Li, Y. et al. Histopathologic and proteogenomic heterogeneity reveals features of clear cell renal cell carcinoma aggressiveness. *Cancer Cell* **41**, 139–163 (2023).
48 Kutikov, A. et al. Anatomic features of enhancing renal masses predict malignant and high-grade pathology: a preoperative nomogram using the RENAL Nephrometry score. *Eur. Urol.* **60**, 241–248 (2011).
49 Zheng, Z., Chen, Z., Xie, Y., Zhong, Q. & Xie, W. Development and validation of a CT-based nomogram for preoperative prediction of clear cell renal cell carcinoma grades. *Eur. Radiol.* **31**, 6078–6086 (2021).
50 Nie, P. et al. A CT-based deep learning radiomics nomogram outperforms the existing prognostic models for outcome prediction in clear cell renal cell carcinoma: a multicenter study. *Eur. Radiol.* **33**, 8858–8868 (2023).
51 Topol, E. J. High-performance medicine: the convergence of human and artificial intelligence. *Nat. Med.* **25**, 44–56 (2019).
52 Rajpurkar, P., Chen, E., Banerjee, O. & Topol, E. J. AI in health and medicine. *Nat. Med.* **28**, 31–38 (2022).
53 Paul, M., Ganguli, S. & Dziugaite, G. K. Deep learning on a data diet: Finding important examples early in training. *Adv. Neural Inf. Process. Syst.* **34**, 20596–20607 (2021).
54 Silverman, S. G. et al. In Pursuit of KI-RADS: toward a single, evidence-based imaging classification of renal masses. *Radiology* **314**, e240308 (2025).
55 Yap, F. Y. et al. Shape and texture-based radiomics signature on CT effectively discriminates benign from malignant renal masses. *Eur. Radiol.* **31**, 1011–1021 (2021).
56 Demirjian, N. L. et al. CT-based radiomics stratification of tumor grade and TNM stage of clear cell renal cell carcinoma. *Eur. Radiol.* **32**, 2552–2563 (2022).
57 Xiong, Y. et al. Artificial intelligence links CT images to pathologic features and survival outcomes of renal masses. *Nat. Commun.* **16**, 1425 (2025).
58 Moor, M. et al. Foundation models for generalist medical artificial intelligence. *Nature* **616**, 259–265 (2023).
59 Tu, T. et al. Towards generalist biomedical AI. *NEJM AI* **1**, AIoa2300138 (2024).
60 Isensee, F., Jaeger, P. F., Kohl, S. A., Petersen, J. & Maier-Hein, K. H. nnU-Net: a self-configuring method for deep learning-based biomedical image segmentation. *Nat. Methods* **18**, 203–211 (2021).
61 Dai, C. et al. Deep learning assessment of small renal masses at contrast-enhanced multiphase CT. *Radiology* **311**, e232178 (2024).
62 Rosenfeld, E., Nakkiran, P., Pouransari, H., Tuzel, O. & Faghri, F. Ape: Aligning pretrained encoders to quickly learn aligned multimodal representations. Preprint at *arXiv* https://doi.org/10.48550/arXiv.2210.03927 (2022).
63 Yang, J., Wu, Z., Zhao, Y. & Ma, Y. Language-image alignment with fixed text encoders. Preprint at *arXiv* https://doi.org/10.48550/arXiv.2506.04209 (2025).
64 Boecking, B. et al. Making the most of text semantics to improve biomedical vision–language processing. In *Proc. Eur. Conf. Comput. Vis.* 1–21 (Springer, 2022).
65 Devlin, J., Chang, M. W., Lee, K. & Toutanova, K. BERT: Pre-training of deep bidirectional transformers for language understanding. In *Proc. Conf. North Am. Chapter Assoc. Comput. Linguist. Hum. Lang. Technol.* **1**, 4171–4186 (2019).
66 Liu, Y. et al. RoBERTa: A robustly optimized BERT pretraining approach. Preprint at *arXiv* https://doi.org/10.48550/arXiv.1907.11692 (2019).
67 Hu, E. J. et al. LoRA: Low-rank adaptation of large language models. In *Proc. Int. Conf. Learn. Represent.* (2022).
68 Johnson, A. E. et al. MIMIC-CXR, a de-identified publicly available database of chest radiographs with free-text reports. *Sci. Data* **6**, 317 (2019).
69 Gao, T., Yao, X. & Chen, D. SimCSE: Simple contrastive learning of sentence embeddings. In *Proc. Conf. Empir. Methods Nat. Lang. Process.* 6894–6910 (2021).
70 Loshchilov, I. & Hutter, F. Decoupled weight decay regularization. Preprint at *arXiv* https://doi.org/10.48550/arXiv.1711.05101 (2017).
71 Chen, T., Xu, B., Zhang, C. & Guestrin, C. Training deep nets with sublinear memory cost. Preprint at *arXiv* https://doi.org/10.48550/arXiv.1604.06174 (2016).
72 Myronenko, A. 3D MRI brain tumor segmentation using autoencoder regularization. In *Proc. Int. MICCAI Brainlesion Workshop* 311–320 (Springer, 2018).
73 Dosovitskiy, A. et al. An image is worth 16x16 words: Transformers for image recognition at scale. Preprint at *arXiv* https://doi.org/10.48550/arXiv.2010.11929 (2020).
74 Harrell, F. E., Califf, R. M., Pryor, D. B., Lee, K. L. & Rosati, R. A. Evaluating the yield of

medical tests. *JAMA* **247**, 2543–2546 (1982).

75 Brier, G. W. Verification of forecasts expressed in terms of probability. *Mon. Weather Rev.* **78**, 1–3 (1950).

76 Papineni, K., Roukos, S., Ward, T. & Zhu, W. J. BLEU: a method for automatic evaluation of machine translation. In *Proc. 40th Annu. Meet. Assoc. Comput. Linguist.* 311–318 (2002).

77 Banerjee, S. & Lavie, A. METEOR: an automatic metric for MT evaluation with improved correlation with human judgments. In *Proc. ACL Workshop Intrinsic Extrinsic Eval. Meas. Mach. Transl. Summarization* 65–72 (2005).

78 Lin, C. Y. ROUGE: a package for automatic evaluation of summaries. In *Text Summarization Branches Out* 74–81 (2004).

79 Kvamme, H., Borgan, Ø. & Scheel, I. Time-to-event prediction with neural networks and Cox regression. *J. Mach. Learn. Res.* **20**, 1–30 (2019).

80 Liu, H., Li, C., Wu, Q. & Lee, Y. J. Visual instruction tuning. *Adv. Neural Inf. Process. Syst. (NeurIPS)* **36**, 34892–34916 (2023).

81 Li, C. et al. LLaVA-Med: Training a large language-and-vision assistant for biomedicine in one day. *Adv. Neural Inf. Process. Syst. (NeurIPS)* **36**, 28541–28564 (2023).

82 Labrak, Y. et al. BioMistral: A collection of open-source pretrained large language models for medical domains. In *Findings Assoc. Comput. Linguist. ACL* 5848–5864 (2024).

83 Alsentzer, E. et al. Publicly available clinical BERT embeddings. In *Proc. 2nd Clinical NLP Workshop* 72–78 (2019).

84 Tao, Y. et al. A disease-centric vision-language foundation model for precision oncology in kidney cancer. https://doi.org/10.5281/zenodo.19565967 (2026).

**Acknowledgments**

We thank pathologists Q. Sun (Fudan University) and H. Lin (the First People's Hospital of Lianyungang) for re-evaluating pathologic slides. We thank radiologists X. Li (Fudan University) and L. Yao (Zhejiang University) for assessing R.E.N.A.L. score. The computations in this research were performed using the CFFF platform of Fudan University. This work was conducted in part during Y.T.'s internship at Microsoft Research Asia.

**Funding**

S.W. discloses support for the research of this work from the National Key Research and Development Program of China [2024YFF1207500] and the International Science and Technology Cooperation Program under the 2023 Shanghai Action Plan for Science [23410710400]. Y.X. discloses support for the research of this work from the Excellent Young Scientists Fund of Fujian Provincial Natural Science Foundation [2026D016]. J.G. discloses support for the research of this work from the National Natural Science Foundation of China [81974393]. J.Y. discloses support for the research of this work from the Hexi University President Fund Innovation Team Project [CXTD2022012]. The funders had no role in the study design, data collection, data analysis, data interpretation or writing of the manuscript.

**Author contributions**

Y. T., Z. Z., Z. W., X. L., and F. C. contributed to data acquisition, analysis, interpretation, statistical analysis, and manuscript drafting. K. W., C. W., X. Z., S. Z., J. Y., X. Jin, X. Jiang, Y. Y., D. L., and L. Q. provided technical and material support and participated in statistical analyses. Y. T., K. W., Z. W., X. L., and S. W. were responsible for algorithm development and statistical analyses. Z. S., J. G., N. Y., Y. X., and S. W. designed the study, interpreted data, drafted and revised the manuscript, secured funding, and supervised the study. All authors critically reviewed the manuscript, approved the final version, and agree to be accountable for all aspects of the work.

**Competing interests**

The authors declare no competing interests.

## Figure legends

**Figure 1: Overview of the RenalCLIP design. a,b,** The study was developed using a large-scale, multi-center dataset. For detailed sample sizes and class distributions for each specific downstream task across all cohorts, please refer to Supplementary Table 2. **a,** Distribution of patients and CT volumes across the pre-training, fine-tuning, and test cohorts (dots, patients; bars, CT volumes). **b,** The cohorts encompass a diverse range of pathologic subtypes and CT phase availability, ensuring comprehensive model training and evaluation. **c,** Schematic of the RenalCLIP framework illustrating the two-stage pre-training pipeline (left and middle panels) and its clinical application (right panel). The left panel depicts uni-modal knowledge enhancement: the image encoder is pre-trained via multi-task learning using structured labels, while the text encoder (Llama3-8B) concurrently undergoes a three-step adaptation (Phase I–III, utilizing LLM2Vec technology) to transform LLM into a text feature extractor. The middle panel illustrates cross-modal alignment, where the pre-trained image and text encoders are aligned using a CLIP-based contrastive learning framework. The right panel demonstrates the subsequent application to the three primary clinical downstream domains: characterization, diagnosis, and prognosis. See Methods for details. **d–f,** Summary of model performance via radar plots, demonstrating the consistently high performance of RenalCLIP against baseline models across these primary domains. The comparison is shown for the internal test cohort (**d**), the combined external test cohort (**e**), and the TCIA cohort (**f**). In these plots, performance on characterization and diagnosis tasks is quantified by the area under the receiver operating characteristic curve (ROC AUC), while prognosis is quantified by Harrell's C-index. Icons in **c** for the report and kidney are by Iconjam and Futuer via www.flaticon.com. Icons representing the expert and prognosis are from Microsoft PowerPoint. Source data are provided as a Source Data file.

**Figure 2: Performance of RenalCLIP on renal mass characterization. a,** Rose plots summarizing the performance on the five R.E.N.A.L. score components show that RenalCLIP outperforms baseline models on most features in both the internal test cohort and the combined external test cohort (comprising the Xiamen, Shandong, Ruijin, Lianyun, and Zhangye cohorts). **b,** Bar charts detailing the performance across the five proprietary external cohorts and the public TCIA cohort further demonstrate the model's generalizability. **c,** Radiology report generation performance on the internal test cohort. Top, RenalCLIP achieves the highest scores across all standard language metrics when benchmarked against several established generative models. Bottom, Expert evaluation of generated reports. Box plots display the distribution of accuracy scores averaged across three radiologists. For the box plots, the center line indicates the median, the box limits represent the 25th and 75th percentiles, and the whiskers extend to the minimum and maximum values. The scoring criteria (0–5 scale) were adapted from HistoGPT[44] (see Methods for details). The jittered data points represent individual case scores, showing that RenalCLIP achieves the highest median rating and consistent clinical alignment compared to baselines. **d,** A representative qualitative example illustrates the high concordance between a report generated by RenalCLIP and the ground-truth report written by a human expert (see Supplementary Fig. 3 for comprehensive comparisons). Unless otherwise stated, all downstream evaluations utilized a standardized input protocol (single aligned contrast-enhanced CT volume, prioritizing the arterial phase). Performance on anatomical characterization (**a, b**) is quantified by ROC AUC. Capped error bars in **b** and **c** (top) represent 95% CIs estimated via 1,000 bootstrap resamples and the centers correspond to the point estimates of each metric. Statistical comparisons (including significance tests) for R.E.N.A.L. scoring are detailed in Supplementary Tables 4–8, while comprehensive results for report generation are provided in Supplementary Tables 12–17. Source data are provided as a Source Data file.

**Figure 3: Performance of RenalCLIP on diagnosis of renal masses. a,** Rose plots summarizing diagnostic performance demonstrate RenalCLIP's superiority in both malignancy and aggressiveness prediction on the internal test and combined external test cohorts. Sample sizes (*n* independent patients) are as follows: *n* = 100 (internal test) and *n* = 1,017 (combined external test) for malignancy prediction; *n* = 99 (internal test) and *n* = 1,005 (combined external test) for aggressiveness prediction. **b,** A comprehensive comparison across five key performance metrics in the combined external cohort further confirms RenalCLIP's robust diagnostic precision. **c,** Bar charts comparing the ROC AUC for malignancy (top) and aggressiveness (bottom) prediction across the five proprietary external cohorts highlight RenalCLIP's consistent performance advantage. **d, e,** Representative ROC curves visualizing the performance margin for malignancy diagnosis on the internal test cohort (**d**) and the combined external test cohort (**e**). **f, g,** Kaplan-Meier analysis in the TCIA cohort validates the clinical relevance of the aggressiveness prediction, showing significant RFS stratification for patients based on RenalCLIP's predictions (**f**) but not for those based on the baseline model, CNN (**g**). **h,** Subgroup analysis on the combined external test cohort, stratified by tumor size (Small Renal Masses (SRMs), ≤ 4 cm vs. large masses, > 4 cm). The left panel displays the performance for malignancy classification, while the right panel shows the performance for aggressiveness classification. **i,** Impact of pre-training data scale. External generalization improves monotonically with increasing pre-training data volume (0–100%), contrasting with the substantial performance drop on external data observed at 0% (CNN (Rand Init) baseline). **j,** Impact of pre-training data diversity. The multi-center (controlled at ~45% data volume) pre-trained model outperforms the single-center (Zhongshan-only) model on external test sets, validating the benefit of source diversity for robust generalization. Hazard ratios (HR) and their 95% CIs were derived from univariate Cox proportional hazards models, while *P*-values for overall curve comparison were calculated using a two-sided log-rank test. Shaded areas in the survival curves and capped error bars in the bar charts represent 95% CIs. For bar charts, CIs were estimated via 1,000 bootstrap resamples, and the centers of the bars correspond to the exact ROC AUC point estimates. Statistical comparisons (including significance tests) for malignancy and aggressiveness diagnosis are detailed in Supplementary Tables 18–19. Source data are provided as a Source Data file.

**Figure 4: Performance of RenalCLIP on prognosis of renal tumors. a,** Rose plots comparing Harrell's C-index for recurrence-free survival (RFS), disease-specific survival (DSS), and overall survival (OS) demonstrate the high prognostic accuracy of RenalCLIP across the internal test cohort, the combined external test cohort, and the TCIA cohort. The total number of independent patients evaluated across these endpoints were $n$ = 52 (internal test), $n$ = 224 (combined external test), and $n$ = 420 (TCIA). For the exact sample sizes per specific survival endpoint, please refer to Supplementary Table 2. **b–e,** Kaplan-Meier analyses for RFS in the TCIA cohort validate the clinical relevance of the derived risk scores. While patient stratification based on risk scores from baseline models, including CNN (**b**), Merlin (**c**), and CT-FM (**d**) failed to show statistically significant survival differences, the stratification by RenalCLIP provided clear and significant prognostic separation between high- and low-risk groups (**e**). **f–h,** Time-dependent analyses in the TCIA cohort further confirm RenalCLIP's durable prognostic power, showing its sustained superiority over baseline models across all evaluated time points as measured by ROC AUC (**f**), C-index (**g**), and Brier score (**h**). **i,** Forest plot illustrating the prognostic performance for RFS specifically within the SRMs subgroup of the TCIA cohort. In the forest plot (**i**), the markers represent the exact Hazard ratios (HR), and the error bars represent the 95% CIs. HR and their 95% CIs shown on survival curves (**b–e**) and forest plot (**i**) were derived from univariate Cox hazards models; $P$-values for **b–e** were from a two-sided log-rank test. Shaded areas in the survival curves (**b–e**) represent 95% CIs of survival probability. Source data are provided as a Source Data file.

**Figure 5: Performance of RenalCLIP on zero-shot transfer and data-efficient fine-tuning. a,** Cross-modal retrieval performance on five cohorts, quantified by Recall@K (K = 1, 3, 5). The plots for text-to-image (t2i, top) and image-to-text (i2t, bottom) retrieval show that RenalCLIP achieves higher performance compared to other vision-language models. **b, c,** Zero-shot classification performance for malignancy (**b**) and aggressiveness (**c**) prediction on three test cohorts. Bar plots show the mean ROC AUC from a stochastic prompt sampling strategy, while star markers indicate the performance using a deterministic maximum similarity strategy. **d–g,** Data-efficient fine-tuning performance on the internal (**d, f**) and combined external (**e, g**) cohorts for malignancy (**d, e**) and aggressiveness (**f, g**) prediction. The plots show model ROC AUC as a function of the fraction of training data used for fine-tuning. Dashed horizontal lines indicate the zero-shot performance of each model for reference. The results highlight both RenalCLIP's high data efficiency (achieving strong performance with only 20–40% of data) and the benefit of its pre-training, as its zero-shot performance frequently surpasses the fully fine-tuned performance of baseline models. Error bars in **a–c** represent 95% CIs estimated via 1,000 bootstrap resamples. The centers correspond to the exact point estimates in a, and the average ROC AUC from the resamples in **b** and **c**. In **d–g**, data points for training fractions less than 100% represent the mean ROC AUC from five random sampling runs, with capped error bars indicating the standard deviation; the 100% data point represents a single run. Source data are provided as a Source Data file.

# Figure 1

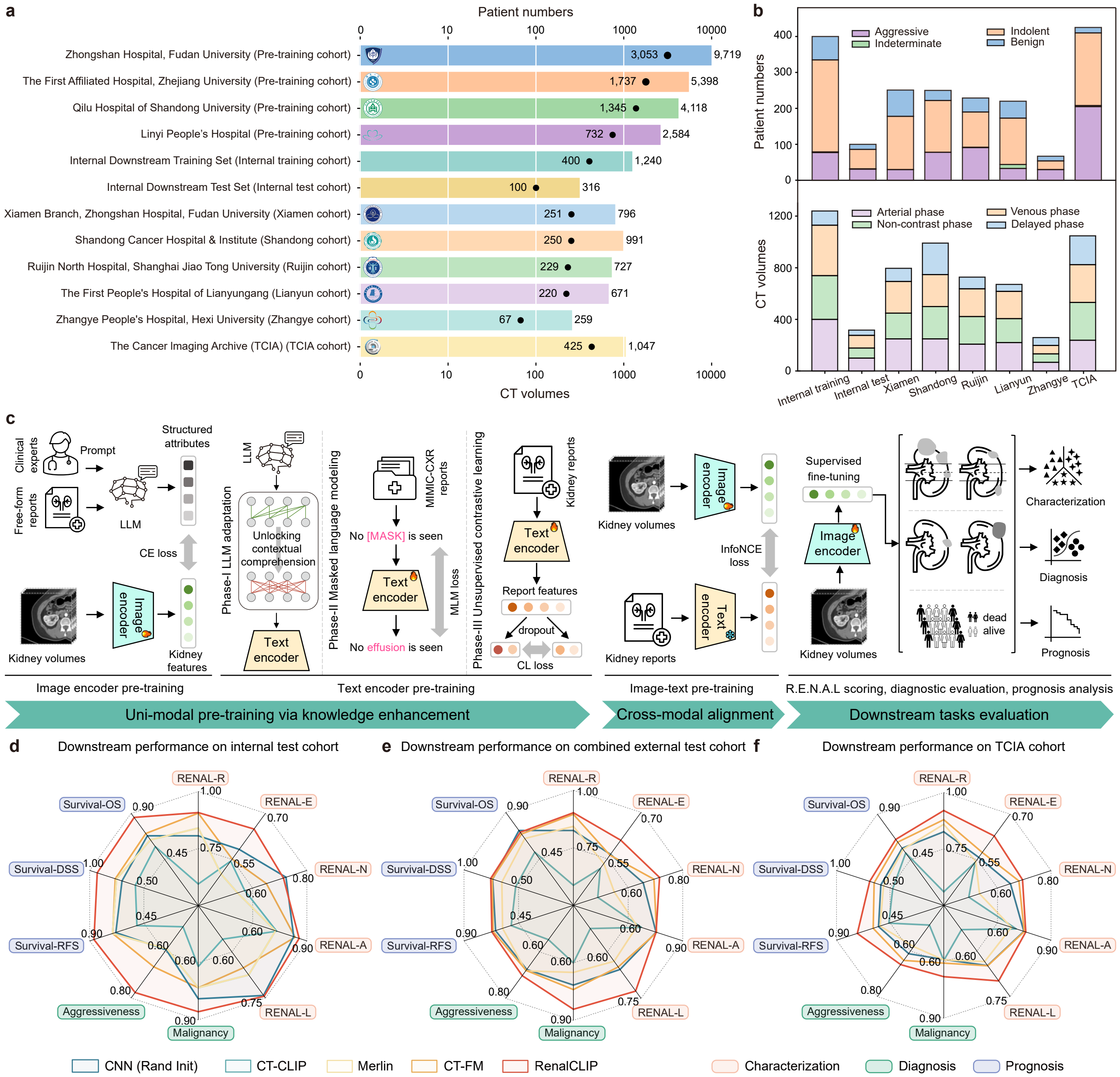
a
Patient numbers
CT volumes
Zhongshan Hospital, Fudan University (Pre-training cohort)
The First Affiliated Hospital, Zhejiang University (Pre-training cohort)
Qilu Hospital of Shandong University (Pre-training cohort)
Linyi People's Hospital (Pre-training cohort)
Internal Downstream Training Set (Internal training cohort)
Internal Downstream Test Set (Internal test cohort)
Xiamen Branch, Zhongshan Hospital, Fudan University (Xiamen cohort)
Shandong Cancer Hospital & Institute (Shandong cohort)
Ruijin North Hospital, Shanghai Jiao Tong University (Ruijin cohort)
The First People's Hospital of Lianyungang (Lianyun cohort)
Zhangye People's Hospital, Hexi University (Zhangye cohort)
The Cancer Imaging Archive (TCIA) (TCIA cohort)
3,053 9,719
1,737 5,398
1,345 4,118
732 2,584
400 1,240
100 316
251 796
250 991
229 727
220 671
67 259
425 1,047
b
Patient numbers
Aggressive
Indeterminate
Indolent
Benign
CT volumes
Arterial phase
Non-contrast phase
Venous phase
Delayed phase
Internal training
Internal test
Xiamen
Shandong
Ruijin
Lianyun
Zhangye
TCIA
c
Clinical experts
Prompt
Free-form reports
LLM
Structured attributes
CE loss
Kidney volumes
Image encoder
Kidney features
Image encoder pre-training
Phase-I LLM adaptation
Unlocking contextual comprehension
Text encoder
Phase-II Masked language modeling
MIMIC-CXR reports
No [MASK] is seen
No effusion is seen
MLM loss
Phase-III Unsupervised contrastive learning
Kidney reports
Report features
dropout
CL loss
Text encoder pre-training
InfoNCE loss
Image-text pre-training
Supervised fine-tuning
Characterization
Diagnosis
dead
alive
Prognosis
R.E.N.A.L scoring, diagnostic evaluation, prognosis analysis
Uni-modal pre-training via knowledge enhancement
Cross-modal alignment
Downstream tasks evaluation
d Downstream performance on internal test cohort
e Downstream performance on combined external test cohort
f Downstream performance on TCIA cohort
RENAL-R
RENAL-E
RENAL-N
RENAL-A
RENAL-L
Malignancy
Aggressiveness
Survival-RFS
Survival-DSS
Survival-OS
CNN (Rand Init)
CT-CLIP
Merlin
CT-FM
RenalCLIP
Characterization
Diagnosis
Prognosis

# Figure 2

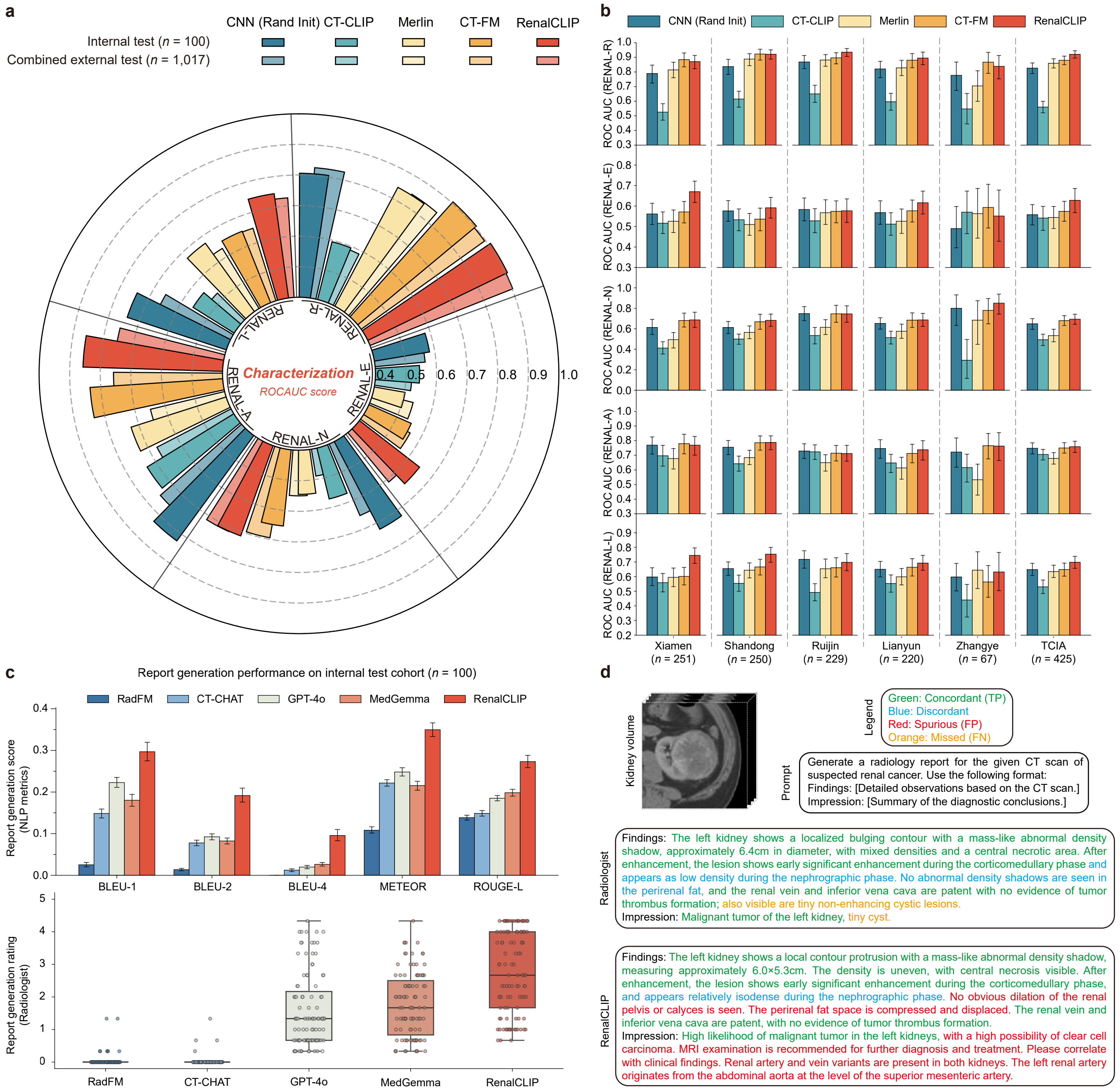

a
CNN (Rand Init) CT-CLIP Merlin CT-FM RenalCLIP
Internal test (n = 100)
Combined external test (n = 1,017)
Characterization
ROCAUC score
RENAL-R
RENAL-E
RENAL-N
RENAL-A
RENAL-L
0.4 0.5 0.6 0.7 0.8 0.9 1.0
b
CNN (Rand Init) CT-CLIP Merlin CT-FM RenalCLIP
ROC AUC (RENAL-R)
ROC AUC (RENAL-E)
ROC AUC (RENAL-N)
ROC AUC (RENAL-A)
ROC AUC (RENAL-L)
Xiamen (n = 251)
Shandong (n = 250)
Ruijin (n = 229)
Lianyun (n = 220)
Zhangye (n = 67)
TCIA (n = 425)
c
Report generation performance on internal test cohort (n = 100)
RadFM CT-CHAT GPT-4o MedGemma RenalCLIP
Report generation score (NLP metrics)
BLEU-1 BLEU-2 BLEU-4 METEOR ROUGE-L
Report generation rating (Radiologist)
RadFM CT-CHAT GPT-4o MedGemma RenalCLIP
d
Kidney volume
Legend
Green: Concordant (TP)
Blue: Discordant
Red: Spurious (FP)
Orange: Missed (FN)
Prompt
Generate a radiology report for the given CT scan of suspected renal cancer. Use the following format:
Findings: [Detailed observations based on the CT scan.]
Impression: [Summary of the diagnostic conclusions.]
Radiologist
Findings: The left kidney shows a localized bulging contour with a mass-like abnormal density shadow, approximately 6.4cm in diameter, with mixed densities and a central necrotic area. After enhancement, the lesion shows early significant enhancement during the corticomedullary phase and appears as low density during the nephrographic phase. No abnormal density shadows are seen in the perirenal fat, and the renal vein and inferior vena cava are patent with no evidence of tumor thrombus formation; also visible are tiny non-enhancing cystic lesions.
Impression: Malignant tumor of the left kidney, tiny cyst.
RenalCLIP
Findings: The left kidney shows a local contour protrusion with a mass-like abnormal density shadow, measuring approximately 6.0×5.3cm. The density is uneven, with central necrosis visible. After enhancement, the lesion shows early significant enhancement during the corticomedullary phase, and appears relatively isodense during the nephrographic phase. No obvious dilation of the renal pelvis or calyces is seen. The perirenal fat space is compressed and displaced. The renal vein and inferior vena cava are patent, with no evidence of tumor thrombus formation.
Impression: High likelihood of malignant tumor in the left kidneys, with a high possibility of clear cell carcinoma. MRI examination is recommended for further diagnosis and treatment. Please correlate with clinical findings. Renal artery and vein variants are present in both kidneys. The left renal artery originates from the abdominal aorta at the level of the superior mesenteric artery.

# Figure 3

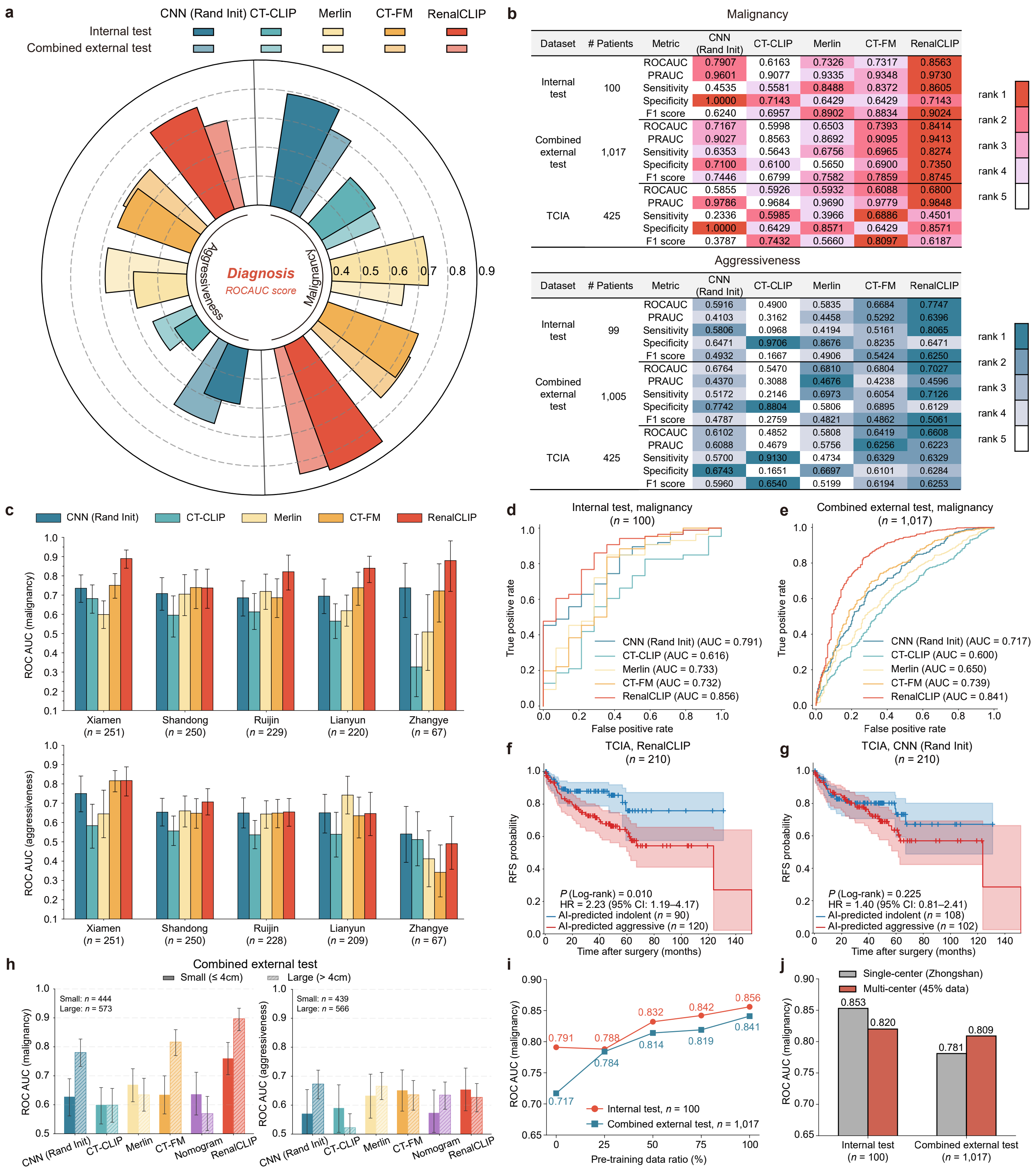

**Malignancy**

| Dataset | # Patients | Metric | CNN (Rand Init) | CT-CLIP | Merlin | CT-FM | RenalCLIP |
|---|---|---|---|---|---|---|---|
| Internal test | 100 | ROCAUC | 0.7907 | 0.6163 | 0.7326 | 0.7317 | 0.8563 |
| | | PRAUC | 0.9601 | 0.9077 | 0.9335 | 0.9348 | 0.9730 |
| | | Sensitivity | 0.4535 | 0.5581 | 0.8488 | 0.8372 | 0.8605 |
| | | Specificity | 1.0000 | 0.7143 | 0.6429 | 0.6429 | 0.7143 |
| | | F1 score | 0.6240 | 0.6957 | 0.8902 | 0.8834 | 0.9024 |
| Combined external test | 1,017 | ROCAUC | 0.7167 | 0.5998 | 0.6503 | 0.7393 | 0.8414 |
| | | PRAUC | 0.9027 | 0.8563 | 0.8692 | 0.9095 | 0.9413 |
| | | Sensitivity | 0.6353 | 0.5643 | 0.6756 | 0.6965 | 0.8274 |
| | | Specificity | 0.7100 | 0.6100 | 0.5650 | 0.6900 | 0.7350 |
| | | F1 score | 0.7446 | 0.6799 | 0.7582 | 0.7859 | 0.8745 |
| TCIA | 425 | ROCAUC | 0.5855 | 0.5926 | 0.5932 | 0.6088 | 0.6800 |
| | | PRAUC | 0.9786 | 0.9684 | 0.9690 | 0.9779 | 0.9848 |
| | | Sensitivity | 0.2336 | 0.5985 | 0.3966 | 0.6886 | 0.4501 |
| | | Specificity | 1.0000 | 0.6429 | 0.8571 | 0.6429 | 0.8571 |
| | | F1 score | 0.3787 | 0.7432 | 0.5660 | 0.8097 | 0.6187 |

**Aggressiveness**

| Dataset | # Patients | Metric | CNN (Rand Init) | CT-CLIP | Merlin | CT-FM | RenalCLIP |
|---|---|---|---|---|---|---|---|
| Internal test | 99 | ROCAUC | 0.5916 | 0.4900 | 0.5835 | 0.6684 | 0.7747 |
| | | PRAUC | 0.4103 | 0.3162 | 0.4458 | 0.5292 | 0.6396 |
| | | Sensitivity | 0.5806 | 0.0968 | 0.4194 | 0.5161 | 0.8065 |
| | | Specificity | 0.6471 | 0.9706 | 0.8676 | 0.8235 | 0.6471 |
| | | F1 score | 0.4932 | 0.1667 | 0.4906 | 0.5424 | 0.6250 |
| Combined external test | 1,005 | ROCAUC | 0.6764 | 0.5470 | 0.6810 | 0.6804 | 0.7027 |
| | | PRAUC | 0.4370 | 0.3088 | 0.4676 | 0.4238 | 0.4596 |
| | | Sensitivity | 0.5172 | 0.2146 | 0.6973 | 0.6054 | 0.7126 |
| | | Specificity | 0.7742 | 0.8804 | 0.5806 | 0.6895 | 0.6129 |
| | | F1 score | 0.4787 | 0.2759 | 0.4821 | 0.4862 | 0.5061 |
| TCIA | 425 | ROCAUC | 0.6102 | 0.4852 | 0.5808 | 0.6419 | 0.6608 |
| | | PRAUC | 0.6088 | 0.4679 | 0.5756 | 0.6256 | 0.6223 |
| | | Sensitivity | 0.5700 | 0.9130 | 0.4734 | 0.6329 | 0.6329 |
| | | Specificity | 0.6743 | 0.1651 | 0.6697 | 0.6101 | 0.6284 |
| | | F1 score | 0.5960 | 0.6540 | 0.5199 | 0.6194 | 0.6253 |

# Figure 4

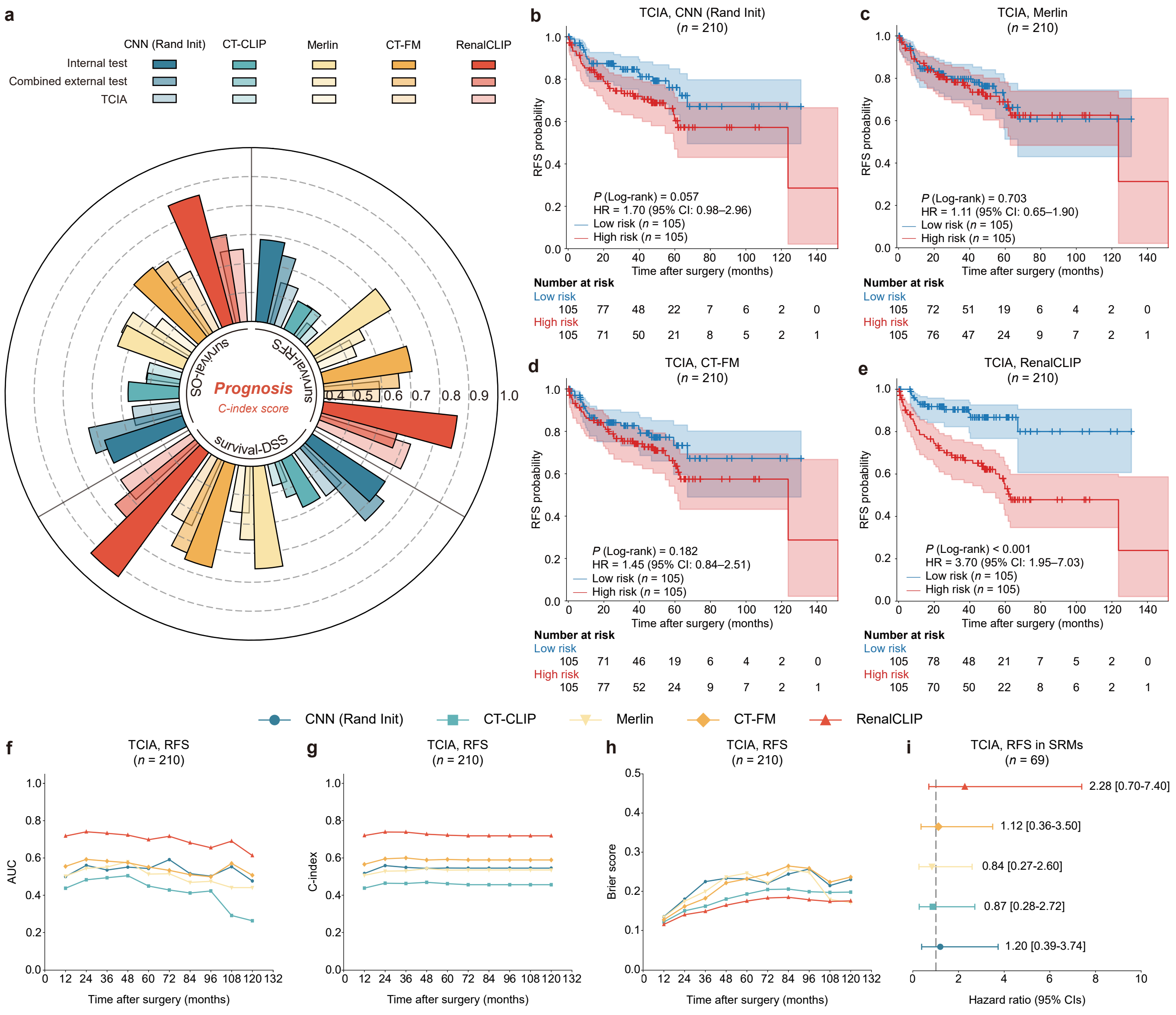

a
CNN (Rand Init)
CT-CLIP
Merlin
CT-FM
RenalCLIP
Internal test
Combined external test
TCIA
survival-OS
survival-RFS
survival-DSS
Prognosis
C-index score
0.4 0.5 0.6 0.7 0.8 0.9 1.0
b
TCIA, CNN (Rand Init)
(n = 210)
RFS probability
Time after surgery (months)
P (Log-rank) = 0.057
HR = 1.70 (95% CI: 0.98–2.96)
Low risk (n = 105)
High risk (n = 105)
Number at risk
Low risk
105 77 48 22 7 6 2 0
High risk
105 71 50 21 8 5 2 1
c
TCIA, Merlin
(n = 210)
RFS probability
Time after surgery (months)
P (Log-rank) = 0.703
HR = 1.11 (95% CI: 0.65–1.90)
Low risk (n = 105)
High risk (n = 105)
Number at risk
Low risk
105 72 51 19 6 4 2 0
High risk
105 76 47 24 9 7 2 1
d
TCIA, CT-FM
(n = 210)
RFS probability
Time after surgery (months)
P (Log-rank) = 0.182
HR = 1.45 (95% CI: 0.84–2.51)
Low risk (n = 105)
High risk (n = 105)
Number at risk
Low risk
105 71 46 19 6 4 2 0
High risk
105 77 52 24 9 7 2 1
e
TCIA, RenalCLIP
(n = 210)
RFS probability
Time after surgery (months)
P (Log-rank) < 0.001
HR = 3.70 (95% CI: 1.95–7.03)
Low risk (n = 105)
High risk (n = 105)
Number at risk
Low risk
105 78 48 21 7 5 2 0
High risk
105 70 50 22 8 6 2 1
CNN (Rand Init)
CT-CLIP
Merlin
CT-FM
RenalCLIP
f
TCIA, RFS
(n = 210)
AUC
Time after surgery (months)
g
TCIA, RFS
(n = 210)
C-index
Time after surgery (months)
h
TCIA, RFS
(n = 210)
Brier score
Time after surgery (months)
i
TCIA, RFS in SRMs
(n = 69)
2.28 [0.70-7.40]
1.12 [0.36-3.50]
0.84 [0.27-2.60]
0.87 [0.28-2.72]
1.20 [0.39-3.74]
Hazard ratio (95% CIs)

# Figure 5

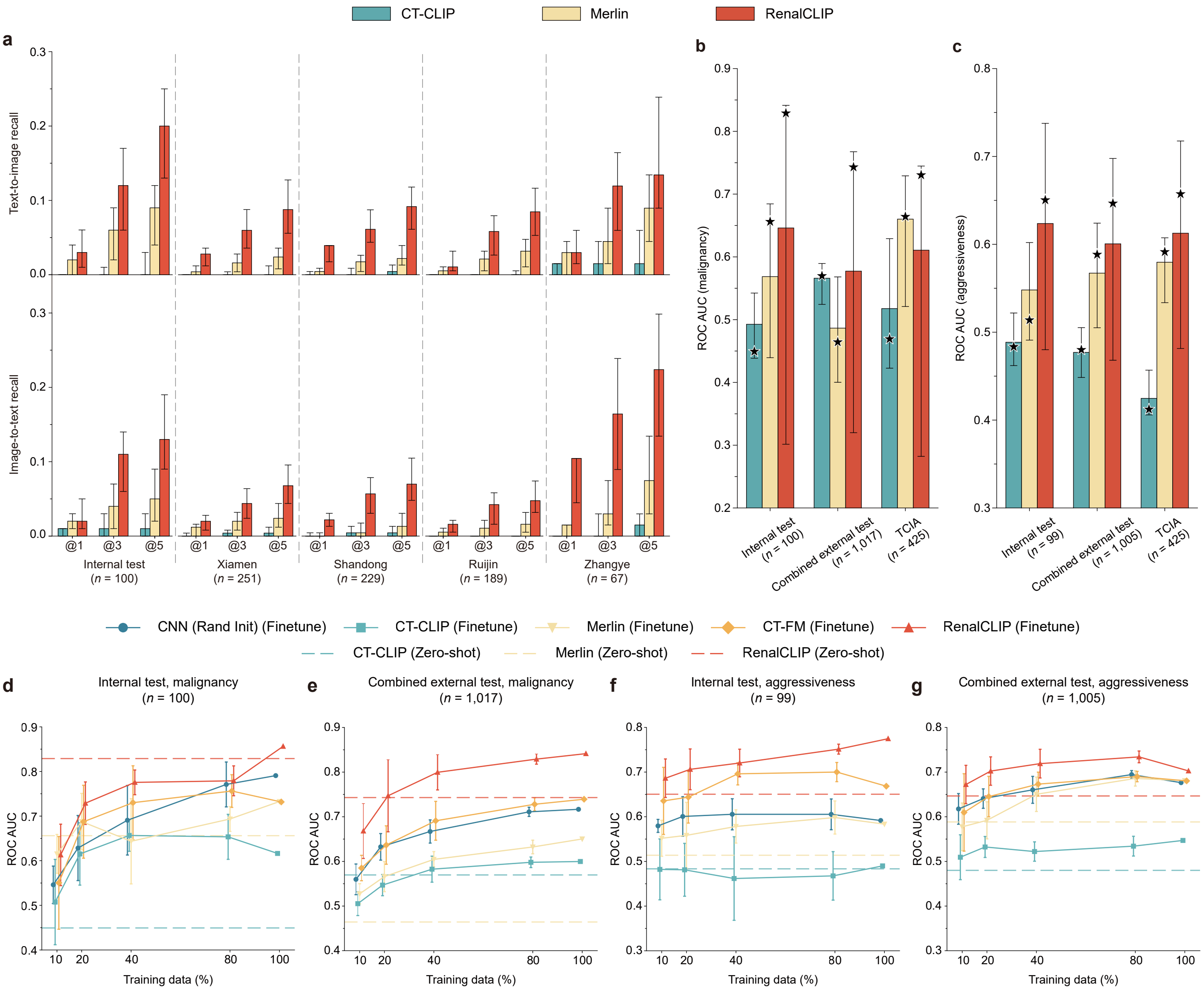
CT-CLIP
Merlin
RenalCLIP
a
Text-to-image recall
Image-to-text recall
@1
@3
@5
Internal test
(n = 100)
Xiamen
(n = 251)
Shandong
(n = 229)
Ruijin
(n = 189)
Zhangye
(n = 67)
b
ROC AUC (malignancy)
Internal test
(n = 100)
Combined external test
(n = 1,017)
TCIA
(n = 425)
c
ROC AUC (aggressiveness)
Internal test
(n = 99)
Combined external test
(n = 1,005)
TCIA
(n = 425)
CNN (Rand Init) (Finetune)
CT-CLIP (Finetune)
Merlin (Finetune)
CT-FM (Finetune)
RenalCLIP (Finetune)
CT-CLIP (Zero-shot)
Merlin (Zero-shot)
RenalCLIP (Zero-shot)
d
Internal test, malignancy
(n = 100)
e
Combined external test, malignancy
(n = 1,017)
f
Internal test, aggressiveness
(n = 99)
g
Combined external test, aggressiveness
(n = 1,005)
ROC AUC
Training data (%)

## Supplementary Materials

# A Disease-Centric Vision-Language Foundation Model for Precision Oncology in Kidney Cancer

### Contents

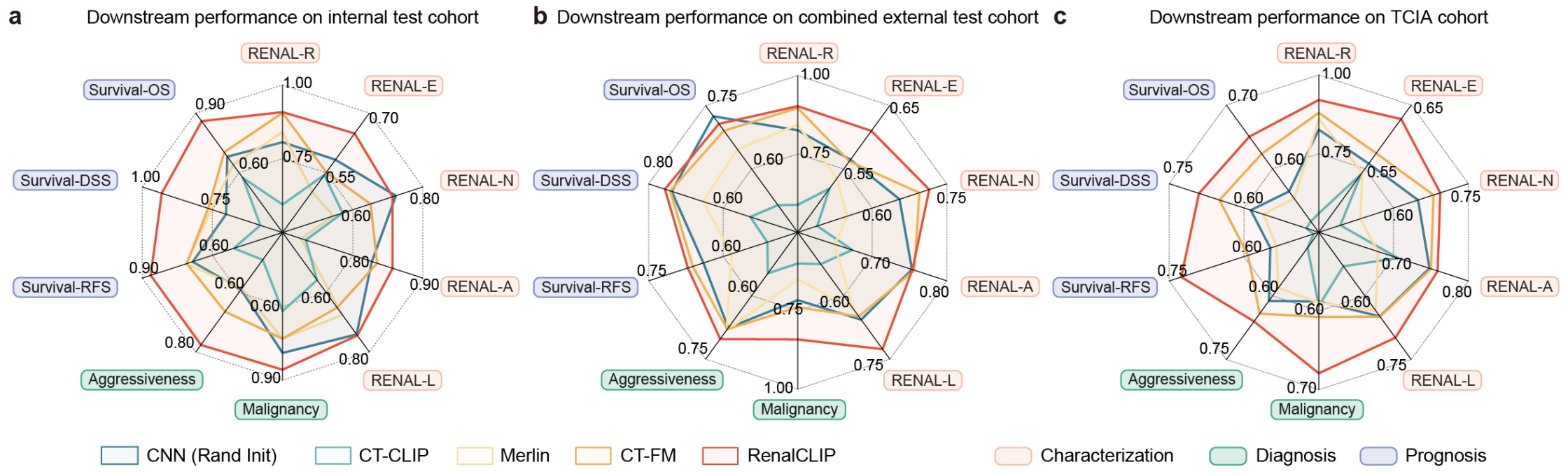

**Supplementary Figure 1: Adjusted-scale radar plots of model performance.** Alternative visualizations of Fig. 1d–f with dynamically scaled axes to minimize visual overlap. **a–c**, Radar plots for the internal test cohort, combined external test cohort and TCIA cohort, respectively. Performance for characterization and diagnosis tasks is quantified by the area under the receiver operating characteristic curve (ROC AUC), whereas prognosis is quantified by Harrell's C-index.

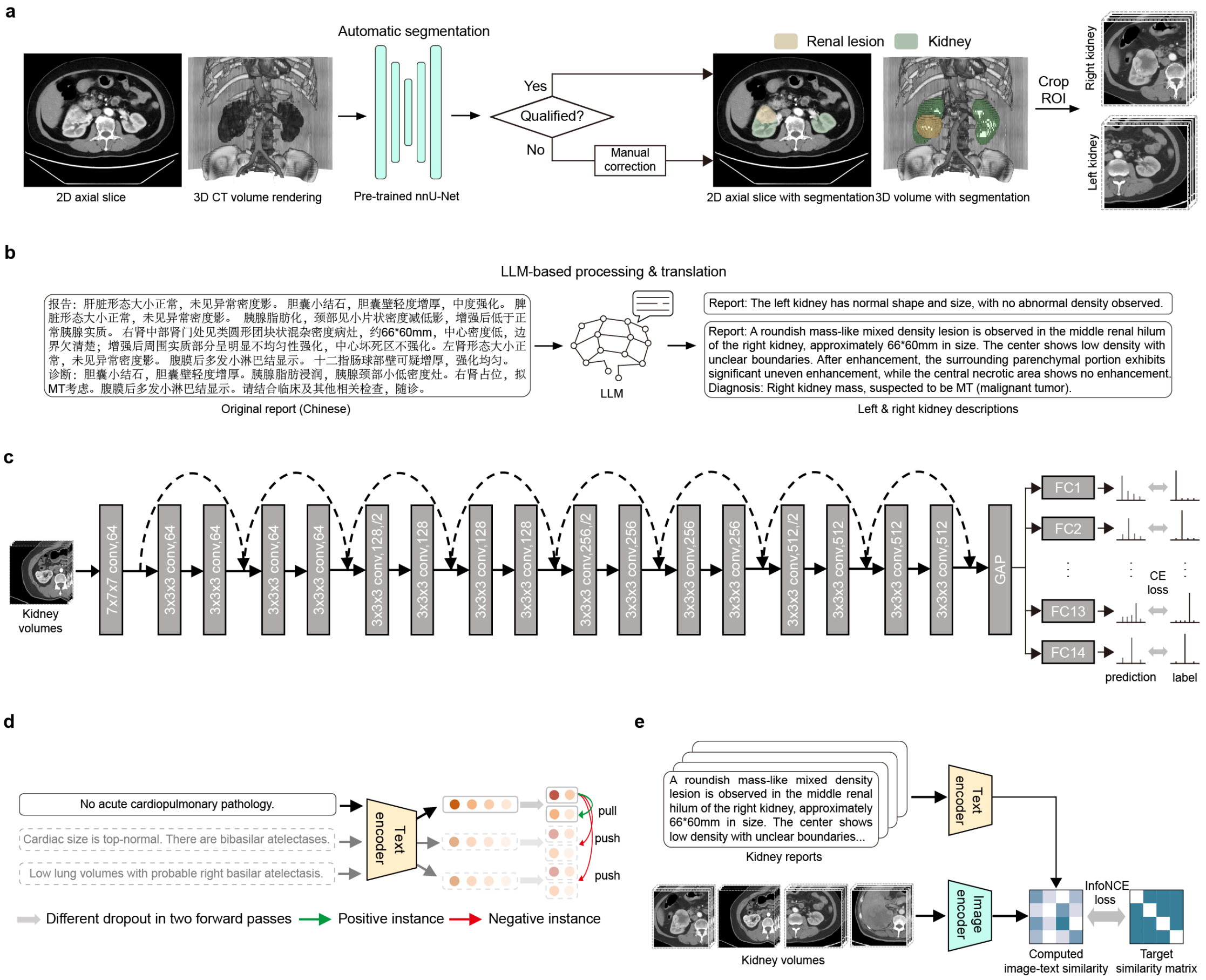

**Supplementary Figure 2: Data processing and RenalCLIP pre-training architecture. a,** Semi-automated region of interest (ROI) localization pipeline: initial nnU-Net segmentations of the kidney and lesion are generated on the contrast-enhanced phase and propagated to other spatially corresponding phases of the same patient. These segmentations undergo expert review and are manually corrected if necessary. Then, a bounding box is computed from the final mask to crop the single-kidney ROI. The 3D volume rendering shown was generated using 3D Slicer. **b,** Processing of original Chinese radiology reports, involving large language model (LLM)-based partitioning into left kidney and right kidney descriptions, followed by programmatic translation into English. **c,** Architecture of the knowledge-enhanced image encoder, featuring a 3D ResNet-18 backbone pre-trained via a multi-task learning framework to predict 14 structured attributes, each with a corresponding fully connected (FC) layer as its prediction head, optimized by cross-entropy (CE) loss. **d,** The SimCSE methodology for text encoder refinement, which maximizes the similarity between two distinct representations of a single input sentence (generated via different dropout masks, positive instance) while minimizing their similarity to other sentences in the batch (negative instance). **e,** The overall vision-language framework, where the knowledge-enhanced image and text encoders project images and their corresponding text into a joint embedding space for alignment via the InfoNCE loss objective.

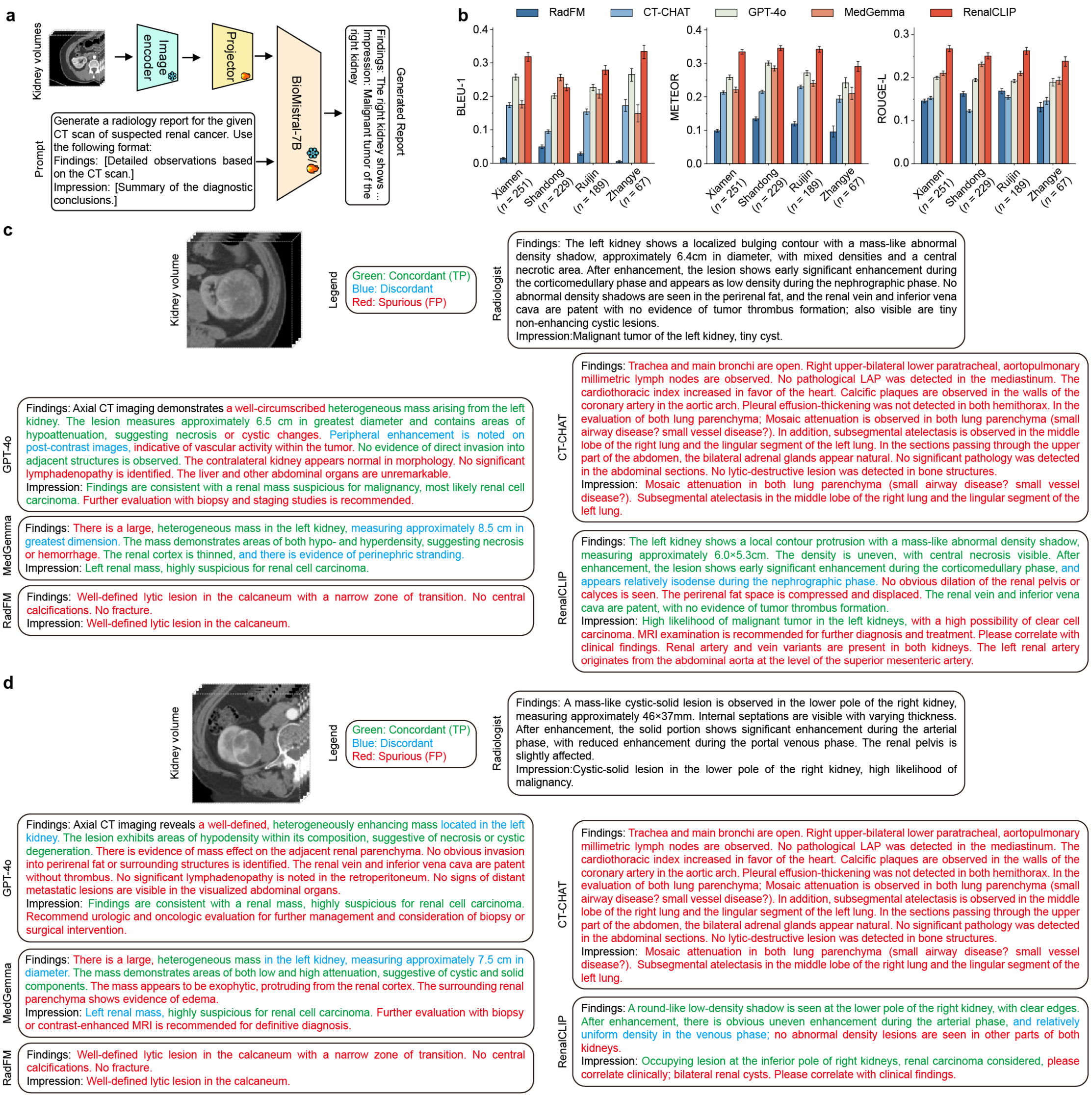

**Supplementary Figure 3: Architecture and evaluation of the radiology report generation model. a,** Schematic of the report generation workflow. The architecture adapts the LLaVA framework, bridging the frozen RenalCLIP image backbone and BioMistral-7B large language model (LLM) via a linear projector that maps visual features into the language embedding space. The two-stage training pipeline comprises (1) initial projector alignment and (2) subsequent Low-Rank Adaptation (LoRA)-based instruction tuning. **b,** Quantitative performance on four external cohorts, measured by standard language metrics (BLEU-1, METEOR, ROUGE-L). RenalCLIP outperforms all baselines, including generalist 2D vision-language models (GPT-4o, MedGemma) and 3D radiology-specific models (RadFM, CT-CHAT). **c,d,** Qualitative comparison of generated reports from two representative cases. Baseline models exhibit distinct failure modes, including non-specific text generation (RadFM), clinically irrelevant findings due to domain mismatch (CT-CHAT), or factual inaccuracies and omissions (GPT-4o, MedGemma). In contrast, RenalCLIP's reports demonstrate the best balance of factual accuracy and preservation of critical details. Capped error bars in **b** represent 95% CIs estimated via 1,000 bootstrap resamples, and the centers of the bars correspond to the exact point estimates of each metric.

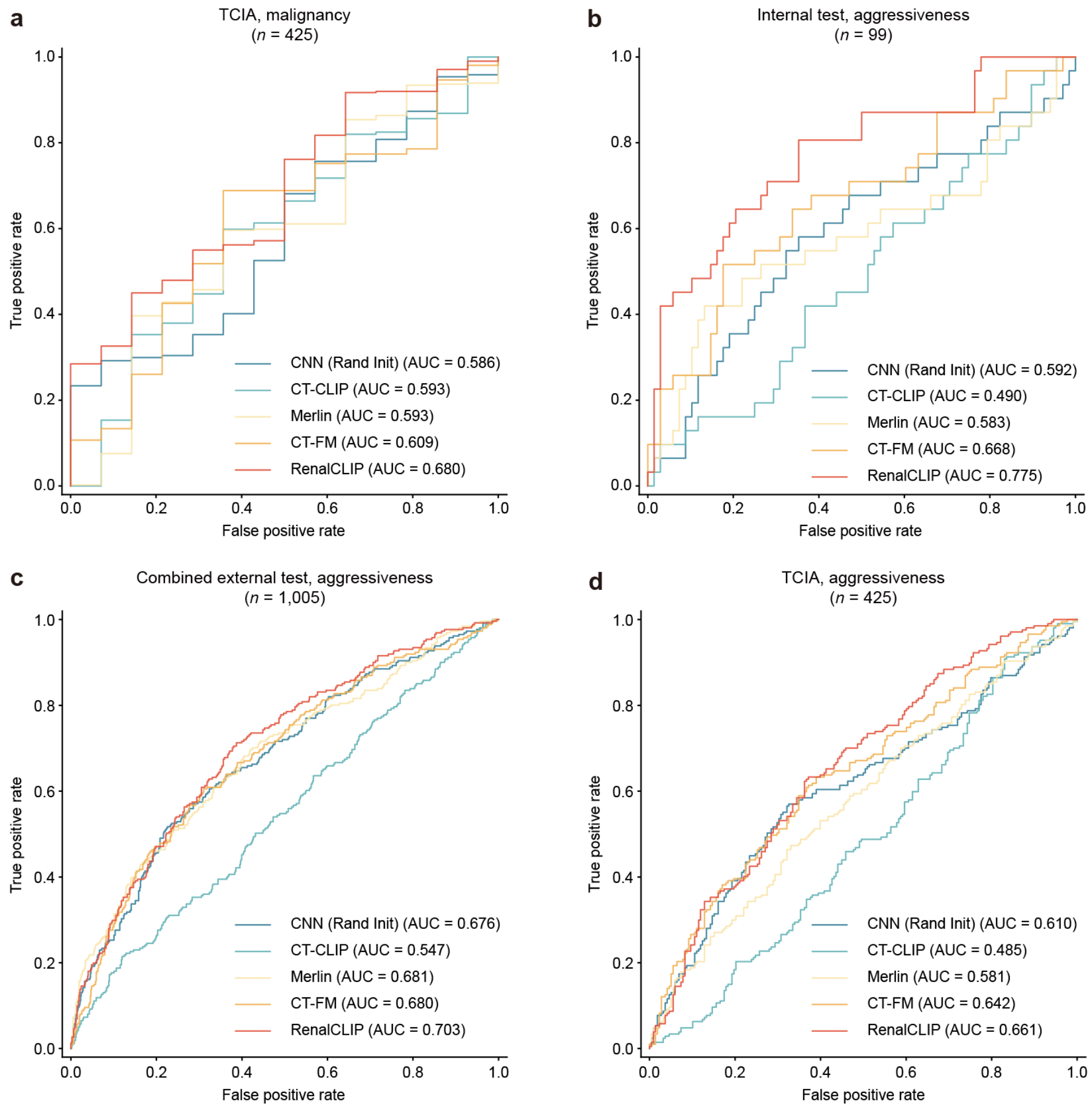

**Supplementary Figure 4: Supplementary ROC curves for diagnostic performance.** Receiver operating characteristic (ROC) curves for the diagnosis of malignancy in the TCIA cohort (**a**), and for the diagnosis of aggressiveness in the internal test cohort (**b**), the combined external test cohort (**c**), and the TCIA cohort (**d**). Across all cohorts and tasks shown, RenalCLIP consistently achieves a higher area under the curve (AUC) compared to the baseline models.

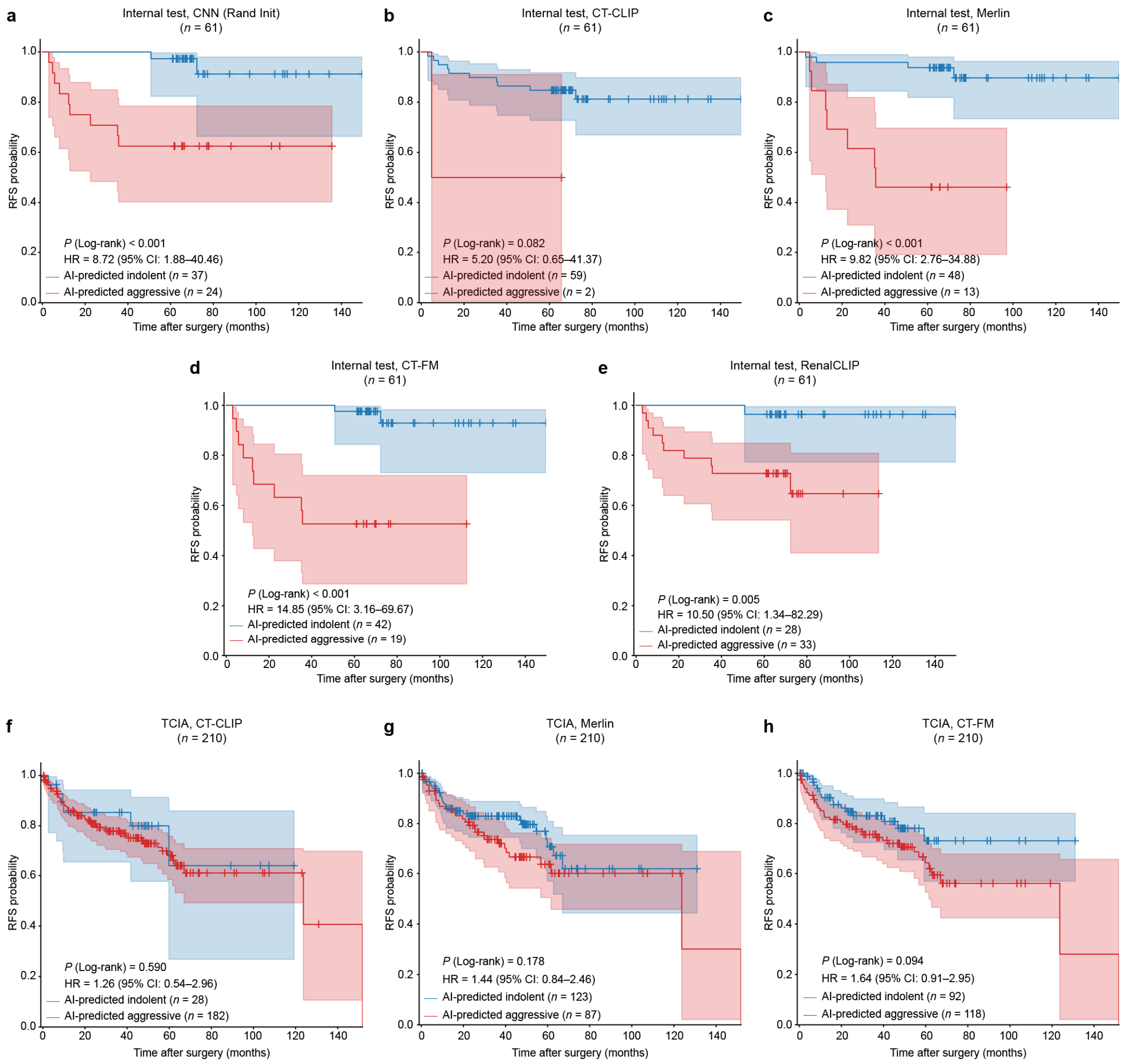

**Supplementary Figure 5: Supplementary Kaplan-Meier analyses for aggressiveness prediction. a–e,** Kaplan-Meier curves for recurrence-free survival (RFS) in the internal test cohort, stratified by the aggressiveness predictions of the four baseline models (**a–d**) and RenalCLIP (**e**). In this setting, predictions from both RenalCLIP and most baseline models show a statistically significant association with patient outcomes. **f–h**, Kaplan-Meier curves for RFS in the TCIA cohort, stratified by the aggressiveness predictions of the remaining baseline models: CT-CLIP (**f**), Merlin (**g**), and CT-FM (**h**). In stark contrast to the result for RenalCLIP (shown in Fig. 3f), none of these baseline models demonstrate statistically significant prognostic stratification in this challenging external cohort. Hazard ratios (HR) and their 95% CIs were derived from univariate Cox proportional hazards models; *P*-values for overall curve comparison were from a two-sided log-rank test. Shaded areas represent 95% CIs of the survival probability.

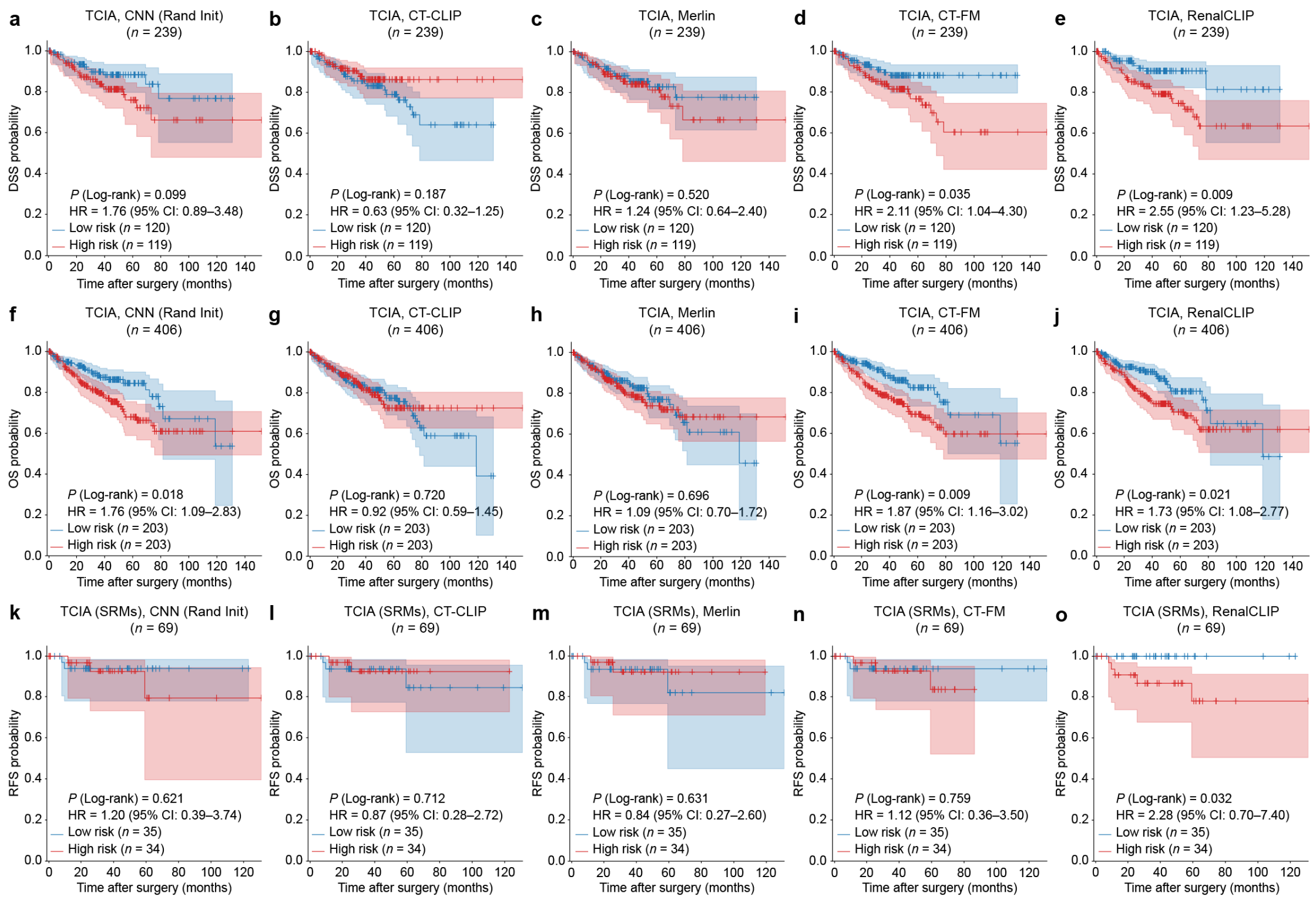

**Supplementary Figure 6: Supplementary Kaplan-Meier analyses for DSS, OS, and RFS (SRMs subgroup) in the TCIA cohort. a–e,** Kaplan-Meier curves for disease-specific survival (DSS) in the TCIA cohort, stratified by the prognostic risk scores derived from RenalCLIP and all baseline models. **f–j,** Kaplan-Meier curves for overall survival (OS), stratified by the same risk scores. **k–o,** Kaplan-Meier curves for recurrence-free survival (RFS) specifically within the small renal masses (SRMs, ≤ 4 cm) subgroup, stratified by the same risk scores. These analyses demonstrate the enhanced prognostic stratification of RenalCLIP for DSS and OS in the general population, and crucially, its capability to predict recurrence risk in SRMs. Hazard ratios (HR) and their 95% CIs were derived from univariate Cox proportional hazards models; *P*-values for overall curve comparison were from a two-sided log-rank test. Shaded areas represent 95% CIs of the survival probability.

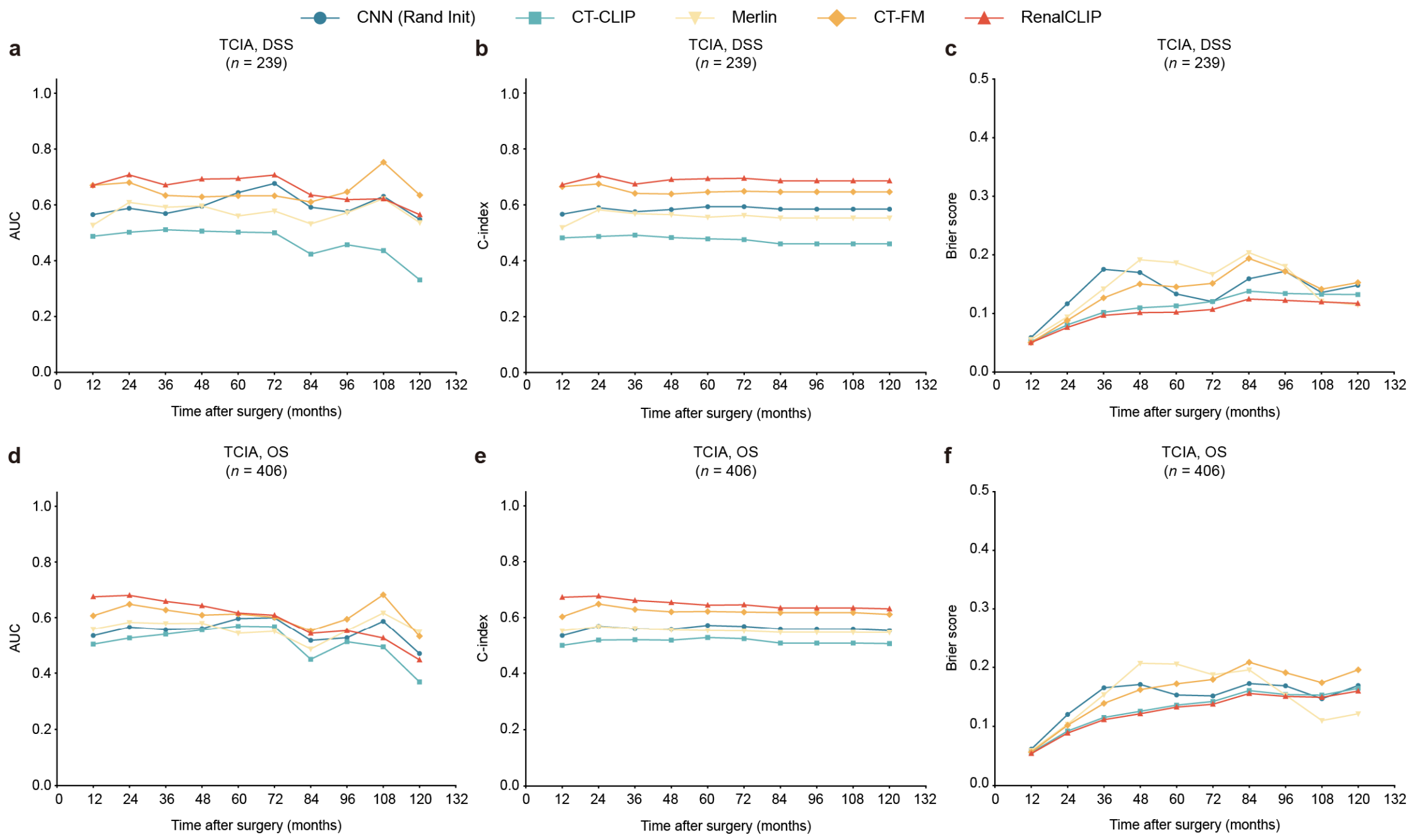

**Supplementary Figure 7: Supplementary time-dependent prognostic analyses for DSS and OS.** Time-dependent prognostic performance for disease-specific survival (DSS; **a–c**) and overall survival (OS; **d–f**) in the TCIA cohort, measured by ROC AUC, C-index, and Brier score. For the C-index for both DSS (**b**) and OS (**e**), and the Brier score for DSS (**c**), RenalCLIP maintains a consistent advantage over all baseline models across the entire follow-up period. For the remaining metrics—ROC AUC for DSS (**a**) and OS (**d**), and the Brier score for OS (**f**)—RenalCLIP demonstrates superior performance for the majority of the follow-up period (up to 7 years), while remaining comparable to the baselines at later time points.

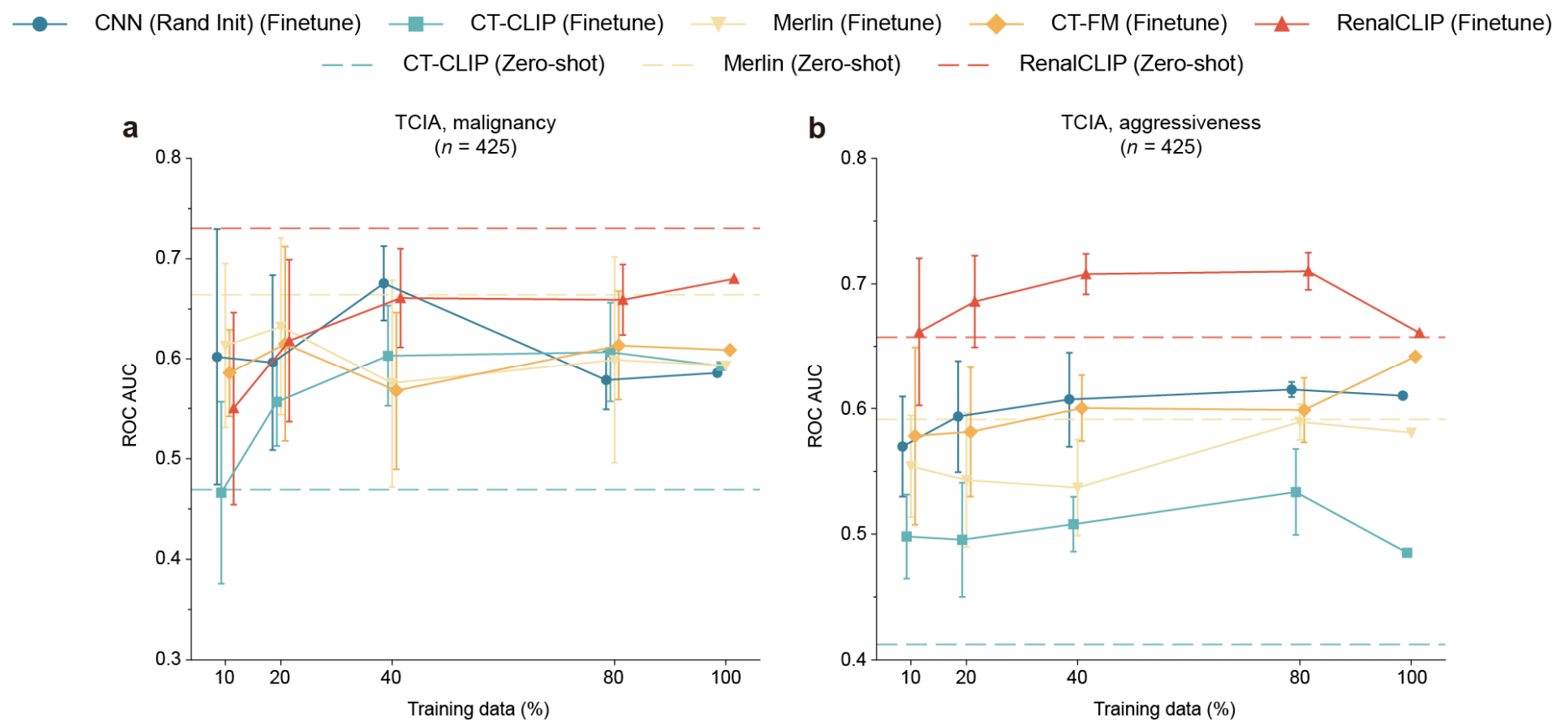

**Supplementary Figure 8: Data-efficient fine-tuning and zero-shot performance on the TCIA cohort.** Data efficiency curves for malignancy prediction (**a**) and aggressiveness prediction (**b**) in the TCIA cohort. The plots show model ROC AUC as a function of the fraction of training data used for fine-tuning, with dashed horizontal lines representing the zero-shot performance of each model. For malignancy prediction (**a**), RenalCLIP's zero-shot performance alone surpasses the peak performance of all other models after they were fully fine-tuned on 100% of the data. For the more challenging aggressiveness task (**b**), RenalCLIP's zero-shot performance is higher than all fully fine-tuned baseline models, while fine-tuning on RenalCLIP itself yields further performance gains. Data points for training fractions less than 100% represent the mean ROC AUC from five random sampling runs, with capped error bars indicating the standard deviation; the 100% data point represents a single run.

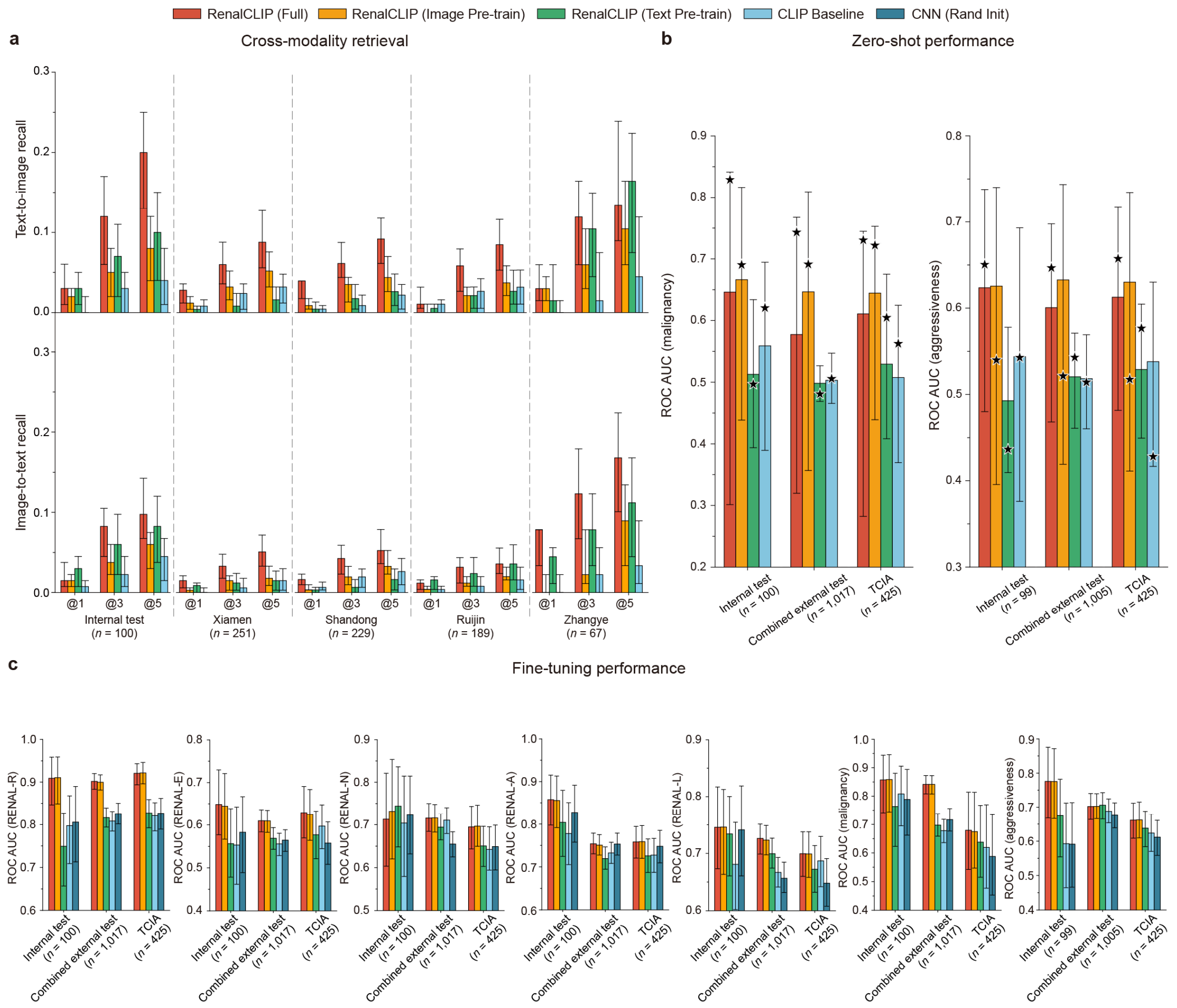

**Supplementary Figure 9: Ablation study on the contributions of domain-specific pre-training components.** Performance of the RenalCLIP (Full) model is benchmarked against four variants: RenalCLIP (Image Pre-train) (initialized with domain-specific image encoder only), RenalCLIP (Text Pre-train) (initialized with domain-specific text encoder only), a CLIP Baseline (generic initialization), and a standard 3D ResNet-18 initialized with random weights (CNN (Rand Init)). Models were evaluated across cross-modal retrieval (**a**; top, text-to-image; bottom, image-to-text), zero-shot classification (**b**; left, malignancy; right, aggressiveness), and fine-tuning on key downstream tasks (**c**). The results establish that while generic pre-training (CLIP Baseline) offers limited advantages over random initialization (CNN (Rand Init)) in fine-tuning tasks, our disease-centric image pre-training provides consistent performance improvements. Text pre-training offers a synergistic benefit, particularly for cross-modal retrieval and deterministic zero-shot classification, where the RenalCLIP (Full) model exceeds uni-modal variants. Error bars in **a–b** represent 95% CIs estimated via 1,000 bootstrap resamples. The centers correspond to the exact point estimates in **a**, and the average ROC AUC from the resamples in **b**.

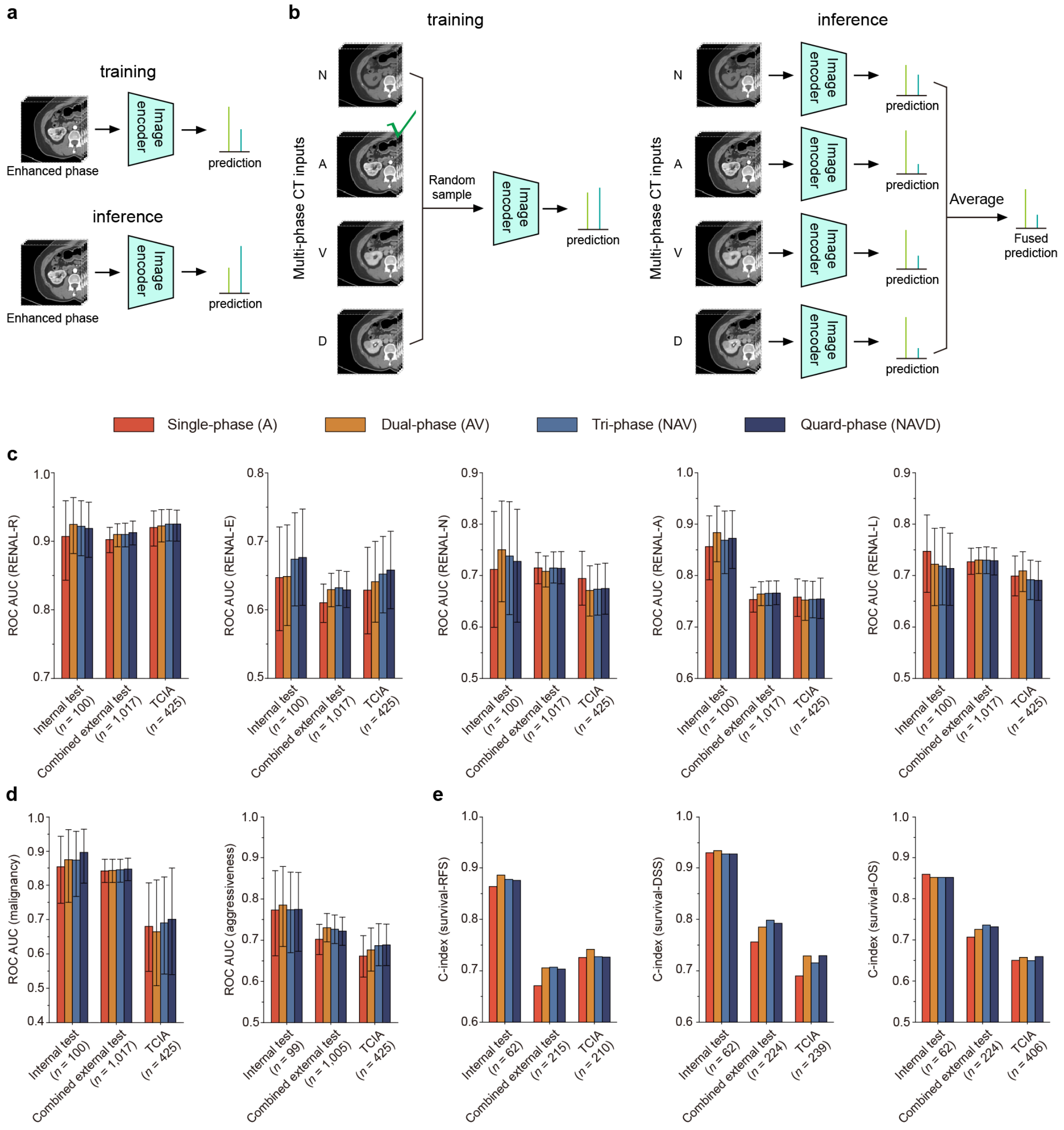

**Supplementary Figure 10: Ablation study on the multi-phase CT fusion strategy. a,b,** Schematics of the two data handling strategies. **a,** The standardized single-phase input strategy used for the main experiments, where a single enhanced phase is used. **b,** The late-fusion strategy, where a model trained on randomly sampled single phases is evaluated at inference time by averaging the logits from all available phases. **c–e,** Performance comparison of four cumulative multi-phase input settings. The comparison is shown for R.E.N.A.L. score prediction (**c**), diagnosis (malignancy and aggressiveness) (**d**), and survival prognosis (**e**). The results reveal a nuanced benefit to using multi-phase data: while fusion strategies generally yield a modest performance gain, the single-phase model provides a consistent baseline, and no single-phase combination is universally optimal across all tasks. Capped error bars in **c** and **d** represent 95% CIs estimated via 1,000 bootstrap resamples, and the centers of the bars correspond to the exact point estimates of each metric.

**Supplementary Table 1. Baseline clinicopathologic and imaging characteristics of the study cohorts.**

| Feature | Pre-training cohort *n* = 6,867 | Internal training cohort *n* = 400 | Internal test cohort *n* = 100 | Xiamen cohort *n* = 251 | Shandong cohort *n* = 250 | Ruijin cohort *n* = 229 | Lianyun cohort *n* = 220 | Zhangye cohort *n* = 67 | TCIA cohort *n* = 425 |
|---|---|---|---|---|---|---|---|---|---|
| **Number of CT volumes** | 21,819 | 1240 | 316 | 796 | 991 | 727 | 671 | 259 | 1,047 |
| **CT phases** (%) | | | | | | | | | |
| Non-contrast phase | 5,756(83.8) | 339(84.8) | 78(78.0) | 199(79.3) | 250(100.0) | 214(93.4) | 186(84.5) | 66(98.5) | 293(68.9) |
| Arterial phase | 6,799(99.0) | 400(100.0) | 100(100.0) | 249(99.2) | 249(99.6) | 208(90.8) | 220(100.0) | 67(100.0) | 238(56.0) |
| Venous phase | 6,642(96.7) | 391(97.8) | 98(98.0) | 246(98.0) | 248(99.2) | 215(93.9) | 211(95.9) | 65(97.0) | 293(68.9) |
| Delayed phase | 2,622(38.2) | 110(27.5) | 40(40.0) | 102(40.6) | 244(97.6) | 90(39.3) | 54(24.5) | 61(91.0) | 223(52.5) |
| **Median age** (IQR) | 56(48,65) | 57(50,65) | 57(48,65) | 53(43,63) | 57(51,65) | 59(50,67) | 57(51,65) | 58(48,66) | 59(51,68) |
| **Male sex** (%) | 4,232(61.6) | 259(64.8) | 62(62.0) | 144(57.4) | 149(59.6) | 142(62.0) | 123(55.9) | 36(53.7) | 261(62.1) |
| **Median tumor size** (IQR) | 3.6(2.5,5.5) | 3.5(2.5,5.5) | 3.8(2.6,5.4) | 3.8(2.5,5.0) | 3.8(2.4,5.0) | 5.0(3.5,7.0) | 4.0(3.0,5.7) | 5.8(3.7,7.0) | 4.5(3.0,6.9) |
| **TNM stage** (%) | | | | | | | | | |
| I | 4,467(65.1) | 288(72.0) | 64(64.0) | 149(59.4) | 179(71.6) | 118(51.5) | 145(65.9) | 31(46.3) | 244(57.4) |
| II | 460(6.7) | 20(5.0) | 9(9.0) | 13(5.2) | 13(5.2) | 33(14.4) | 20(9.1) | 15(22.4) | 33(7.8) |
| III | 594(8.7) | 26(6.5) | 12(12.0) | 13(5.2) | 28(11.2) | 25(10.9) | 7(3.2) | 7(10.4) | 89(20.9) |
| IV | 86(1.3) | 1(0.3) | 1(1.0) | 3(1.2) | 2(0.8) | 14(6.1) | 1(0.5) | 1(1.5) | 45(10.6) |
| Not applicable | 1,260(18.3) | 65(16.3) | 14(14.0) | 73(29.1) | 28(11.2) | 39(17.0) | 47(21.4) | 13(19.4) | 14(3.3) |
| **WHO/ISUP Grade** [a] (%) | | | | | | | | | |
| Low (1 & 2) | 3,745(54.5) | 232(58.1) | 51(51.0) | 128(51.0) | 149(59.6) | 88(38.4) | 115(52.2) | 20(29.9) | 184(43.3) |
| High (3 & 4) | 1,020(14.8) | 56(14.1) | 22(22.0) | 17(6.8) | 59(23.6) | 63(27.5) | 17(7.8) | 28(41.8) | 191(44.9) |
| Absence or not applicable | 2,102(30.7) | 112(28.0) | 27(27.0) | 106(42.2) | 42(16.8) | 78(34.1) | 88(40.0) | 19(28.4) | 50(11.8) |
| **Necrosis** [a] (%) | 610(8.9) | 41(10.3) | 12(12.0) | 14(5.6) | 30(12.0) | 28(12.2) | 12(5.5) | 17(25.4) | 72(16.9) |
| **Sarcomatoid differentiation** [a] (%) | 115(1.7) | 7(1.8) | 7(7.0) | 6(2.4) | 2(0.8) | 3(1.3) | 2(0.9) | 2(3.0) | 16(3.8) |
| **Histologic classification** (%) | | | | | | | | | |
| **Benign and indolent** (%) | 1,260(18.3) | 65(16.3) | 14(14.0) | 73(29.1) | 28(11.2) | 39(17.0) | 47(21.4) | 13(19.4) | 14(3.3) |
| Angiomyolipoma | 674(9.8) | 43(10.8) | 8(8.0) | 53(21.1) | 17(6.8) | 28(12.2) | 17(7.7) | 6(9.0) | 3(0.7) |

| Feature | Pre-training cohort *n* = 6,867 | Internal training cohort *n* = 400 | Internal test cohort *n* = 100 | Xiamen cohort *n* = 251 | Shandong cohort *n* = 250 | Ruijin cohort *n* = 229 | Lianyun cohort *n* = 220 | Zhangye cohort *n* = 67 | TCIA cohort *n* = 425 |
|---|---|---|---|---|---|---|---|---|---|
| Complex renal cyst | 436(6.3) | 11(2.8) | 2(2.0) | 17(6.8) | 7(2.8) | 6(2.6) | 23(10.5) | 5(7.5) | 0 |
| Oncocytoma | 67(1.0) | 5(1.3) | 3(3.0) | 2(0.8) | 4(1.6) | 4(1.7) | 6(2.7) | 0 | 9(2.1) |
| Mixed epithelial and stromal renal tumors | 20(0.3) | 0 | 0 | 0 | 0 | 0 | 1(0.5) | 0 | 2(0.5) |
| Metanephric tumors | 10(0.1) | 2(0.5) | 0 | 1(0.4) | 0 | 0 | 0 | 0 | 0 |
| Other benign masses [b] | 53(0.8) | 4(1.0) | 1(1.0) | 0 | 0 | 1(1.6) | 0 | 2(3.0) | 0 |
| **Malignant and indolent** (%) | 3,837(55.9) | 256(64.0) | 54(54.0) | 148(59.0) | 142(56.8) | 97(42.4) | 130(59.1) | 22(32.8) | 204(48.0) |
| Indolent clear cell RCC [c] | 3,158(46.0) | 206(51.5) | 42(42.0) | 117(46.6) | 119(47.6) | 79(34.6) | 97(44.1) | 15(22.4) | 148(34.8) |
| Indolent papillary RCC [c] | 195(2.8) | 12(3.0) | 3(3.0) | 4(1.6) | 13(5.2) | 2(0.9) | 10(4.5) | 1(1.5) | 24(5.6) |
| Chromophobe RCC | 280(4.1) | 25(6.3) | 9(9.0) | 19(7.6) | 10(4.0) | 9(3.9) | 12(5.5) | 6(9.0) | 30(7.1) |
| Clear cell papillary renal cell tumor | 30(0.4) | 3(0.8) | 0 | 3(1.2) | 0 | 0 | 3(1.4) | 0 | 0 |
| Multilocular cystic renal neoplasm of low malignant potential | 57(0.8) | 3(0.8) | 0 | 1(0.4) | 0 | 7(3.1) | 3(1.4) | 0 | 1(0.2) |
| Epithelioid angiomyolipoma | 46(0.7) | 6(1.5) | 0 | 3(1.2) | 0 | 0 | 1(0.5) | 0 | 0 |
| Other indolent tumors [d] | 71(1.0) | 1(0.3) | 0 | 1(0.4) | 0 | 0 | 4(1.8) | 0 | 1(0.2) |
| **Malignant and aggressive** (%) | 1,680(24.5) | 77(19.3) | 31(31.0) | 30(12.0) | 80(32.0) | 92(40.2) | 31(14.1) | 32(47.8) | 207(48.7) |
| Aggressive clear cell RCC | 1,373(20.0) | 62(15.5) | 27(27.0) | 20(8.0) | 68(27.2) | 75(32.8) | 22(10) | 28(41.8) | 192(45.2) |
| Aggressive papillary RCC | 103(1.5) | 8(2.0) | 1(1.0) | 4(1.6) | 8(3.2) | 7(3.1) | 3(1.4) | 4(6.0) | 11(2.6) |
| TFE3-rearranged RCC [e] | 52(0.8) | 1(0.3) | 1(1.0) | 5(2.0) | 3(1.2) | 1(0.4) | 4(1.8) | 0 | 0 |
| Renal cell carcinoma, NOS | 49(0.7) | 3(0.8) | 1(1.0) | 1(0.4) | 1(0.4) | 4(1.7) | 1(0.5) | 0 | 2(0.5) |
| Sarcomas [f] | 17(0.2) | 3(0.8) | 1(1.0) | 0 | 0 | 1(0.4) | 0 | 0 | 0 |
| Other aggressive tumors [g] | 86(1.3) | 0 | 0 | 0 | 0 | 4(1.7) | 1(0.5) | 0 | 2(0.5) |
| **Malignant and indeterminate** [h] (%) | 90(1.3) | 2(0.5) | 1(1.0) | 0 | 0 | 1(0.4) | 12(5.5) | 0 | 0 |

a. WHO/ISUP grade was evaluated in clear cell RCC and papillary RCC only. Necrosis and sarcomatoid differentiation were evaluated in malignant tumors. Any sarcomatoid differentiation led to the classification of malignant tumors as aggressive. b. Other benign masses included hemangioma, leiomyoma, schwannoma, etc. c. Clear cell RCC and papillary RCC with high-grade (3-4) components, sarcomatoid differentiation or invasion into major veins or perinephric tissues were classified as aggressive. The presence of coagulative necrosis also led to the classification as aggressive for clear cell RCC but not for papillary RCC. d. Other

| Feature | Pre-training cohort *n* = 6,867 | Internal training cohort *n* = 400 | Internal test cohort *n* = 100 | Xiamen cohort *n* = 251 | Shandong cohort *n* = 250 | Ruijin cohort *n* = 229 | Lianyun cohort *n* = 220 | Zhangye cohort *n* = 67 | TCIA cohort *n* = 425 |
|---|---|---|---|---|---|---|---|---|---|

indolent tumors included mucinous tubular and spindle cell carcinoma, tubulocystic RCC, eosinophilic solid and cystic RCC, etc. e. Xp11 translocation RCC and melanotic Xp11 translocation RCC followed an aggressive course. f. Sarcomas included rhabdomyosarcoma, synovial sarcoma, Ewing sarcoma, etc. g. Other aggressive tumors included collecting carcinoma, renal medullary carcinoma, indolent subtypes with sarcomatoid differentiation, etc. h. The aggressiveness for some of the tumors with missing information of tumor grade, immunohistochemistry or genetic testing were unclear.

**Supplementary Table 2. Sample sizes and class distributions for each downstream clinical task across cohorts.** *n* represents the total number of eligible patients with complete labels for each specific task. For R.E.N.A.L. components (R, E, N, L), data are presented as *n* (%) for categorical distribution and Mean (Standard Deviation (SD)) for the numerical score (scale 1–3). Component A represents location and is reported as *n* (%) only. For the aggressiveness diagnosis task, a small subset of patients was excluded from evaluation due to indeterminate pathological grades. For prognosis and cross-modal (specifically to retrieval and report generation tasks, which require paired radiology reports for evaluation) tasks, – indicates that the specific clinical labels (e.g., survival outcomes) or radiology reports were not available for that cohort.

| **Clinical application** | **Variable** | **Internal training cohort *n* = 400** | **Internal test cohort *n* = 100** | **Xiamen cohort *n* = 251** | **Shandong cohort *n* = 250** | **Ruijin cohort *n* = 229** | **Lianyun cohort *n* = 220** | **Zhangye cohort *n* = 67** | **Combined external test cohort *n* = 1,017** | **TCIA cohort *n* = 425** |
|---|---|---|---|---|---|---|---|---|---|---|
| **Characterization** [a] | **(R) Radius** (%) | *n* = 400 | *n* = 100 | *n* = 251 | *n* = 250 | *n* = 229 | *n* = 220 | *n* = 67 | *n* = 1,017 | *n* = 425 |
| | Mean score (SD) | 1.5(0.7) | 1.6(0.7) | 1.6(0.7) | 1.5(0.7) | 2.0(0.7) | 1.8(0.7) | 2.2(0.7) | 1.7(0.7) | 1.8(0.8) |
| | Score 1 | 226(56.5) | 50(50.0) | 137(54.6) | 145(58.0) | 61(26.6) | 88(40.0) | 13(19.4) | 444(43.7) | 173(40.7) |
| | Score 2 | 130(32.5) | 36(36.0) | 85(33.9) | 74(29.6) | 106(46.3) | 94(42.7) | 29(43.3) | 388(38.2) | 143(33.7) |
| | Score 3 | 44(11.0) | 14(14.0) | 29(11.5) | 31(12.4) | 62(27.1) | 38(17.3) | 25(37.3) | 185(18.1) | 109(25.6) |
| | **(E) Exophytic** (%) | *n* = 400 | *n* = 100 | *n* = 251 | *n* = 250 | *n* = 229 | *n* = 220 | *n* = 67 | *n* = 1,017 | *n* = 425 |
| | Mean score (SD) | 2.0(0.7) | 2.1(0.7) | 1.8(0.8) | 1.9(0.8) | 1.8(0.7) | 1.8(0.7) | 1.6(0.7) | 1.8(0.8) | 1.5(0.6) |
| | Score 1 | 110(27.5) | 25(25.0) | 106(42.2) | 94(37.6) | 91(39.7) | 95(43.2) | 38(56.7) | 424(41.7) | 246(57.9) |
| | Score 2 | 194(48.5) | 44(44.0) | 91(36.3) | 95(38.0) | 96(41.9) | 85(38.6) | 20(29.9) | 387(38.1) | 156(36.7) |
| | Score 3 | 96(24.0) | 31(31.0) | 54(21.5) | 61(24.4) | 42(18.4) | 40(18.2) | 9(13.4) | 206(20.2) | 23(5.4) |
| | **(N) Nearness** (%) | *n* = 400 | *n* = 100 | *n* = 251 | *n* = 250 | *n* = 229 | *n* = 220 | *n* = 67 | *n* = 1,017 | *n* = 425 |
| | Mean score (SD) | 2.3(0.9) | 2.0(1.0) | 2.0(1.0) | 2.2(0.9) | 2.6(0.7) | 2.3(0.9) | 2.7(0.7) | 2.3(0.9) | 2.7(0.6) |
| | Score 1 | 120(30.0) | 46(46.0) | 125(49.8) | 94(37.6) | 33(14.4) | 62(28.2) | 11(16.4) | 325(32.0) | 44(10.4) |
| | Score 2 | 34(8.5) | 7(7.0) | 12(4.8) | 19(7.6) | 16(7.0) | 22(10.0) | 0(0.0) | 69(6.8) | 41(9.6) |
| | Score 3 | 246(61.5) | 47(47.0) | 114(45.4) | 137(54.8) | 180(78.6) | 136(61.8) | 56(83.6) | 623(61.2) | 340(80.0) |
| | **(A) Anterior/Posterior** (%) | *n* = 400 | *n* = 100 | *n* = 251 | *n* = 250 | *n* = 229 | *n* = 220 | *n* = 67 | *n* = 1,017 | *n* = 425 |
| | Anterior (a) | 164(41.0) | 33(33.0) | 116(46.2) | 98(39.2) | 89(38.9) | 126(57.3) | 18(26.9) | 447(44.0) | 125(29.4) |
| | Posterior (p) | 146(36.5) | 42(42.0) | 123(49.0) | 95(38.0) | 94(41.0) | 74(33.6) | 23(34.3) | 409(40.2) | 140(32.9) |
| | Indeterminate (x) | 90(22.5) | 25(25.0) | 12(4.8) | 57(22.8) | 46(20.1) | 20(9.1) | 26(38.8) | 161(15.8) | 160(37.7) |
| | **(L) Location** (%) | *n* = 400 | *n* = 100 | *n* = 251 | *n* = 250 | *n* = 229 | *n* = 220 | *n* = 67 | *n* = 1,017 | *n* = 425 |
| | Mean score (SD) | 2.1(0.9) | 2.0(0.9) | 1.9(1.0) | 1.8(0.9) | 2.3(0.9) | 2.0(0.9) | 2.5(0.8) | 2.0(0.9) | 2.3(0.9) |
| | Score 1 | 125(31.3) | 38(38.0) | 131(52.2) | 134(53.6) | 63(27.5) | 88(40.0) | 13(19.4) | 429(42.2) | 115(27.1) |
| | Score 2 | 103(25.7) | 28(28.0) | 21(8.4) | 36(14.4) | 30(13.1) | 41(18.6) | 6(9.0) | 134(13.2) | 66(15.5) |

| Clinical application | Variable | Internal training cohort *n* = 400 | Internal test cohort *n* = 100 | Xiamen cohort *n* = 251 | Shandong cohort *n* = 250 | Ruijin cohort *n* = 229 | Lianyun cohort *n* = 220 | Zhangye cohort *n* = 67 | Combined external test cohort *n* = 1,017 | TCIA cohort *n* = 425 |
|---|---|---|---|---|---|---|---|---|---|---|
| | Score 3 | 172(43.0) | 34(34.0) | 99(39.4) | 80(32.0) | 136(59.4) | 91(41.4) | 48(71.6) | 454(44.6) | 244(57.4) |
| **Diagnosis** [b] (applies to fine-tuning & zero-shot) | **Malignancy** (%) | *n* = 400 | *n* = 100 | *n* = 251 | *n* = 250 | *n* = 229 | *n* = 220 | *n* = 67 | *n* = 1,017 | *n* = 425 |
| | Benign | 65(16.3) | 14(14.0) | 73(29.1) | 28(11.2) | 39(17.0) | 47(21.4) | 13(19.4) | 200(19.7) | 14(3.3) |
| | Malignant | 335(83.8) | 86(86.0) | 178(70.9) | 222(88.8) | 190(83.0) | 173(78.6) | 54(80.6) | 817(80.3) | 411(96.7) |
| | **Aggressiveness** (%) | *n* = 398 | *n* = 99 | *n* = 251 | *n* = 250 | *n* = 228 | *n* = 209 | *n* = 67 | *n* = 1,005 | *n* = 425 |
| | Indolent | 321(80.7) | 68(68.7) | 221(88.0) | 170(68.0) | 137(60.1) | 178(85.2) | 38(56.7) | 744(74.0) | 218(51.3) |
| | Aggressive | 77(19.3) | 31(31.3) | 30(12.0) | 80(32.0) | 91(39.9) | 31(14.8) | 29(43.3) | 261(26.0) | 207(48.7) |
| **Prognosis** [c] | **RFS outcome** (%) | *n* = 274 | *n* = 62 | *n* = 85 | – | *n* = 78 | – | *n* = 52 | *n* = 215 | *n* = 210 |
| | Censored | 241(88.0) | 51(82.3) | 80(94.1) | – | 64(82.1) | – | 41(78.8) | 185(86.0) | 155(73.8) |
| | Recurrence | 33(12.0) | 11(17.7) | 5(5.9) | – | 14(17.9) | – | 11(21.2) | 30(14.0) | 55(26.2) |
| | **DSS outcome** (%) | *n* = 274 | *n* = 62 | *n* = 87 | – | *n* = 85 | – | *n* = 52 | *n* = 224 | *n* = 239 |
| | Censored | 252(92.0) | 54(87.1) | 83(95.4) | – | 72(84.7) | – | 42(80.8) | 197(87.9) | 203(84.9) |
| | Death of kidney cancer | 22(8.0) | 8(12.9) | 4(4.6) | – | 13(15.3) | – | 10(19.2) | 27(12.1) | 36(15.1) |
| | **OS outcome** (%) | *n* = 274 | *n* = 62 | *n* = 87 | – | *n* = 85 | – | *n* = 52 | *n* = 224 | *n* = 406 |
| | Censored | 245(89.4) | 52(83.9) | 82(94.3) | – | 65(76.5) | – | 42(80.8) | 189(84.4) | 330(81.3) |
| | Death from all causes | 29(10.6) | 10(16.1) | 5(5.7) | – | 20(23.5) | – | 10(19.2) | 35(15.6) | 76(18.7) |
| **Cross-modal** (retrieval & report generation) | **Available reports** | *n* = 400 | *n* = 100 | *n* = 251 | *n* = 229 | *n* = 189 | – | *n* = 67 | *n* = 736 | – |

a. Manually annotated by 8 experienced radiologists divided into 4 groups (2 per group). The 1,942 downstream scans were distributed among these groups, where each group jointly evaluated the cases and assigned final scores through collaborative discussion and consensus. b. Determined strictly based on post-operative histopathology reports (see Methods for details). c. Ascertained through rigorous post-operative clinical follow-up.

**Supplementary Table 3: R.E.N.A.L. nephrometry score criteria.** Components R, E, N, and L are assigned a score of 1–3 points. Component A (Anterior/Posterior) serves as a descriptive suffix and is not included in the numerical score.

| Component | Description | Criteria |
|---|---|---|
| R | Radius (cm, largest diameter in any single plane) | Score 1. ≤4<br>Score 2. 4-7<br>Score 3. ≥7 |
| E | Exophytic/endophytic (% mass exophytic (protruding out) vs. endophytic (contained) relative to renal parenchyma) | Score 1. ≥ 50% exophytic<br>Score 2. < 50% exophytic<br>Score 3. 100% endophytic |
| N | Nearness to collecting system or renal sinus (mm, measure shortest distance) | Score 1. ≥7<br>Score 2. 4-7<br>Score 3. ≤4 |
| A | Anterior/posterior (primary location of tumor relative to coronal plane at level of hilar vessels) | A. Anterior<br>P. Posterior<br>X. Neither |
| L | Location relative to polar lines | Score 1. Entirely below inferior or above superior pole<br>Score 2. Mass crosses the polar line<br>Score 3. > 50% of mass lies across the polar line, or is entirely between the polar lines, or crosses the axial midline |

**Supplementary Table 4: ROC AUC on the RENAL-R task across different models and cohorts.** Best performing model for each cohort is bolded. 95% CI is included in parentheses. * indicates $P < 0.05$ (two-sided paired permutation test vs. CNN Rand Init). The sample size (*n*) represents the exact number of independent patients evaluated in each cohort.

| Cohort | CNN (Rand Init) | CT-CLIP | Merlin | CT-FM | RenalCLIP |
|---|---|---|---|---|---|
| Internal test | 0.806 | 0.593 | 0.838 | 0.904 | **0.908** |
| (*n* = 100) | (0.713-0.890) | (0.506-0.680) | (0.764-0.906) | (0.825-0.959) | **(0.848-0.958)** |
| Xiamen | 0.788 | 0.525 | 0.814 | **0.883** | 0.873* |
| (*n* = 251) | (0.725-0.847) | (0.471-0.583) | (0.760-0.867) | **(0.835-0.930)** | (0.822-0.913) |
| Shandong | 0.836 | 0.615 | 0.888 | **0.922** | 0.921* |
| (*n* = 250) | (0.783-0.886) | (0.559-0.668) | (0.843-0.925) | **(0.881-0.955)** | (0.888-0.951) |
| Ruijin | 0.869 | 0.651 | 0.881 | 0.896 | **0.935*** |
| (*n* = 229) | (0.823-0.913) | (0.592-0.709) | (0.839-0.919) | (0.856-0.932) | **(0.906-0.961)** |
| Lianyun | 0.821 | 0.596 | 0.829 | 0.881 | **0.895*** |
| (*n* = 220) | (0.761-0.873) | (0.538-0.654) | (0.775-0.880) | (0.829-0.926) | **(0.850-0.936)** |
| Zhangye | 0.778 | 0.548 | 0.703 | **0.868** | 0.839 |
| (*n* = 67) | (0.674-0.868) | (0.445-0.653) | (0.600-0.808) | **(0.790-0.934)** | (0.753-0.913) |
| Combined external test | 0.825 | 0.587 | 0.844 | 0.896 | **0.902*** |
| (*n* = 1,017) | (0.801-0.850) | (0.559-0.613) | (0.819-0.866) | (0.877-0.916) | **(0.883-0.920)** |
| TCIA | 0.826 | 0.559 | 0.861 | 0.880 | **0.920*** |
| (*n* = 425) | (0.789-0.862) | (0.521-0.600) | (0.827-0.890) | (0.850-0.908) | **(0.895-0.945)** |

**Supplementary Table 5: ROC AUC on the RENAL-E task across different models and cohorts.** Best performing model for each cohort is bolded. 95% CI is included in parentheses. * indicates $P < 0.05$ (two-sided paired permutation test vs. CNN Rand Init). The sample size (*n*) represents the exact number of independent patients evaluated in each cohort.

| Cohort | CNN (Rand Init) | CT-CLIP | Merlin | CT-FM | RenalCLIP |
|---|---|---|---|---|---|
| Internal test | 0.583 | 0.546 | 0.502 | 0.549 | **0.646** |
| (*n* = 100) | (0.489-0.667) | (0.456-0.633) | (0.420-0.581) | (0.466-0.628) | **(0.573-0.721)** |
| Xiamen | 0.562 | 0.516 | 0.526 | 0.570 | **0.671*** |
| (*n* = 251) | (0.510-0.614) | (0.466-0.571) | (0.471-0.581) | (0.518-0.623) | **(0.617-0.722)** |
| Shandong | 0.576 | 0.533 | 0.509 | 0.537 | **0.592** |
| (*n* = 250) | (0.526-0.627) | (0.479-0.586) | (0.457-0.564) | (0.479-0.590) | **(0.541-0.642)** |
| Ruijin | **0.583** | 0.528 | 0.567 | 0.574 | 0.578 |
| (*n* = 229) | **(0.528-0.639)** | (0.470-0.587) | (0.510-0.631) | (0.518-0.625) | (0.520-0.635) |
| Lianyun | 0.569 | 0.512 | 0.525 | 0.577 | **0.617** |
| (*n* = 220) | (0.511-0.625) | (0.457-0.567) | (0.467-0.585) | (0.522-0.629) | **(0.561-0.673)** |
| Zhangye | 0.487 | 0.569 | 0.563 | **0.594** | 0.549 |
| (*n* = 67) | (0.396-0.598) | (0.465-0.673) | (0.443-0.686) | **(0.493-0.706)** | (0.428-0.678) |
| Combined external test | 0.564 | 0.519 | 0.535 | 0.562 | **0.610*** |
| (*n* = 1,017) | (0.538-0.589) | (0.493-0.546) | (0.508-0.563) | (0.536-0.587) | **(0.583-0.637)** |
| TCIA | 0.559 | 0.541 | 0.544 | 0.575 | **0.629** |
| (*n* = 425) | (0.508-0.607) | (0.481-0.599) | (0.490-0.598) | (0.525-0.623) | **(0.568-0.686)** |

**Supplementary Table 6: ROC AUC on the RENAL-N task across different models and cohorts.** Best performing model for each cohort is bolded. 95% CI is included in parentheses. * indicates $P < 0.05$ (two-sided paired permutation test vs. CNN Rand Init). The sample size (*n*) represents the exact number of independent patients evaluated in each cohort.

| Cohort | CNN (Rand Init) | CT-CLIP | Merlin | CT-FM | RenalCLIP |
|---|---|---|---|---|---|
| Internal test | **0.723** | 0.571 | 0.544 | 0.651 | 0.713 |
| (*n* = 100) | **(0.632-0.814)** | (0.455-0.676) | (0.439-0.656) | (0.556-0.748) | (0.600-0.833) |
| Xiamen | 0.614 | 0.412 | 0.494 | 0.681 | **0.688** |
| (*n* = 251) | (0.544-0.692) | (0.354-0.474) | (0.413-0.563) | (0.617-0.753) | **(0.614-0.762)** |
| Shandong | 0.613 | 0.499 | 0.566 | 0.670 | **0.684*** |
| (*n* = 250) | (0.553-0.672) | (0.449-0.548) | (0.504-0.628) | (0.595-0.742) | **(0.624-0.744)** |
| Ruijin | **0.748** | 0.535 | 0.612 | 0.744 | 0.743 |
| (*n* = 229) | **(0.680-0.817)** | (0.453-0.614) | (0.544-0.689) | (0.663-0.830) | (0.664-0.825) |
| Lianyun | 0.652 | 0.513 | 0.574 | 0.684 | **0.687** |
| (*n* = 220) | (0.597-0.708) | (0.452-0.576) | (0.507-0.646) | (0.613-0.756) | **(0.623-0.750)** |
| Zhangye | 0.796 | 0.299 | 0.688 | 0.779 | **0.851** |
| (*n* = 67) | (0.633-0.932) | (0.115-0.495) | (0.448-0.874) | (0.649-0.896) | **(0.744-0.939)** |
| Combined external test | 0.655 | 0.490 | 0.549 | 0.694 | **0.715*** |
| (*n* = 1,017) | (0.625-0.683) | (0.456-0.518) | (0.516-0.582) | (0.660-0.730) | **(0.683-0.747)** |
| TCIA | 0.648 | 0.494 | 0.533 | 0.678 | **0.694** |
| (*n* = 425) | (0.594-0.699) | (0.437-0.547) | (0.473-0.595) | (0.632-0.728) | **(0.644-0.742)** |

**Supplementary Table 7: ROC AUC on the RENAL-A task across different models and cohorts.** Best performing model for each cohort is bolded. 95% CI is included in parentheses. * indicates $P < 0.05$ (two-sided paired permutation test vs. CNN Rand Init). The sample size (*n*) represents the exact number of independent patients evaluated in each cohort.

| Cohort | CNN (Rand Init) | CT-CLIP | Merlin | CT-FM | RenalCLIP |
|---|---|---|---|---|---|
| Internal test | 0.827 | 0.731 | 0.725 | 0.833 | **0.857** |
| (*n* = 100) | (0.758-0.891) | (0.658-0.806) | (0.642-0.805) | (0.766-0.897) | **(0.796-0.916)** |
| Xiamen | 0.767 | 0.695 | 0.677 | **0.779** | 0.767 |
| (*n* = 251) | (0.709-0.825) | (0.626-0.768) | (0.606-0.746) | **(0.710-0.844)** | (0.701-0.829) |
| Shandong | 0.755 | 0.643 | 0.683 | 0.785 | **0.787** |
| (*n* = 250) | (0.705-0.801) | (0.592-0.694) | (0.636-0.734) | (0.735-0.833) | **(0.738-0.833)** |
| Ruijin | **0.730** | 0.724 | 0.649 | 0.716 | 0.712 |
| (*n* = 229) | **(0.678-0.779)** | (0.671-0.771) | (0.592-0.704) | (0.661-0.767) | (0.660-0.768) |
| Lianyun | 0.747 | 0.648 | 0.611 | 0.713 | **0.739** |
| (*n* = 220) | (0.680-0.806) | (0.582-0.707) | (0.537-0.686) | (0.649-0.775) | **(0.668-0.800)** |
| Zhangye | 0.725 | 0.613 | 0.533 | **0.768** | 0.763 |
| (*n* = 67) | (0.619-0.818) | (0.517-0.708) | (0.428-0.640) | **(0.676-0.849)** | (0.665-0.855) |
| Combined external test | 0.753 | 0.674 | 0.650 | **0.757** | 0.754 |
| (*n* = 1,017) | (0.727-0.779) | (0.647-0.699) | (0.623-0.676) | **(0.731-0.782)** | (0.728-0.778) |
| TCIA | 0.748 | 0.707 | 0.680 | 0.751 | **0.758** |
| (*n* = 425) | (0.709-0.785) | (0.668-0.741) | (0.640-0.718) | (0.715-0.786) | **(0.719-0.796)** |

**Supplementary Table 8: ROC AUC on the RENAL-L task across different models and cohorts.** Best performing model for each cohort is bolded. 95% CI is included in parentheses. * indicates $P < 0.05$ (two-sided paired permutation test vs. CNN Rand Init). The sample size (*n*) represents the exact number of independent patients evaluated in each cohort.

| Cohort | CNN (Rand Init) | CT-CLIP | Merlin | CT-FM | RenalCLIP |
|---|---|---|---|---|---|
| Internal test | 0.739 | 0.561 | 0.677 | 0.649 | **0.747** |
| (*n* = 100) | (0.661-0.819) | (0.467-0.648) | (0.588-0.755) | (0.569-0.730) | **(0.668-0.816)** |
| Xiamen | 0.597 | 0.558 | 0.597 | 0.601 | **0.746*** |
| (*n* = 251) | (0.539-0.661) | (0.496-0.622) | (0.531-0.659) | (0.543-0.663) | **(0.691-0.797)** |
| Shandong | 0.654 | 0.554 | 0.644 | 0.666 | **0.754*** |
| (*n* = 250) | (0.609-0.701) | (0.500-0.611) | (0.591-0.694) | (0.612-0.719) | **(0.698-0.800)** |
| Ruijin | **0.718** | 0.492 | 0.655 | 0.660 | 0.698 |
| (*n* = 229) | **(0.658-0.777)** | (0.434-0.552) | (0.593-0.720) | (0.598-0.729) | (0.640-0.757) |
| Lianyun | 0.649 | 0.553 | 0.597 | 0.663 | **0.694** |
| (*n* = 220) | (0.598-0.705) | (0.494-0.612) | (0.544-0.656) | (0.604-0.723) | **(0.644-0.745)** |
| Zhangye | 0.597 | 0.440 | **0.643** | 0.561 | 0.633 |
| (*n* = 67) | (0.503-0.691) | (0.324-0.547) | **(0.519-0.770)** | (0.459-0.676) | (0.504-0.766) |
| Combined external test | 0.657 | 0.525 | 0.631 | 0.649 | **0.727*** |
| (*n* = 1,017) | (0.632-0.685) | (0.499-0.552) | (0.605-0.659) | (0.621-0.677) | **(0.702-0.751)** |
| TCIA | 0.648 | 0.529 | 0.634 | 0.649 | **0.699*** |
| (*n* = 425) | (0.608-0.691) | (0.485-0.577) | (0.594-0.678) | (0.604-0.693) | **(0.657-0.739)** |

**Supplementary Table 9: Sample sizes and class distributions for characterization task across cohorts, stratified by tumor size.** *n* represents the total number of eligible patients with complete labels for each specific task.

| Clinical application | Variable | Internal training cohort N = 400 | | | Internal test cohort N = 100 | | | Combined external test cohort N = 1,017 | | | TCIA cohort N = 425 | | |
|---|---|---|---|---|---|---|---|---|---|---|---|---|---|
| | | **≤ 4 cm** *n* **= 226** | **4 – 7 cm** *n* **= 130** | **> 7 cm** *n* **= 44** | **≤ 4 cm** *n* **= 50** | **4 – 7 cm** *n* **= 36** | **> 7 cm** *n* **= 14** | **≤ 4 cm** *n* **= 444** | **4 – 7 cm** *n* **= 388** | **> 7 cm** *n* **= 185** | **≤ 4 cm** *n* **= 173** | **4 – 7 cm** *n* **= 143** | **> 7 cm** *n* **= 109** |
| **Characterization** | **(E) Exophytic** (%) | | | | | | | | | | | | |
| | Score 1 | 56(24.8) | 40(30.8) | 14(31.8) | 15(30.0) | 7(19.4) | 3(21.4) | 155(34.9) | 160(41.2) | 109(58.9) | 83(48.0) | 83(58.0) | 80(73.4) |
| | Score 2 | 102(45.1) | 71(54.6) | 21(47.7) | 19(38.0) | 16(44.4) | 9(64.3) | 158(35.6) | 175(45.1) | 54(29.2) | 71(41.0) | 57(39.9) | 28(25.7) |
| | Score 3 | 68(30.1) | 19(14.6) | 9(20.5) | 16(32.0) | 13(36.2) | 2(14.3) | 131(29.5) | 53(13.7) | 22(11.9) | 19(11.0) | 3(2.1) | 1(0.9) |
| | **(N) Nearness** (%) | | | | | | | | | | | | |
| | Score 1 | 93(41.2) | 23(17.7) | 4(9.1) | 33(66.0) | 11(30.6) | 2(14.3) | 254(57.2) | 53(13.7) | 18(9.7) | 38(22.0) | 4(2.8) | 2(1.8) |
| | Score 2 | 24(10.6) | 9(6.9) | 1(2.3) | 6(12.0) | 1(2.8) | 0(0.0) | 41(9.2) | 27(7.0) | 1(0.5) | 26(15.0) | 13(9.1) | 2(1.8) |
| | Score 3 | 109(48.2) | 98(75.4) | 39(88.6) | 11(22.0) | 24(66.6) | 12(85.7) | 149(33.6) | 308(79.3) | 166(89.8) | 109(63.0) | 126(88.1) | 105(96.4) |
| | **(A) Anterior/Posterior**(%) | | | | | | | | | | | | |
| | Anterior (a) | 103(45.6) | 48(36.9) | 13(29.6) | 18(36.0) | 11(30.6) | 4(28.6) | 212(47.7) | 166(42.8) | 69(37.3) | 63(36.4) | 41(28.7) | 21(19.3) |
| | Posterior (p) | 92(40.7) | 47(36.2) | 7(15.9) | 28(56.0) | 11(30.6) | 3(21.4) | 189(42.6) | 169(43.6) | 51(27.6) | 70(40.5) | 54(37.7) | 16(14.7) |
| | Indeterminate (x) | 31(13.7) | 35(26.9) | 24(54.5) | 4(8.0) | 14(38.8) | 7(50.0) | 43(9.7) | 53(13.6) | 65(35.1) | 40(23.1) | 48(33.6) | 72(66.0) |
| | **(L) Location** (%) | | | | | | | | | | | | |
| | Score 1 | 95(42.0) | 26(20.0) | 4(9.1) | 27(54.0) | 11(30.6) | 0(0.0) | 297(66.9) | 110(28.4) | 22(11.9) | 78(45.1) | 30(21.0) | 7(6.4) |
| | Score 2 | 58(25.7) | 38(29.2) | 7(15.9) | 15(30.0) | 11(30.6) | 2(14.3) | 59(13.3) | 63(16.2) | 12(6.5) | 26(15.0) | 30(21.0) | 10(9.2) |
| | Score 3 | 73(32.3) | 66(50.8) | 33(75.0) | 8(16.0) | 14(38.8) | 12(85.7) | 88(19.8) | 215(55.4) | 151(81.6) | 69(39.9) | 83(58.0) | 92(84.4) |

**Supplementary Table 10: ROC AUC of different models for predicting RENAL score components (E, N, A, L) stratified by tumor size subgroups (≤4 cm, 4–7 cm, and >7 cm).** The analysis was performed on the combined external test cohort (*n* = 1,017). Best performing model for each cohort is bolded and 95% CI is included in parentheses.

| Task | Subgroup | CNN (Rand Init) | CT-CLIP | Merlin | CT-FM | RenalCLIP |
|---|---|---|---|---|---|---|
| RENAL-E | ≤4 cm | 0.557 (0.517-0.596) | 0.536 (0.496-0.574) | 0.534 (0.493-0.572) | 0.555 (0.517-0.590) | **0.606 (0.571-0.643)** |
| | 4–7 cm | 0.580 (0.540-0.620) | 0.519 (0.474-0.565) | 0.513 (0.469-0.559) | 0.566 (0.519-0.608) | **0.594 (0.548-0.634)** |
| | >7 cm | 0.592 (0.518-0.661) | 0.487 (0.421-0.558) | 0.500 (0.426-0.580) | **0.618 (0.555-0.682)** | 0.509 (0.436-0.581) |
| RENAL-N | ≤4 cm | 0.584 (0.541-0.629) | 0.476 (0.433-0.517) | 0.527 (0.483-0.574) | 0.622 (0.577-0.671) | **0.641 (0.591-0.691)** |
| | 4–7 cm | 0.631 (0.579-0.693) | 0.488 (0.432-0.542) | 0.505 (0.442-0.562) | 0.671 (0.606-0.736) | **0.679 (0.618-0.739)** |
| | >7 cm | 0.587 (0.526-0.647) | 0.563 (0.535-0.597) | 0.632 (0.553-0.712) | 0.556 (0.430-0.659) | **0.654 (0.548-0.760)** |
| RENAL-A | ≤4 cm | 0.754 (0.717-0.793) | 0.648 (0.603-0.691) | 0.643 (0.597-0.688) | 0.745 (0.704-0.786) | **0.770 (0.729-0.813)** |
| | 4–7 cm | 0.730 (0.689-0.770) | 0.666 (0.624-0.707) | 0.650 (0.604-0.695) | 0.756 (0.712-0.796) | **0.758 (0.717-0.799)** |
| | >7 cm | **0.652 (0.591-0.716)** | 0.644 (0.589-0.697) | 0.550 (0.488-0.612) | 0.605 (0.541-0.667) | 0.581 (0.517-0.647) |
| RENAL-L | ≤4 cm | 0.620 (0.577-0.662) | 0.517 (0.472-0.561) | 0.540 (0.497-0.586) | 0.607 (0.565-0.648) | **0.660 (0.618-0.703)** |
| | 4–7 cm | 0.576 (0.533-0.620) | 0.501 (0.460-0.544) | 0.590 (0.543-0.636) | 0.579 (0.536-0.622) | **0.637 (0.595-0.679)** |
| | >7 cm | 0.598 (0.501-0.689) | 0.504 (0.415-0.593) | 0.608 (0.503-0.696) | **0.612 (0.515-0.707)** | 0.603 (0.497-0.706) |

**Supplementary Table 11: Comparison of inter-observer variability between RenalCLIP and six human experts on the R.E.N.A.L. nephrometry scoring task.** The evaluation was performed on a randomly selected subset of 50 cases from the internal test cohort. The human expert panel consisted of three board-certified urologists (Experts 1, 3, 4) and three board-certified radiologists (Experts 2, 5, 6). Performance metrics include accuracy, Cohen's kappa, and the relative rank of each rater (determined by Cohen's kappa, 1 = Best) against the original consensus ground truth.

| Component | Rater | Accuracy | Cohen's kappa | Rank |
|---|---|---|---|---|
| RENAL-R | Expert 1 | 0.800 | 0.679 | 3 |
| | Expert 2 | 0.800 | 0.651 | 5 |
| | Expert 3 | 0.720 | 0.520 | 7 |
| | Expert 4 | 0.800 | 0.673 | 4 |
| | Expert 5 | 0.880 | 0.795 | 1 |
| | Expert 6 | 0.780 | 0.637 | 6 |
| | RenalCLIP | 0.740 | 0.736 | 2 |
| RENAL-E | Expert 1 | 0.500 | 0.250 | 7 |
| | Expert 2 | 0.520 | 0.306 | 5 |
| | Expert 3 | 0.620 | 0.424 | 3 |
| | Expert 4 | 0.660 | 0.491 | 1 |
| | Expert 5 | 0.540 | 0.313 | 4 |
| | Expert 6 | 0.640 | 0.461 | 2 |
| | RenalCLIP | 0.520 | 0.290 | 6 |
| RENAL-N | Expert 1 | 0.720 | 0.495 | 1 |
| | Expert 2 | 0.620 | 0.354 | 6 |
| | Expert 3 | 0.640 | 0.356 | 5 |
| | Expert 4 | 0.580 | 0.265 | 7 |
| | Expert 5 | 0.680 | 0.426 | 3 |
| | Expert 6 | 0.700 | 0.480 | 2 |
| | RenalCLIP | 0.660 | 0.383 | 4 |
| RENAL-A | Expert 1 | 0.840 | 0.760 | 1 |
| | Expert 2 | 0.600 | 0.390 | 6 |
| | Expert 3 | 0.560 | 0.361 | 7 |
| | Expert 4 | 0.620 | 0.444 | 4 |
| | Expert 5 | 0.620 | 0.437 | 5 |
| | Expert 6 | 0.720 | 0.578 | 3 |
| | RenalCLIP | 0.740 | 0.607 | 2 |
| RENAL-L | Expert 1 | 0.460 | 0.228 | 6 |
| | Expert 2 | 0.680 | 0.504 | 1 |
| | Expert 3 | 0.680 | 0.504 | 1 |
| | Expert 4 | 0.460 | 0.193 | 7 |
| | Expert 5 | 0.640 | 0.464 | 3 |
| | Expert 6 | 0.600 | 0.404 | 4 |
| | RenalCLIP | 0.580 | 0.369 | 5 |

**Supplementary Table 12: Expert evaluation of report generation accuracy on the internal test cohort (*n* = 100).** Scores (range 0–5) were averaged across three radiologists for each case. The inter-rater agreement among the three radiologists (Fleiss' kappa = 0.43) supports the reliability of the observed ranking. Data are presented as mean ± standard deviation (SD) and median [interquartile range (IQR)]. The best performing model is bolded. *P*-values were calculated using a two-sided paired permutation test comparing baselines against RenalCLIP.

| Model | Mean ± SD | Median [IQR] | *P*-value |
|---|---|---|---|
| RadFM | 0.05 ± 0.20 | 0.00 [0.00-0.00] | < 0.001 |
| CT-CHAT | 0.03 ± 0.17 | 0.00 [0.00-0.00] | < 0.001 |
| GPT-4o | 1.57 ± 1.14 | 1.33 [0.67-2.17] | < 0.001 |
| MedGemma | 1.75 ± 0.99 | 1.67 [0.83-2.50] | < 0.001 |
| RenalCLIP | **2.74 ± 1.26** | **2.67 [1.67-4.00]** | - |

**Supplementary Table 13: Quantitative evaluation of report generation using linguistic metrics on the internal test cohort (*n* = 100).** Best performing model for each metric is bolded. 95% CI is included in parentheses.

| Metric | RadFM | CT-CHAT | GPT-4o | MedGemma | RenalCLIP |
|---|---|---|---|---|---|
| BLEU-1 | 0.025 (0.020-0.031) | 0.148 (0.138-0.159) | 0.223 (0.211-0.235) | 0.180 (0.166-0.194) | **0.297 (0.275-0.319)** |
| BLEU-2 | 0.014 (0.011-0.017) | 0.078 (0.071-0.084) | 0.093 (0.085-0.100) | 0.082 (0.075-0.090) | **0.192 (0.176-0.209)** |
| BLEU-4 | 0.000 (0.000-0.000) | 0.013 (0.009-0.016) | 0.020 (0.016-0.024) | 0.026 (0.022-0.031) | **0.096 (0.083-0.110)** |
| METEOR | 0.108 (0.101-0.117) | 0.222 (0.214-0.230) | 0.248 (0.238-0.258) | 0.216 (0.204-0.226) | **0.350 (0.333-0.366)** |
| ROUGE-L | 0.138 (0.133-0.144) | 0.149 (0.142-0.156) | 0.185 (0.179-0.191) | 0.199 (0.190-0.207) | **0.273 (0.259-0.288)** |

**Supplementary Table 14: Quantitative evaluation of report generation using linguistic metrics on the Xiamen cohort (*n* = 251).** Best performing model for each metric is bolded. 95% CI is included in parentheses.

| Metric | RadFM | CT-CHAT | GPT-4o | MedGemma | RenalCLIP |
|---|---|---|---|---|---|
| BLEU-1 | 0.015<br>(0.012-0.017) | 0.173<br>(0.166-0.181) | 0.257<br>(0.249-0.266) | 0.176<br>(0.165-0.187) | **0.318**<br>**(0.305-0.331)** |
| BLEU-2 | 0.007<br>(0.006-0.009) | 0.091<br>(0.087-0.096) | 0.112<br>(0.107-0.117) | 0.079<br>(0.074-0.085) | **0.192**<br>**(0.183-0.201)** |
| BLEU-4 | 0.000<br>(0.000-0.000) | 0.020<br>(0.017-0.022) | 0.025<br>(0.022-0.029) | 0.024<br>(0.021-0.027) | **0.076**<br>**(0.071-0.081)** |
| METEOR | 0.098<br>(0.094-0.102) | 0.212<br>(0.208-0.218) | 0.258<br>(0.251-0.265) | 0.221<br>(0.213-0.229) | **0.334**<br>**(0.327-0.341)** |
| ROUGE-L | 0.146<br>(0.141-0.150) | 0.153<br>(0.150-0.157) | 0.200<br>(0.196-0.203) | 0.210<br>(0.205-0.216) | **0.267**<br>**(0.261-0.275)** |

**Supplementary Table 15: Quantitative evaluation of report generation using linguistic metrics on the Shandong cohort (*n* = 229).** Best performing model for each metric is bolded. 95% CI is included in parentheses.

| Metric | RadFM | CT-CHAT | GPT-4o | MedGemma | RenalCLIP |
|---|---|---|---|---|---|
| BLEU-1 | 0.049 (0.044-0.055) | 0.095 (0.089-0.100) | 0.201 (0.194-0.209) | 0.256 (0.246-0.266) | **0.226 (0.217-0.236)** |
| BLEU-2 | 0.027 (0.024-0.030) | 0.045 (0.042-0.047) | 0.087 (0.083-0.091) | 0.109 (0.103-0.115) | **0.135 (0.129-0.142)** |
| BLEU-4 | 0.000 (0.000-0.000) | 0.009 (0.007-0.010) | 0.019 (0.017-0.021) | 0.030 (0.027-0.034) | **0.056 (0.052-0.060)** |
| METEOR | 0.134 (0.128-0.140) | 0.215 (0.210-0.220) | 0.301 (0.294-0.307) | 0.285 (0.277-0.292) | **0.345 (0.338-0.353)** |
| ROUGE-L | 0.163 (0.157-0.168) | 0.122 (0.119-0.126) | 0.195 (0.191-0.198) | 0.231 (0.226-0.236) | **0.250 (0.244-0.258)** |

**Supplementary Table 16: Quantitative evaluation of report generation using linguistic metrics on the Ruijin cohort (*n* = 189).** Best performing model for each metric is bolded. 95% CI is included in parentheses.

| Metric | RadFM | CT-CHAT | GPT-4o | MedGemma | RenalCLIP |
|---|---|---|---|---|---|
| BLEU-1 | 0.029<br>(0.024-0.034) | 0.154<br>(0.146-0.162) | 0.226<br>(0.217-0.236) | 0.207<br>(0.194-0.220) | **0.279**<br>**(0.266-0.292)** |
| BLEU-2 | 0.016<br>(0.013-0.019) | 0.086<br>(0.081-0.091) | 0.096<br>(0.090-0.102) | 0.089<br>(0.082-0.095) | **0.167**<br>**(0.159-0.176)** |
| BLEU-4 | 0.000<br>(0.000-0.000) | 0.022<br>(0.019-0.024) | 0.020<br>(0.016-0.023) | 0.028<br>(0.024-0.032) | **0.068**<br>**(0.064-0.074)** |
| METEOR | 0.119<br>(0.113-0.125) | 0.230<br>(0.224-0.235) | 0.271<br>(0.264-0.278) | 0.240<br>(0.232-0.249) | **0.342**<br>**(0.333-0.350)** |
| ROUGE-L | 0.169<br>(0.163-0.175) | 0.155<br>(0.150-0.159) | 0.192<br>(0.189-0.196) | 0.211<br>(0.205-0.215) | **0.263**<br>**(0.255-0.270)** |

**Supplementary Table 17: Quantitative evaluation of report generation using linguistic metrics on the Zhangye cohort (*n* = 67).** Best performing model for each metric is bolded. 95% CI is included in parentheses.

| Metric | RadFM | CT-CHAT | GPT-4o | MedGemma | RenalCLIP |
|---|---|---|---|---|---|
| BLEU-1 | 0.005 (0.003-0.009) | 0.172 (0.155-0.190) | 0.265 (0.245-0.283) | 0.149 (0.123-0.175) | **0.343 (0.313-0.352)** |
| BLEU-2 | 0.003 (0.002-0.005) | 0.084 (0.074-0.095) | 0.120 (0.110-0.130) | 0.067 (0.056-0.079) | **0.199 (0.180-0.205)** |
| BLEU-4 | 0.000 (0.000-0.000) | 0.018 (0.014-0.022) | 0.027 (0.021-0.032) | 0.018 (0.014-0.023) | **0.076 (0.066-0.082)** |
| METEOR | 0.095 (0.079-0.113) | 0.193 (0.184-0.204) | 0.241 (0.227-0.257) | 0.210 (0.192-0.229) | **0.291 (0.276-0.305)** |
| ROUGE-L | 0.132 (0.121-0.143) | 0.146 (0.138-0.154) | 0.189 (0.181-0.198) | 0.193 (0.185-0.202) | **0.238 (0.227-0.249)** |

**Supplementary Table 18: ROC AUC on the malignancy diagnosis task across different models and cohorts.** Best performing model for each cohort is bolded. 95% CI is included in parentheses. * indicates $P < 0.05$ (two-sided paired permutation test vs. CNN Rand Init). The sample size ($n$) represents the exact number of independent patients evaluated in each cohort.

| Cohort | CNN (Rand Init) | CT-CLIP | Merlin | CT-FM | RenalCLIP |
|---|---|---|---|---|---|
| Internal test | 0.791 | 0.616 | 0.733 | 0.732 | **0.856** |
| ($n$ = 100) | (0.662-0.894) | (0.453-0.769) | (0.561-0.878) | (0.561-0.886) | **(0.741-0.944)** |
| Xiamen | 0.736 | 0.682 | 0.599 | 0.751 | **0.890*** |
| ($n$ = 251) | (0.662-0.805) | (0.606-0.753) | (0.526-0.670) | (0.684-0.812) | **(0.840-0.934)** |
| Shandong | 0.707 | 0.597 | 0.704 | 0.738 | **0.738** |
| ($n$ = 250) | (0.621-0.791) | (0.482-0.696) | (0.593-0.806) | (0.630-0.831) | **(0.631-0.835)** |
| Ruijin | 0.685 | 0.611 | 0.718 | 0.682 | **0.824*** |
| ($n$ = 229) | (0.591-0.774) | (0.521-0.708) | (0.626-0.809) | (0.583-0.785) | **(0.726-0.908)** |
| Lianyun | 0.693 | 0.566 | 0.619 | 0.738 | **0.838*** |
| ($n$ = 220) | (0.604-0.783) | (0.474-0.655) | (0.538-0.700) | (0.647-0.819) | **(0.765-0.902)** |
| Zhangye | 0.738 | 0.326 | 0.507 | 0.719 | **0.880*** |
| ($n$ = 67) | (0.584-0.865) | (0.172-0.496) | (0.308-0.702) | (0.560-0.863) | **(0.719-0.982)** |
| Combined external test | 0.717 | 0.600 | 0.650 | 0.739 | **0.841*** |
| ($n$ = 1,017) | (0.676-0.756) | (0.555-0.638) | (0.607-0.690) | (0.698-0.779) | **(0.808-0.873)** |
| TCIA | 0.586 | 0.593 | 0.593 | 0.609 | **0.680** |
| ($n$ = 425) | (0.453-0.734) | (0.430-0.733) | (0.437-0.739) | (0.455-0.755) | **(0.542-0.814)** |

**Supplementary Table 19: ROC AUC on the aggressiveness diagnosis task across different models and cohorts.** Best performing model for each cohort is bolded. 95% CI is included in parentheses. * indicates $P < 0.05$ (two-sided paired permutation test vs. CNN Rand Init). The sample size ($n$) represents the exact number of independent patients evaluated in each cohort.

| Cohort | CNN (Rand Init) | CT-CLIP | Merlin | CT-FM | RenalCLIP |
|---|---|---|---|---|---|
| Internal test | 0.592 | 0.490 | 0.584 | 0.668 | **0.775*** |
| ($n$ = 99) | (0.466-0.713) | (0.356-0.619) | (0.440-0.717) | (0.544-0.782) | **(0.673-0.878)** |
| Xiamen | 0.751 | 0.586 | 0.644 | 0.817 | **0.817** |
| ($n$ = 251) | (0.655-0.841) | (0.467-0.694) | (0.521-0.767) | (0.759-0.869) | **(0.738-0.888)** |
| Shandong | 0.653 | 0.556 | 0.661 | 0.646 | **0.707** |
| ($n$ = 250) | (0.582-0.725) | (0.481-0.634) | (0.577-0.737) | (0.570-0.723) | **(0.639-0.775)** |
| Ruijin | 0.650 | 0.535 | 0.643 | 0.649 | **0.654** |
| ($n$ = 228) | (0.572-0.728) | (0.464-0.611) | (0.570-0.713) | (0.573-0.719) | **(0.581-0.724)** |
| Lianyun | 0.651 | 0.543 | **0.740** | 0.636 | 0.645 |
| ($n$ = 209) | (0.546-0.746) | (0.421-0.653) | **(0.644-0.840)** | (0.520-0.732) | (0.532-0.756) |
| Zhangye | **0.541** | 0.513 | 0.413 | 0.345 | 0.484 |
| ($n$ = 67) | **(0.393-0.683)** | (0.376-0.656) | (0.267-0.557) | (0.214-0.484) | (0.358-0.631) |
| Combined external test | 0.676 | 0.547 | 0.681 | 0.680 | **0.703** |
| ($n$ = 1,005) | (0.639-0.712) | (0.508-0.588) | (0.642-0.719) | (0.640-0.716) | **(0.667-0.738)** |
| TCIA | 0.610 | 0.485 | 0.581 | 0.642 | **0.661** |
| ($n$ = 425) | (0.559-0.661) | (0.434-0.539) | (0.527-0.633) | (0.589-0.696) | **(0.611-0.708)** |

**Supplementary Table 20: Sample sizes and class distributions for diagnosis and prognosis tasks across cohorts, stratified by tumor size.** Patients were stratified into Small Renal Masses (SRMs, ≤ 4 cm) and large masses (> 4 cm) to assess the impact of tumor size on model performance. *n* represents the total number of eligible patients with complete labels for each specific task. For the aggressiveness diagnosis task, a small subset of patients was excluded from evaluation due to indeterminate pathological grades.

| **Clinical application** | **Variable** | **Internal training cohort *n* = 400** | | **Internal test cohort *n* = 100** | | **Combined external test cohort *n* = 1,017** | | **TCIA cohort *n* = 425** | |
|---|---|---|---|---|---|---|---|---|---|
| | | **≤ 4 cm** | **> 4 cm** | **≤ 4 cm** | **> 4 cm** | **≤ 4 cm** | **> 4 cm** | **≤ 4 cm** | **> 4 cm** |
| **Diagnosis** (applies to fine-tuning & zero-shot) | **Malignancy** (%) | *n* = 226 | *n* = 174 | *n* = 50 | *n* = 50 | *n* = 444 | *n* = 573 | *n* = 173 | *n* = 252 |
| | Benign | 36(15.9) | 29(16.7) | 10(20.0) | 4(8.0) | 82(18.5) | 118(20.6) | 10(5.8) | 4(1.6) |
| | Malignant | 190(84.1) | 145(83.3) | 40(80.0) | 46(92.0) | 362(81.5) | 455(79.4) | 163(94.2) | 248(98.4) |
| | **Aggressiveness** (%) | *n* = 225 | *n* = 173 | *n* = 49 | *n* = 50 | *n* = 439 | *n* = 566 | *n* = 173 | *n* = 252 |
| | Indolent | 206(91.6) | 115(66.5) | 40(81.6) | 28(56.0) | 385(87.7) | 359(63.4) | 129(74.6) | 89(35.3) |
| | Aggressive | 19(8.4) | 58(33.5) | 9(18.4) | 22(44.0) | 54(12.3) | 207(36.6) | 44(25.4) | 163(64.7) |
| **Prognosis** | **RFS outcome** (%) | *n* = 161 | *n* = 113 | *n* = 28 | *n* = 34 | *n* = 80 | *n* = 135 | *n* = 69 | *n* = 141 |
| | Censored | 155(96.3) | 86(76.1) | 27(96.4) | 24(70.6) | 79(98.8) | 106(78.5) | 64(92.8) | 91(64.5) |
| | Recurrence | 6(3.7) | 27(23.9) | 1(3.6) | 10(29.4) | 1(1.2) | 29(21.5) | 5(7.2) | 50(35.5) |
| | **DSS outcome** (%) | *n* = 161 | *n* = 113 | *n* = 28 | *n* = 34 | *n* = 81 | *n* = 143 | *n* = 75 | *n* = 164 |
| | Censored | 157(97.5) | 95(84.1) | 28(100.0) | 26(76.5) | 81(100.0) | 116(81.1) | 74(98.7) | 129(78.7) |
| | Death of kidney cancer | 4(2.5) | 18(15.9) | 0(0.0) | 8(23.5) | 0(0.0) | 27(18.9) | 1(1.3) | 35(21.3) |
| | **OS outcome** (%) | *n* = 161 | *n* = 113 | *n* = 28 | *n* = 34 | *n* = 81 | *n* = 143 | *n* = 161 | *n* = 245 |
| | Censored | 151(93.8) | 94(83.2) | 27(96.4) | 25(73.5) | 77(95.1) | 112(78.3) | 150(93.2) | 180(73.5) |
| | Death from all causes | 10(6.2) | 19(16.8) | 1(3.6) | 9(26.5) | 4(4.9) | 31(21.7) | 11(6.8) | 65(26.5) |

**Supplementary Table 21: ROC AUC of different models for diagnostic classification (Malignancy and Aggressiveness) stratified by tumor size subgroups (≤4 cm and >4 cm).** The analysis was performed on the combined external test cohort. Best performing model for each cohort is bolded and 95% CI is included in parentheses. The sample size (*n*) represents the exact number of independent patients evaluated in each subgroup.

| Task | Subgroup | CNN (Rand Init) | CT-CLIP | Merlin | CT-FM | Nomogram | RenalCLIP |
|---|---|---|---|---|---|---|---|
| Malignancy | ≤4 cm | 0.630 | 0.601 | 0.671 | 0.636 | 0.636 | **0.762** |
| | (*n* = 444) | (0.561-0.690) | (0.534-0.659) | (0.610-0.724) | (0.568-0.700) | (0.565-0.712) | **(0.704-0.815)** |
| | >4 cm | 0.780 | 0.598 | 0.634 | 0.816 | 0.569 | **0.896** |
| | (*n* = 573) | (0.733-0.827) | (0.540-0.656) | (0.578-0.691) | (0.770-0.859) | (0.511-0.629) | **(0.856-0.933)** |
| Aggressiveness | ≤4 cm | 0.571 | 0.589 | 0.632 | 0.653 | 0.575 | **0.655** |
| | (*n* = 439) | (0.494-0.654) | (0.506-0.670) | (0.555-0.707) | (0.579-0.721) | (0.506-0.652) | **(0.581-0.728)** |
| | >4 cm | **0.673** | 0.523 | 0.667 | 0.636 | 0.635 | 0.627 |
| | (*n* = 566) | **(0.622-0.721)** | (0.472-0.570) | (0.619-0.713) | (0.583-0.685) | (0.586-0.680) | (0.577-0.675) |

**Supplementary Table 22: C-index on the prognosis (RFS endpoint) task across different models and cohorts.** Best performing model for each cohort is bolded. The sample size (*n*) represents the exact number of independent patients evaluated in each cohort.

| Cohort | CNN (Rand Init) | CT-CLIP | Merlin | CT-FM | RenalCLIP |
|---|---|---|---|---|---|
| Internal test (*n* = 62) | 0.684 | 0.509 | 0.711 | 0.709 | **0.864** |
| Xiamen (*n* = 85) | 0.668 | 0.760 | 0.641 | 0.519 | **0.828** |
| Ruijin (*n* = 78) | 0.456 | 0.529 | 0.538 | **0.609** | 0.523 |
| Zhangye (*n* = 52) | **0.794** | 0.420 | 0.605 | 0.726 | 0.741 |
| Combined external test (*n* = 215) | 0.640 | 0.511 | 0.583 | 0.661 | **0.671** |
| TCIA (*n* = 210) | 0.547 | 0.461 | 0.535 | 0.592 | **0.726** |

**Supplementary Table 23: C-index on the prognosis (DSS endpoint) task across different models and cohorts.** Best performing model for each cohort is bolded. The sample size (*n*) represents the exact number of independent patients evaluated in each cohort.

| Cohort | CNN (Rand Init) | CT-CLIP | Merlin | CT-FM | RenalCLIP |
|---|---|---|---|---|---|
| Internal test (*n* = 62) | 0.702 | 0.580 | 0.754 | 0.765 | **0.930** |
| Xiamen (*n* = 87) | 0.804 | 0.754 | 0.862 | **0.919** | 0.873 |
| Ruijin (*n* = 85) | 0.663 | 0.587 | 0.634 | **0.708** | 0.700 |
| Zhangye (*n* = 52) | **0.780** | 0.421 | 0.572 | 0.694 | 0.720 |
| Combined external test (*n* = 224) | 0.738 | 0.527 | 0.654 | 0.742 | **0.756** |
| TCIA (*n* = 239) | 0.586 | 0.475 | 0.561 | 0.649 | **0.690** |

**Supplementary Table 24: C-index on the prognosis (OS endpoint) task across different models and cohorts.** Best performing model for each cohort is bolded. The sample size (*n*) represents the exact number of independent patients evaluated in each cohort.

| Cohort | CNN (Rand Init) | CT-CLIP | Merlin | CT-FM | RenalCLIP |
|---|---|---|---|---|---|
| Internal test (*n* = 62) | 0.682 | 0.576 | 0.641 | 0.704 | **0.859** |
| Xiamen (*n* = 87) | 0.797 | 0.803 | **0.872** | 0.763 | 0.691 |
| Ruijin (*n* = 85) | 0.645 | 0.537 | 0.629 | 0.635 | **0.648** |
| Zhangye (*n* = 52) | **0.780** | 0.421 | 0.572 | 0.694 | 0.720 |
| Combined external test (*n* = 224) | **0.724** | 0.515 | 0.646 | 0.690 | 0.707 |
| TCIA (*n* = 406) | 0.564 | 0.517 | 0.553 | 0.623 | **0.650** |

**Supplementary Table 25. Cohort- and endpoint-specific median risk score thresholds used to stratify patients into high- and low-risk groups across all benchmarked models.**

For each cohort and endpoint, the cutoff was calculated as the median risk score among patients with nonmissing follow-up time, event status, and risk score for that endpoint. Patients with scores above the cutoff were assigned to the high-risk group; the remaining patients were assigned to the low-risk group. n denotes the number of eligible patients.

### A. Recurrence-free survival (RFS)

| Cohort | n | CNN (Rand Init) | CT-CLIP | Merlin | CT-FM | RenalCLIP |
|---|---|---|---|---|---|---|
| Internal test | 62 | -4.419897 | -0.346553 | -9.831388 | -18.442986 | -0.386112 |
| Xiamen | 85 | -4.700004 | 0.097675 | -0.886516 | -18.673702 | -0.317465 |
| Ruijin | 78 | -3.570477 | -0.346653 | -7.156956 | -18.113005 | -0.137933 |
| Zhangye | 52 | -4.261896 | -0.112493 | 1.590336 | -16.836731 | -0.072771 |
| Combined external | 215 | -4.285084 | -0.316392 | -0.617850 | -18.189079 | -0.190636 |
| TCIA | 210 | -3.218741 | -0.193463 | -3.045284 | -17.560242 | -0.185488 |

### B. Disease-specific survival (DSS)

| Cohort | n | CNN (Rand Init) | CT-CLIP | Merlin | CT-FM | RenalCLIP |
|---|---|---|---|---|---|---|
| Internal test | 62 | -4.419897 | -0.346553 | -9.831388 | -18.442986 | -0.386112 |
| Xiamen | 87 | -4.652021 | 0.160324 | -0.568154 | -18.611826 | -0.294776 |
| Ruijin | 85 | -3.502184 | -0.346485 | -4.673376 | -17.925455 | -0.081096 |
| Zhangye | 52 | -4.261896 | -0.112493 | 1.590336 | -16.836731 | -0.072771 |
| Combined external | 224 | -4.254960 | -0.318576 | -0.612673 | -18.113005 | -0.188484 |
| TCIA | 239 | -3.192495 | -0.243618 | -3.201100 | -17.680422 | -0.209033 |

### C. Overall survival (OS)

| Cohort | n | CNN (Rand Init) | CT-CLIP | Merlin | CT-FM | RenalCLIP |
|---|---|---|---|---|---|---|
| Internal test | 62 | -4.419897 | -0.346553 | -9.831388 | -18.442986 | -0.386112 |
| Xiamen | 87 | -4.652021 | 0.160324 | -0.568154 | -18.611826 | -0.294776 |
| Ruijin | 85 | -3.502184 | -0.346485 | -4.673376 | -17.925455 | -0.081096 |
| Zhangye | 52 | -4.261896 | -0.112493 | 1.590336 | -16.836731 | -0.072771 |
| Combined external | 224 | -4.254960 | -0.318576 | -0.612673 | -18.113005 | -0.188484 |
| TCIA | 406 | -3.567001 | -0.092568 | -2.580337 | -18.408607 | -0.260263 |

Thresholds are displayed to six decimal places. The accompanying CSV provides full-precision cutoffs and group sizes. RFS, DSS and OS cutoffs may differ because the eligible patient subsets differ.

**Supplementary Table 26: C-index of different models for prognosis (RFS endpoint) task stratified by tumor size subgroups (≤ 4 cm and > 4 cm).** The analysis was performed on the TCIA cohort. Best performing model for each cohort is bolded. The sample size (*n*) represents the exact number of independent patients evaluated in each subgroup.

| Subgroup | CNN (Rand Init) | CT-CLIP | Merlin | CT-FM | RenalCLIP |
|---|---|---|---|---|---|
| ≤4 cm (*n* = 69) | 0.527 | 0.321 | 0.399 | 0.399 | **0.856** |
| >4 cm (*n* = 141) | 0.488 | 0.484 | 0.528 | 0.549 | **0.649** |

**Supplementary Table 27: Univariate and multivariate analysis of RFS in TCIA cohort.** *P*-values were calculated using the Wald test. Asterisk (*) indicates statistical significance ($p < 0.05$).

| Variables | Univariate analysis | | | Multivariate analysis | | |
|---|---|---|---|---|---|---|
| | Hazard ratio | 95% CI | *P*-value | Hazard ratio | 95% CI | *P*-value |
| **CNN (Rand Init)** | | | | | | |
| **TNM Stage** | | | | | | |
| Stage II vs Stage I | 2.80 | 0.79-9.96 | 0.111 | 2.39 | 0.64-8.90 | 0.193 |
| Stage III vs Stage I | 10.09 | 4.09-24.9 | <0.001* | 9.67 | 3.86-24.2 | < 0.001* |
| Stage IV vs Stage I | 40.14 | 16.07-100.23 | <0.001* | 30.57 | 11.60-80.52 | < 0.001* |
| **WHO/ISUP Grade** | | | | | | |
| Grade 2 vs Grade 1 | 0.60 | 0.08-4.59 | 0.623 | 0.558 | 0.07-4.36 | 0.578 |
| Grade 3 vs Grade 1 | 1.63 | 0.22-12.02 | 0.635 | 0.936 | 0.12-7.08 | 0.949 |
| Grade 4 vs Grade 1 | 4.43 | 0.69-33.46 | 0.149 | 1.893 | 0.24-15.22 | 0.548 |
| **Risk score** | 1.04 | 0.95-1.14 | 0.360 | 0.956 | 0.87-1.05 | 0.371 |
| **CT-CLIP** | | | | | | |
| **TNM Stage** | | | | | | |
| Stage II vs Stage I | 2.80 | 0.79-9.96 | 0.111 | 2.20 | 0.60-8.02 | 0.234 |
| Stage III vs Stage I | 10.09 | 4.09-24.9 | < 0.001* | 9.22 | 3.70-22.95 | < 0.001* |
| Stage IV vs Stage I | 40.14 | 16.07-100.23 | < 0.001* | 28.05 | 10.76-73.14 | < 0.001* |
| **WHO/ISUP Grade** | | | | | | |
| Grade 2 vs Grade 1 | 0.60 | 0.08-4.59 | 0.623 | 0.49 | 0.06-3.81 | 0.497 |
| Grade 3 vs Grade 1 | 1.63 | 0.22-12.02 | 0.635 | 0.87 | 0.11-6.59 | 0.891 |
| Grade 4 vs Grade 1 | 4.43 | 0.69-33.46 | 0.149 | 1.61 | 0.20-12.75 | 0.653 |
| **Risk score** | 0.76 | 0.36-1.59 | 0.464 | 0.869 | 0.41-1.85 | 0.717 |
| **Merlin** | | | | | | |
| **TNM Stage** | | | | | | |
| Stage II vs Stage I | 2.80 | 0.79-9.96 | 0.111 | 2.04 | 0.55-7.53 | 0.283 |
| Stage III vs Stage I | 10.09 | 4.09-24.9 | < 0.001* | 8.72 | 3.49-21.81 | < 0.001* |
| Stage IV vs Stage I | 40.14 | 16.07-100.23 | < 0.001* | 29.01 | 11.12-75.72 | < 0.001* |
| **WHO/ISUP Grade** | | | | | | |
| Grade 2 vs Grade 1 | 0.60 | 0.08-4.59 | 0.623 | 0.55 | 0.07-4.26 | 0.564 |
| Grade 3 vs Grade 1 | 1.63 | 0.22-12.02 | 0.635 | 0.95 | 0.13-7.23 | 0.963 |
| Grade 4 vs Grade 1 | 4.43 | 0.69-33.46 | 0.149 | 1.76 | 0.22-14.00 | 0.592 |
| **Risk score** | 1.02 | 0.97-1.07 | 0.408 | 1.02 | 0.97-1.06 | 0.436 |
| **CT-FM** | | | | | | |
| **TNM Stage** | | | | | | |
| Stage II vs Stage I | 2.80 | 0.79-9.96 | 0.111 | 2.15 | 0.58-7.98 | 0.251 |
| Stage III vs Stage I | 10.09 | 4.09-24.9 | < 0.001* | 9.06 | 3.58-22.92 | < 0.001* |
| Stage IV vs Stage I | 40.14 | 16.07-100.23 | < 0.001* | 28.33 | 10.84-74.03 | < 0.001* |
| **WHO/ISUP Grade** | | | | | | |
| Grade 2 vs Grade 1 | 0.60 | 0.08-4.59 | 0.623 | 0.50 | 0.06-4.05 | 0.515 |
| Grade 3 vs Grade 1 | 1.63 | 0.22-12.02 | 0.635 | 0.89 | 0.11-6.87 | 0.908 |
| Grade 4 vs Grade 1 | 4.43 | 0.69-33.46 | 0.149 | 1.62 | 0.18-14.37 | 0.664 |
| **Risk score** | 1.058 | 1.02-1.10 | 0.007* | 1.00 | 0.96-1.05 | 0.969 |
| **RenalCLIP** | | | | | | |
| **TNM Stage** | | | | | | |
| Stage II vs Stage I | 2.80 | 0.79-9.96 | 0.111 | 1.23 | 0.32-4.78 | 0.765 |
| Stage III vs Stage I | 10.09 | 4.09-24.9 | < 0.001* | 5.90 | 2.23-15.65 | < 0.001* |
| Stage IV vs Stage I | 40.14 | 16.07-100.23 | < 0.001* | 22.13 | 8.27-59.28 | < 0.001* |

| Variables | Univariate analysis | | | Multivariate analysis | | |
|---|---|---|---|---|---|---|
| | **Hazard ratio** | **95% CI** | ***P*-value** | **Hazard ratio** | **95% CI** | ***P*-value** |
| **WHO/ISUP Grade** | | | | | | |
| Grade 2 vs Grade 1 | 0.60 | 0.08-4.59 | 0.623 | 0.45 | 0.06-3.50 | 0.445 |
| Grade 3 vs Grade 1 | 1.63 | 0.22-12.02 | 0.635 | 0.81 | 0.11-6.10 | 0.836 |
| Grade 4 vs Grade 1 | 4.43 | 0.69-33.46 | 0.149 | 1.14 | 0.14-9.22 | 0.904 |
| **Risk score** | 4.31 | 2.46-7.54 | < 0.001* | 2.27 | 1.17-4.42 | 0.016* |

**Supplementary Table 28: Performance of different models on the cross-modal retrieval task for the internal test cohort (*n* = 100).** Best performing model for each metric is bolded. 95% CI is included in parentheses.

| Metric | Random Guess | CT-CLIP | Merlin | RenalCLIP |
|---|---|---|---|---|
| **Text-to-image** | | | | |
| Recall@1 | 0.010 | 0.000 (0.000-0.000) | 0.020 (0.000-0.040) | **0.030 (0.010-0.060)** |
| Recall@3 | 0.030 | 0.000 (0.000-0.010) | 0.060 (0.020-0.090) | **0.120 (0.060-0.170)** |
| Recall@5 | 0.050 | 0.000 (0.000-0.030) | 0.090 (0.040-0.120) | **0.200 (0.130-0.250)** |
| **Image-to-text** | | | | |
| Recall@1 | 0.010 | 0.010 (0.000-0.010) | 0.020 (0.010-0.030) | **0.020 (0.010-0.050)** |
| Recall@3 | 0.030 | 0.010 (0.000-0.030) | 0.040 (0.010-0.070) | **0.110 (0.060-0.140)** |
| Recall@5 | 0.050 | 0.010 (0.000-0.030) | 0.050 (0.020-0.090) | **0.130 (0.090-0.190)** |

**Supplementary Table 29: Performance of different models on the cross-modal retrieval task for the Xiamen cohort (*n* = 251).** Best performing model for each metric is bolded. 95% CI is included in parentheses.

| Metric | Random Guess | CT-CLIP | Merlin | RenalCLIP |
|---|---|---|---|---|
| **Text-to-image** | | | | |
| Recall@1 | 0.004 | 0.000 (0.000-0.000) | 0.004 (0.000-0.012) | **0.028 (0.012-0.036)** |
| Recall@3 | 0.012 | 0.000 (0.000-0.004) | 0.016 (0.004-0.028) | **0.060 (0.036-0.088)** |
| Recall@5 | 0.020 | 0.000 (0.000-0.012) | 0.024 (0.008-0.036) | **0.088 (0.056-0.127)** |
| **Image-to-text** | | | | |
| Recall@1 | 0.004 | 0.000 (0.000-0.004) | 0.012 (0.000-0.016) | **0.020 (0.008-0.028)** |
| Recall@3 | 0.012 | 0.004 (0.000-0.008) | 0.020 (0.008-0.032) | **0.044 (0.024-0.064)** |
| Recall@5 | 0.020 | 0.004 (0.000-0.012) | 0.024 (0.012-0.044) | **0.068 (0.044-0.096)** |

**Supplementary Table 30: Performance of different models on the cross-modal retrieval task for the Shandong cohort (*n* = 229).** Best performing model for each metric is bolded. 95% CI is included in parentheses.

| Metric | Random Guess | CT-CLIP | Merlin | RenalCLIP |
|---|---|---|---|---|
| **Text-to-image** | | | | |
| Recall@1 | 0.004 | 0.000<br>(0.000-0.004) | 0.004<br>(0.000-0.009) | **0.039**<br>**(0.017-0.039)** |
| Recall@3 | 0.013 | 0.000<br>(0.000-0.009) | 0.017<br>(0.004-0.026) | **0.061**<br>**(0.044-0.087)** |
| Recall@5 | 0.022 | 0.004<br>(0.000-0.013) | 0.022<br>(0.013-0.039) | **0.092**<br>**(0.061-0.118)** |
| **Image-to-text** | | | | |
| Recall@1 | 0.004 | 0.000<br>(0.000-0.004) | 0.000<br>(0.000-0.004) | **0.022**<br>**(0.013-0.031)** |
| Recall@3 | 0.013 | 0.004<br>(0.000-0.013) | 0.004<br>(0.000-0.017) | **0.057**<br>**(0.031-0.079)** |
| Recall@5 | 0.022 | 0.004<br>(0.000-0.013) | 0.013<br>(0.000-0.031) | **0.070**<br>**(0.048-0.105)** |

**Supplementary Table 31: Performance of different models on the cross-modal retrieval task for the Ruijin cohort ($n$ = 189).** Best performing model for each metric is bolded. 95% CI is included in parentheses.

| Metric | Random Guess | CT-CLIP | Merlin | RenalCLIP |
|---|---|---|---|---|
| **Text-to-image** | | | | |
| Recall@1 | 0.005 | 0.000 (0.000-0.000) | 0.005 (0.000-0.011) | **0.011 (0.005-0.032)** |
| Recall@3 | 0.016 | 0.000 (0.000-0.000) | 0.021 (0.005-0.032) | **0.058 (0.026-0.079)** |
| Recall@5 | 0.026 | 0.000 (0.000-0.005) | 0.032 (0.011-0.048) | **0.085 (0.053-0.116)** |
| **Image-to-text** | | | | |
| Recall@1 | 0.005 | 0.000 (0.000-0.000) | 0.005 (0.000-0.011) | **0.016 (0.005-0.021)** |
| Recall@3 | 0.016 | 0.000 (0.000-0.000) | 0.011 (0.000-0.021) | **0.042 (0.016-0.058)** |
| Recall@5 | 0.026 | 0.000 (0.000-0.000) | 0.016 (0.005-0.032) | **0.048 (0.032-0.074)** |

**Supplementary Table 32: Performance of different models on the cross-modal retrieval task for the Zhangye cohort (*n* = 67).** Best performing model for each metric is bolded. 95% CI is included in parentheses.

| Metric | Random Guess | CT-CLIP | Merlin | RenalCLIP |
|---|---|---|---|---|
| **Text-to-image** | | | | |
| Recall@1 | 0.015 | 0.015 (0.000-0.015) | 0.030 (0.000-0.045) | **0.030 (0.015-0.060)** |
| Recall@3 | 0.045 | 0.015 (0.000-0.045) | 0.045 (0.015-0.090) | **0.119 (0.060-0.164)** |
| Recall@5 | 0.075 | 0.015 (0.000-0.060) | 0.090 (0.045-0.134) | **0.134 (0.090-0.239)** |
| **Image-to-text** | | | | |
| Recall@1 | 0.015 | 0.000 (0.000-0.000) | 0.015 (0.000-0.015) | **0.104 (0.045-0.104)** |
| Recall@3 | 0.045 | 0.000 (0.000-0.030) | 0.030 (0.015-0.075) | **0.164 (0.090-0.239)** |
| Recall@5 | 0.075 | 0.015 (0.000-0.030) | 0.075 (0.030-0.134) | **0.224 (0.134-0.299)** |

**Supplementary Table 33: Generic prompt templates for zero-shot classification.** This table lists the 20 generic sentence templates used to construct the full prompt ensemble. To generate a complete prompt, the {} placeholder within each template is populated with a specific class descriptor. The full sets of class descriptors for the malignancy and aggressiveness tasks are detailed in Supplementary Tables 34-35.

| |
|---|
| A soft tissue mass that is {} is visible in the kidney. |
| A density shadow that is {} is observed in the kidney. |
| An area that is {} is visualized at the pole of the kidney. |
| A mass-like shadow that is {} is noted in the kidney. |
| An occupying lesion that is {} is seen in the kidney. |
| An abnormality that is {} is detected in the kidney. |
| A region of interest that is {} is found in the kidney. |
| A structural irregularity that is {} is visible in the kidney. |
| A mixed-density lesion that is {} is apparent in the kidney. |
| A nodule that is {} is observed in the kidney. |
| A shadow-like lesion that is {} is identified in the kidney. |
| A well-defined region that is {} is noted in the kidney. |
| An abnormal soft tissue that is {} is detected in the kidney. |
| A feature that is {} is recognized in the kidney. |
| A density lesion that is {} is revealed in the kidney. |
| The kidney shows a soft tissue mass that is {}. |
| The kidney reveals a density shadow that is {}. |
| The kidney demonstrates a lesion that is {}. |
| The kidney presents an occupying abnormality that is {}. |
| The kidney displays a structural irregularity that is {}. |

**Supplementary Table 34: Representative text prompts for malignancy zero-shot classification.**

| Class | Representative Class Prompts |
|---|---|
| Benign | benign, described as a mass containing fatty soft tissue, consistent with an angiomyolipoma |
| | benign, presenting as a round cystic lesion with a clear boundary and no enhancement after contrast |
| | benign, a finding of a nodule with uniform density, smooth borders, and no significant enhancement |
| | benign, a complex cyst with fine septa showing only mild, non-suspicious enhancement |
| | benign, a lesion showing mild and uniform progressive enhancement after contrast, without washout |
| Malignant | malignant, showing early significant uneven enhancement during the corticomedullary phase, and reduced density in the nephrographic phase |
| | malignant, presenting as a mass-like abnormal density shadow which is uneven and contains a central necrotic area |
| | malignant, a cystic-solid occupying lesion with thickened, irregular walls and enhancing solid nodules |
| | malignant, described as a solid tumor showing rapid enhancement and subsequent washout of contrast |
| | malignant, an irregular mass with unclear boundaries and heterogeneous density |

**Supplementary Table 35: Representative text prompts for aggressiveness zero-shot classification.**

| Class | Representative Class Prompts |
|---|---|
| Indolent | non-invasive, as the lesion is confined by the renal capsule and has not breached it |
| | non-invasive, with the perirenal fat spaces appearing clear and uninvolved |
| | non-invasive, showing no signs of involvement of the renal pelvis or calyces |
| | non-invasive, with no evidence of tumor thrombus in the renal vein or inferior vena cava |
| | non-invasive, presenting as a well-circumscribed lesion with sharp, distinct borders |
| Aggressive | invasive, as the tumor demonstrates extrarenal extension into the perirenal fat |
| | invasive, as the tumor appears to have breached the renal capsule |
| | invasive, showing clear infiltration of the renal pelvis or collecting system |
| | invasive, with a filling defect consistent with a tumor thrombus in the renal vein or IVC |
| | invasive, presenting as an infiltrative mass with poorly defined or blurred borders |

**Supplementary Table 36: ROC AUC on the zero-shot malignancy diagnosis task across different models and cohorts.** Results are grouped by zero-shot strategy. Best performing model for each cohort is bolded. 95% CI is included in parentheses. The sample size (*n*) represents the exact number of independent patients evaluated in each cohort.

| Cohort | CT-CLIP | Merlin | RenalCLIP |
|---|---|---|---|
| **Stochastic prompt sampling** | | | |
| Internal test | 0.493 | 0.569 | **0.646** |
| (*n* = 100) | (0.439-0.542) | (0.439-0.684) | **(0.302-0.841)** |
| Combined external test | 0.566 | 0.486 | **0.577** |
| (*n* = 1,017) | (0.524-0.590) | (0.400-0.568) | **(0.320-0.768)** |
| TCIA | 0.518 | **0.660** | 0.611 |
| (*n* = 425) | (0.423-0.629) | **(0.521-0.729)** | (0.282-0.745) |
| **Maximum similarity ensemble** | | | |
| Internal test (*n* = 100) | 0.449 | 0.656 | **0.829** |
| Combined external test (*n* = 1,017) | 0.570 | 0.464 | **0.743** |
| TCIA (*n* = 425) | 0.469 | 0.664 | **0.730** |

**Supplementary Table 37: ROC AUC on the zero-shot aggressiveness diagnosis task across different models and cohorts.** Results are grouped by zero-shot strategy. Best performing model for each cohort is bolded. 95% CI is included in parentheses. The sample size (*n*) represents the exact number of independent patients evaluated in each cohort.

| Cohort | CT-CLIP | Merlin | RenalCLIP |
|---|---|---|---|
| **Stochastic prompt sampling** | | | |
| Internal test | 0.489 | 0.548 | **0.624** |
| (*n* = 99) | (0.462-0.522) | (0.491-0.602) | **(0.480-0.738)** |
| Combined external test | 0.477 | 0.567 | **0.601** |
| (*n* = 1,005) | (0.448-0.505) | (0.505-0.624) | **(0.468-0.698)** |
| TCIA | 0.425 | 0.580 | **0.613** |
| (*n* = 425) | (0.406-0.457) | (0.534-0.607) | **(0.481-0.718)** |
| **Maximum similarity ensemble** | | | |
| Internal test (*n* = 99) | 0.483 | 0.514 | **0.650** |
| Combined external test (*n* = 1,005) | 0.480 | 0.588 | **0.647** |
| TCIA (*n* = 425) | 0.412 | 0.591 | **0.657** |

**Supplementary Table 38: ROC AUC of different models for zero-shot diagnostic classification (Malignancy and Aggressiveness) stratified by tumor size subgroups (≤4 cm and >4 cm).** The analysis was performed on the combined external test cohort. Predictions were generated using the maximum similarity ensemble strategy. Best performing model for each cohort is bolded. The sample size (*n*) represents the exact number of independent patients evaluated in each subgroup.

| Task | Subgroup | CT-CLIP | Merlin | RenalCLIP |
|---|---|---|---|---|
| Malignancy | ≤4 cm (*n* = 444) | 0.622 | 0.490 | **0.681** |
| | >4 cm (*n* = 573) | 0.525 | 0.453 | **0.807** |
| Aggressiveness | ≤4 cm (*n* = 439) | 0.475 | **0.587** | 0.539 |
| | >4 cm (*n* = 566) | 0.518 | 0.559 | **0.612** |

**Supplementary Table 39: Model performance on malignancy diagnosis with varying proportions of training data.** All values represent ROC AUC. For training data proportions less than 1.0, values are the mean ± standard deviation of $n = 5$ independent runs with different random seeds. The 1.0 proportion represents a single run on the full training dataset. The best-performing model for each cohort and proportion is highlighted in bold.

| Cohort | CNN (Rand Init) | CT-CLIP | Merlin | CT-FM | RenalCLIP |
|---|---|---|---|---|---|
| **10% Training data** | | | | | |
| Internal test | 0.546 ± 0.042 | 0.507 ± 0.096 | **0.615 ± 0.044** | 0.550 ± 0.104 | 0.613 ± 0.069 |
| Combined external test | 0.560 ± 0.034 | 0.505 ± 0.027 | 0.527 ± 0.023 | 0.585 ± 0.028 | **0.668 ± 0.061** |
| TCIA | 0.602 ± 0.128 | 0.466 ± 0.091 | **0.613 ± 0.082** | 0.586 ± 0.043 | 0.550 ± 0.096 |
| **20% Training data** | | | | | |
| Internal test | 0.628 ± 0.073 | 0.615 ± 0.070 | 0.688 ± 0.063 | 0.687 ± 0.082 | **0.729 ± 0.048** |
| Combined external test | 0.633 ± 0.030 | 0.547 ± 0.024 | 0.566 ± 0.034 | 0.637 ± 0.043 | **0.747 ± 0.081** |
| TCIA | 0.596 ± 0.087 | 0.557 ± 0.044 | **0.632 ± 0.088** | 0.615 ± 0.097 | 0.618 ± 0.081 |
| **40% Training data** | | | | | |
| Internal test | 0.691 ± 0.078 | 0.657 ± 0.036 | 0.644 ± 0.096 | 0.730 ± 0.083 | **0.776 ± 0.028** |
| Combined external test | 0.667 ± 0.026 | 0.583 ± 0.029 | 0.605 ± 0.017 | 0.691 ± 0.043 | **0.800 ± 0.039** |
| TCIA | **0.676 ± 0.037** | 0.603 ± 0.050 | 0.575 ± 0.103 | 0.568 ± 0.078 | 0.661 ± 0.049 |
| **80% Training data** | | | | | |
| Internal test | 0.771 ± 0.050 | 0.654 ± 0.050 | 0.694 ± 0.028 | 0.756 ± 0.037 | **0.779 ± 0.033** |
| Combined external test | 0.711 ± 0.010 | 0.598 ± 0.012 | 0.632 ± 0.015 | 0.728 ± 0.015 | **0.829 ± 0.011** |
| TCIA | 0.578 ± 0.029 | 0.607 ± 0.049 | 0.599 ± 0.103 | 0.613 ± 0.054 | **0.659 ± 0.035** |
| **100% Training data** | | | | | |
| Internal test | 0.791 | 0.616 | 0.733 | 0.732 | **0.856** |
| Combined external test | 0.717 | 0.600 | 0.650 | 0.739 | **0.841** |
| TCIA | 0.586 | 0.593 | 0.593 | 0.609 | **0.680** |

**Supplementary Table 40: Model performance on aggressiveness diagnosis with varying proportions of training data.** All values represent ROC AUC. For training data proportions less than 1.0, values are the mean ± standard deviation of n = 5 independent runs with different random seeds. The 1.0 proportion represents a single run on the full training dataset. The best-performing model for each cohort and proportion is highlighted in bold.

| Cohort | CNN (Rand Init) | CT-CLIP | Merlin | CT-FM | RenalCLIP |
|---|---|---|---|---|---|
| **10% Training data** | | | | | |
| Internal test | 0.580 ± 0.015 | 0.482 ± 0.068 | 0.552 ± 0.040 | 0.636 ± 0.075 | **0.686 ± 0.043** |
| Combined external test | 0.618 ± 0.034 | 0.509 ± 0.050 | 0.578 ± 0.053 | 0.610 ± 0.087 | **0.672 ± 0.043** |
| TCIA | 0.570 ± 0.040 | 0.498 ± 0.034 | 0.554 ± 0.041 | 0.578 ± 0.071 | **0.661 ± 0.059** |
| **20% Training data** | | | | | |
| Internal test | 0.601 ± 0.044 | 0.481 ± 0.059 | 0.558 ± 0.074 | 0.644 ± 0.058 | **0.706 ± 0.046** |
| Combined external test | 0.641 ± 0.022 | 0.532 ± 0.023 | 0.592 ± 0.041 | 0.645 ± 0.045 | **0.702 ± 0.032** |
| TCIA | 0.594 ± 0.044 | 0.496 ± 0.045 | 0.543 ± 0.053 | 0.581 ± 0.052 | **0.686 ± 0.037** |
| **40% Training data** | | | | | |
| Internal test | 0.606 ± 0.035 | 0.462 ± 0.093 | 0.578 ± 0.037 | 0.696 ± 0.025 | **0.720 ± 0.031** |
| Combined external test | 0.660 ± 0.030 | 0.522 ± 0.022 | 0.650 ± 0.038 | 0.673 ± 0.027 | **0.719 ± 0.032** |
| TCIA | 0.607 ± 0.038 | 0.508 ± 0.022 | 0.537 ± 0.038 | 0.600 ± 0.026 | **0.708 ± 0.016** |
| **80% Training data** | | | | | |
| Internal test | 0.605 ± 0.034 | 0.468 ± 0.054 | 0.598 ± 0.037 | 0.700 ± 0.021 | **0.751 ± 0.011** |
| Combined external test | 0.694 ± 0.009 | 0.534 ± 0.022 | 0.686 ± 0.017 | 0.690 ± 0.012 | **0.734 ± 0.013** |
| TCIA | 0.615 ± 0.006 | 0.534 ± 0.034 | 0.589 ± 0.014 | 0.599 ± 0.026 | **0.710 ± 0.015** |
| **100% Training data** | | | | | |
| Internal test | 0.592 | 0.490 | 0.584 | 0.668 | **0.775** |
| Combined external test | 0.676 | 0.547 | 0.681 | 0.680 | **0.703** |
| TCIA | 0.610 | 0.485 | 0.581 | 0.642 | **0.661** |

**Supplementary Table 41: LLM prompt for kidney-specific report extraction.**

**# Context**

You are a radiologist tasked with extracting information from radiology reports. The report you will be given is for a contrast-enhanced abdominal CT. Your task is to separately extract the descriptive findings pertaining to the left kidney and the right kidney. Ensure that descriptions from all contrast phases are included. If there are any descriptions pertaining to both kidneys (i.e., bilateral findings), extract this information and appropriately place it in the fields for both the left and right kidneys. The output must be in JSON format.

**# Example**

Input:

Report Findings:

Liver surface is smooth, lobes are proportional, and liver parenchyma shows no abnormal density; Spleen is not enlarged, with uniform density; Pancreas is unremarkable. A punctate high-density focus is seen in the right kidney. A small, low-density nodule measuring approximately 11x8mm is seen at the lower pole of the left kidney, protruding from the renal surface. It shows slight enhancement after contrast administration, with heterogeneous density and an internal low-density area. Multiple tiny cystic non-enhancing foci are also seen in the left kidney; The medial limb and root of the left adrenal gland are slightly thickened; No free fluid in the abdominal cavity.

Impression:

Nodule at the lower pole of the left kidney, likely a benign lesion, MRI recommended if necessary; Small cysts in the left kidney; Small stone in the right kidney.

**JSON Output:**{
    "report_findings": {
        "left_kidney": "A small, low-density nodule measuring approximately 11x8mm is seen at the lower pole of the left kidney, protruding from the renal surface. It shows slight enhancement after contrast administration, with heterogeneous density and an internal low-density area. Multiple tiny cystic non-enhancing foci are also seen in the left kidney; The medial limb and root of the left adrenal gland are slightly thickened;",
        "right_kidney": "A punctate high-density focus is seen in the right kidney."},
    "impression": {
        "left_kidney": "Nodule at the lower pole of the left kidney, likely a benign lesion, MRI recommended if necessary; Small cysts in the left kidney;",
        "right_kidney": "Small stone in the right kidney."}
}

**# Your Turn**

**Input:** Report Findings: {}, Impression: {}

**Output:** {}

**Supplementary Table 42: LLM prompt for report translation.**

| |
|---|
| **# Task** |
| Accurately translate the following imaging report findings and impression into English. The output must be in JSON format. |
| **# Input:** {} |
| **# Output:** {} |

**Supplementary Table 43: Questionnaire for LLM-based renal mass feature extraction.**

| No. | Questions | Options |
|---|---|---|
| 1. | Overall Nature and Number of Lesions | A. Solitary lesion<br>B. Multiple/Disseminated lesions<br>C. Other<br>D. Not mentioned/No mass |
| 2. | Location of the Renal Mass | A. Upper pole<br>B. Mid pole<br>C. Lower pole<br>D. Not mentioned/No mass |
| 3. | Size of the Renal Mass (assessed by maximum transverse diameter) | A. $\leq 4$ cm<br>B. $> 4$ cm and $\leq 7$ cm<br>C. $> 7$ cm and $\leq 10$ cm<br>D. $> 10$ cm<br>E. Not mentioned/No mass |
| 4. | Exophytic Renal Mass? | A. Yes<br>B. Not described |
| 5. | Morphology of the Renal Mass? | A. Round, ovoid, or oval-shaped<br>B. Irregular, ill-defined<br>C. Mass-like<br>D. Lobulated<br>E. Not mentioned/No mass |
| 6. | Cystic or Solid Nature of the Renal Mass? | A. Cystic<br>B. Cystic and solid<br>C. Solid<br>D. Not mentioned/No mass |
| 7. | Density/Signal Intensity of the Renal Mass? | A. Hypodense/Low signal<br>B. Isodense/Isointense<br>C. Hyperdense/High signal<br>D. Heterogeneous density/signal<br>E. Fat density (Angiomyolipoma (AML))<br>F. Mixed density<br>G. Not mentioned/No mass |
| 8. | Imaging Evidence of Necrosis within the Renal Mass? | A. Yes<br>B. No<br>C. Not mentioned/No mass |
| 9. | Enhancement Pattern of the Renal Mass after Contrast Administration? | A. Non-enhancing<br>B. Mild enhancement, apparent enhancement<br>C. Moderate enhancement<br>D. Marked enhancement<br>E. Heterogeneous enhancement<br>F. Cyst wall/septal enhancement<br>G. Not mentioned/No mass |
| 10. | Enhancement of the Renal Mass during the Arterial Phase? | A. No enhancement<br>B. Mild enhancement<br>C. Marked enhancement<br>D. Not mentioned/No mass |
| 11. | Clarity of Renal Mass Margins? | A. Well-defined (clear margins)<br>B. Ill-defined (poorly defined margins) |

| No. | Questions | Options |
|---|---|---|
| | | C. Not mentioned/No mass |
| 12. | Is the Renal Mass Likely Benign or Malignant? | A. Highly likely benign<br>B. Highly likely malignant<br>C. More likely benign<br>D. More likely malignant<br>E. Not mentioned/No mass |
| 13. | What is the Most Likely Nature of the Renal Mass? | A. Malignant Tumor<br>B. Angiomyolipoma (AML)<br>C. Cyst<br>D. Other Benign Lesion<br>E. Not mentioned/No mass |
| 14. | Invasion of Renal Pelvis/Calyces/Renal Sinus/Peripelvic Fat? | A. Yes<br>B. No<br>C. Not mentioned/No mass |

## Supplementary Table 44: LLM prompt for renal mass feature extraction.

**# Context**

You are a radiologist assisting in extracting information from imaging reports. Please carefully read the following imaging report and provide appropriate answers to the questions.

**# Task**

Imaging Report: {}

**# Question**

Note: If the answer to any question cannot be obtained from the imaging report, please select DK (Don't Know).

0. Is a renal mass mentioned in the report (including cysts, benign lesions, or malignant lesions)? If so, is it located in the left kidney, right kidney, or bilateral?

A. Left kidney B. Right kidney C. Bilateral kidneys D. Not mentioned

Then, for each identified renal mass, please answer the following questions (full details and options for each question are provided in Supplementary Table 43):

1. Overall Nature and Number
2. Location within Kidney
3. Size
4. Exophytic Type?
5. Morphology
6. Cystic or Solid Nature
7. Density/Signal Intensity
8. Evidence of Necrosis?
9. Enhancement Pattern (Overall)
10. Enhancement (Arterial Phase)
11. Margins Clarity
12. Likely Benign or Malignant?
13. Most Likely Nature?
14. Invasion of Renal Pelvis/Sinus/Peripelvic Fat?

Prefix the question number with L (e.g., L1) for answers pertaining to the left kidney mass. Prefix the question number with R (e.g., R1) for answers pertaining to the right kidney mass.

**# Format**

Please provide your answer in JSON format, structured as follows:

```
{
0: "...",
L1: "...",
...,
L14: "...",
R1: "...",
...,
R14: "..."
}
```

Only provide the letter corresponding to the answer choice; do not include the question description or answer options.

**# Result**

**Supplementary Table 45: Hyperparameters for image backbone pre-training.** Training was conducted on a single 80GB NVIDIA A100 GPU for up to 200 epochs. The best-performing model checkpoint was selected based on the highest macro-averaged AUC on the pre-training dataset's validation set, reflecting its performance across multiple classification tasks.

| **Hyperparameter** | **Value** |
|---|---|
| Batch size | 300 |
| Learning rate | 5e-4 |
| Weight decay | 5e-3 |
| AdamW β | (0.9,0.999) |
| Learning rate schedule | Cosine |
| Warmup ratio | 10% |
| Epochs | 200 |

**Supplementary Table 46: Hyperparameters for text encoder pre-training in the MLM phase.** This phase was conducted on a single 80GB NVIDIA A100 GPU for up to 1,000 iterations.

| **Hyperparameter** | **Value** |
|---|---|
| Layers | 32 |
| Heads | 32 |
| Embedding dimension | 4,096 |
| Hidden dimension | 14,336 |
| Position embedding | RoPE |
| Vocabulary size | 128,256 |
| Batch size | 32 |
| Learning rate | 5e-5 |
| MLM probability | 0.2 |
| LoRA alpha | 16 |
| LoRA scale | 32 |
| LoRA dropout | 0.05 |
| Weight decay | 0 |
| AdamW $\beta$ | (0.9,0.999) |
| Learning rate schedule | Linear |
| Automatic mixed precision | Bf16 |
| Gradient checkpointing | √ |
| Iterations | 1,000 |

**Supplementary Table 47: Hyperparameters for text encoder pre-training in the SimCSE phase.** This phase was conducted on a single 80GB NVIDIA A100 GPU for up to 1,000 iterations.

| Hyperparameter | Value |
|---|---|
| Layers | 32 |
| Heads | 32 |
| Embedding dimension | 4,096 |
| Hidden dimension | 14,336 |
| Position embedding | RoPE |
| Vocabulary size | 128,256 |
| Batch size | 128 |
| Learning rate | 3e-5 |
| SimCSE dropout | 0.2 |
| LoRA alpha | 16 |
| LoRA scale | 32 |
| LoRA dropout | 0.05 |
| Weight decay | 0 |
| AdamW $\beta$ | (0.9,0.999) |
| Learning rate schedule | Linear |
| Automatic mixed precision | Bf16 |
| Gradient checkpointing | √ |
| Iterations | 1,000 |

**Supplementary Table 48: Training hyperparameters for cross-modal pre-training.** Experiments were conducted on four NVIDIA A100 80GB GPUs, utilizing a local batch size of 1,024 per GPU with mixed-precision training for enhanced efficiency. Training proceeded for up to 100 epochs, with the best-performing model checkpoint selected based on the highest cross-modal retrieval recall observed on the pre-training dataset's validation set.

| **Hyperparameter** | **Value** |
|---|---|
| Batch size per GPU | 1,024 |
| Image backbone learning rate | 1e-5 |
| Projection head learning rate | 5e-4 |
| Weight decay | 1e-2 |
| AdamW β | (0.9,0.98) |
| Learning rate schedule | Cosine |
| Warmup ratio | 10% |
| Epochs | 100 |
| Automatic mixed precision | Bf16 |
| Temperature | Learnable |
| Gradient clipping max norm | 0.2 |

**Supplementary Table 49: Data augmentation techniques for 3D CT volumes.**

| Augmentation | Description | Probability | Parameters/Range |
|---|---|---|---|
| Random Crop | Randomly crop a sub-volume from the input. | 1.0 | From (140, 140, 32) to (128, 128, 32) |
| Translation | Randomly shift the 3D data along the X and Y axes. | 1.0 | [−10, 10] pixels |
| Rotation | Randomly rotate the 3D data around the X, Y, and Z axes. | 1.0 | [−10/π ,10/π] radians |
| Scaling | Randomly scale the 3D data along the X and Y axes. | 1.0 | [0.9, 1.1] |
| Horizontal Flip | Flip the 3D data along the X-axis.[a] | 0.5 | - |
| Intensity Scaling | Randomly scale image intensity. | 1.0 | [0.9, 1.1] |
| Intensity Shift | Randomly shift image intensity by adding an offset. | 1.0 | [−0.1, 0.1] |
| Contrast Adjustment | Randomly change image intensity using gamma transformation. | 0.5 | [0.5,2.0] |

a. To maintain semantic consistency in image-text pairs (e.g., preserving left/right anatomical distinctions), horizontal flipping was applied only during fine-tuning for specific downstream tasks and excluded from the pre-training phase.

**Supplementary Table 50: Summary of baseline models, sources, and training protocols.** Clinical diagnosis includes anatomical assessment, diagnostic classification, and survival prediction. Fine-tuned indicates that the model weights were updated on the downstream training set. Frozen indicates evaluation using official pre-trained weights without updates. – indicates not applicable.

| Model | Model type | Source | Downstream tasks | | |
|---|---|---|---|---|---|
| | | | Clinical diagnosis | Retrieval & zero-shot | Report generation |
| ResNet-18 | 3D CNN | Random Init | Fine-tuned | – | – |
| CT-FM | 3D CNN | Public repository[1] | Fine-tuned | – | – |
| CT-CLIP | 3D VLM | Public repository[2] | Fine-tuned | Frozen | – |
| Merlin | 3D VLM | Public repository[3] | Fine-tuned | Frozen | – |
| RadFM | 3D VLM | Public repository[4] | – | – | Frozen |
| CT-CHAT | 3D VLM | Public repository[5] | – | – | Frozen |
| GPT-4o | 2D VLM | Proprietary API[6] | – | – | Frozen |
| MedGemma | 2D VLM | Public repository[7] | – | – | Frozen |
| RenalCLIP | 3D VLM | Self-pre-trained | Fine-tuned | Frozen | Fine-tuned |

[1] https://github.com/project-lighter/CT-FM
[2] https://github.com/ibrahimethemhamamci/CT-CLIP
[3] https://github.com/StanfordMIMI/Merlin
[4] https://github.com/chaoyi-wu/RadFM
[5] https://github.com/ibrahimethemhamamci/CT-CHAT
[6] https://ai.azure.com/catalog/models/gpt-4o
[7] https://huggingface.co/google/medgemma-4b-it

**Supplementary Table 51: Training hyperparameters for radiology report generation (Step 1).** For the initial training step, which focused solely on optimizing the multi-modal projector while the image encoder and LLM remained frozen, a single 80GB NVIDIA A100 GPU was utilized. Training proceeded for up to 20 epochs, with the best-performing model checkpoint selected based on the lowest validation loss observed on the pre-training dataset's validation set.

| **Hyperparameter** | **Value** |
|---|---|
| Batch size | 30 |
| Learning rate | 1e-3 |
| Weight decay | 5e-2 |
| AdamW β | (0.9,0.999) |
| Learning rate schedule | Cosine |
| Warmup ratio | 10% |
| Epochs | 20 |

**Supplementary Table 52: Training hyperparameters for radiology report generation (Step 2).** This fine-tuning step built upon the initial pre-training setup, focusing on optimizing the multi-modal projector and fine-tuning the LLM (BioMistral-7B) using LoRA. Training was conducted on a single 80GB NVIDIA A100 GPU. The maximum sequence length for captions was set to 256, a length chosen as the majority of reports in the dataset were shorter than this.

| **Hyperparameter** | **Value** |
|---|---|
| Layers | 32 |
| Heads | 32 |
| Embedding dimension | 4,096 |
| Hidden dimension | 14,336 |
| Position embedding | RoPE |
| Vocabulary size | 32,000 |
| Max sequence length | 256 |
| Batch size | 30 |
| Projector learning rate | 5e-4 |
| LoRA weights learning rate | 2e-6 |
| LoRA alpha | 4 |
| LoRA scale | 4 |
| LoRA dropout | 0.2 |
| Weight decay | 5e-2 |
| AdamW β | (0.9,0.999) |
| Learning rate schedule | Cosine |
| Warmup ratio | 10% |
| Epochs | 2 |